\documentclass[a4paper,fleqn,usenatbib]{mnras}

\usepackage[T1]{fontenc}
\usepackage{ae,aecompl}
\usepackage[normalem]{ulem}

\usepackage{csquotes}

\usepackage{graphicx}	
\usepackage{amsmath}	
\usepackage{amssymb}	

\usepackage{newtxtext,newtxmath}

\def\gs{\mathrel{\raise0.35ex\hbox{$\scriptstyle >$}\kern-0.6em \lower0.40ex\hbox{{$\scriptstyle \sim$}}}}
\def\ls{\mathrel{\raise0.35ex\hbox{$\scriptstyle <$}\kern-0.6em \lower0.40ex\hbox{{$\scriptstyle \sim$}}}}
\def\app#1#2{\mathrel{\setbox0=\hbox{$#1\sim$}\setbox2=\hbox{\rlap{\hbox{$#1\propto$}}\lower1.1\ht0\box0}\raise0.25\ht2\box2}}
\def\proptosim{\mathpalette\app\relax}

\newcommand{\cpropstoo}{\texttt{CPROPS{\scriptsize TOO}}}
\newcommand{\astrodendro}{\texttt{ASTRODENDRO}}

\title[WISDOM Project -- XII.\ Clump-clump collisions in NGC~404]{WISDOM Project -- XII.\
 Clump properties and turbulence regulated by clump-clump collisions in the  dwarf galaxy NGC~404
}

\author[L.\ Liu et al.]{Lijie Liu,$^{1,2,3}$\thanks{E-mail: ljliu.astro@gmail.com}
Martin Bureau,$^{1,4}$
Guang-Xing Li,$^{5}$ 
Timothy A. Davis,$^{6}$ 
Dieu D. Nguyen,$^{7}$
\newauthor
Fu-Heng Liang,$^{1}$
Woorak Choi,$^{4}$ 
Mark R. Smith$^{1}$
and Satoru Iguchi $^{8,9}$
\\
\\
$^{1}$Sub-department of Astrophysics, Department of Physics,
University of Oxford, Keble Road, Oxford OX1 3RH, UK \\
$^{2}$Cosmic Dawn Center (DAWN) \\
$^{3}$DTU-Space, Technical University of Denmark, Elektrovej 327,
DK2800 Kgs.\ Lyngby, Denmark \\
$^{4}$Yonsei Frontier Lab and Department of Astronomy, Yonsei
University, 50 Yonsei-ro, Seodaemun-gu, Seoul 03722, Republic of Korea \\
$^{5}$South-Western Institute for Astronomy Research, Yunnan
University, Chenggong District, Kunming 650091, P.\ R.\ China \\
$^{6}$School of Physics \& Astronomy, Cardiff University, Queens
 Buildings, The Parade, Cardiff, CF24 3AA, UK \\
 $^{7}$Universit\'{e} de Lyon 1, Ens de Lyon, CNRS, Centre de Recherche Astrophysique de Lyon (CRAL) 
 UMR5574, F-69230 Saint-Genis-Laval, France   \\
 $^{8}$Department of Astronomical Science, SOKENDAI (The Graduate
University of Advanced Studies), Mitaka, Tokyo 181-8588, Japan \\
$^{9}$ National Astronomical Observatory of Japan, National
Institutes of Natural Sciences, Mitaka, Tokyo 181-8588, Japan 
}

\date{Accepted XXX. Received YYY; in original form ZZZ}

\pubyear{2021}

\begin{document}
\label{firstpage}
\pagerange{\pageref{firstpage}--\pageref{lastpage}}
\maketitle

\begin{abstract}
  We present a study of molecular structures (clumps and clouds) in
  the dwarf galaxy NGC~404 using high-resolution
  ($\approx0.86\times0.51$~pc$^2$) Atacama Large
  Millimeter/sub-millimeter Array $^{12}$CO(2-1) observations. We find
  two distinct regions in NGC~404: a gravitationally-stable central
  region (Toomre parameter $Q=3-30$) and a gravitationally-unstable
  molecular ring ($Q\lesssim1$). The molecular structures in the
  central region have a steeper size -- linewidth relation and larger
  virial parameters than those in the molecular ring, suggesting gas
  is more turbulent in the former. In the molecular ring, clumps
  exhibit a shallower mass -- size relation and larger virial
  parameters than clouds, implying density structures and dynamics are
  regulated by different physical mechanisms at different spatial
  scales. We construct an analytical model of clump-clump collisions
  to explain the results in the molecular ring. We propose that
  clump-clump collisions are driven by gravitational instabilities
  coupled with galactic shear, that lead to a population of clumps
  whose accumulation lengths (i.e.\ average separations) are
  approximately equal to their tidal radii. Our model-predicted clump
  masses and sizes (and mass -- size relation) and turbulence energy
  injection rates (and size -- linewidth relation) match the
  observations in the molecular ring very well, suggesting clump-clump
  collisions is the main mechanism regulating clump properties and gas
  turbulence in that region. As expected, our collision model does
  not apply to the central region, where turbulence is likely driven
  by clump migration.
\end{abstract}


\section{Introduction}

%
Dwarf galaxies are low-mass systems that are often different from
present-day large spiral galaxies like the Milky Way (MW). They have
overabundant atomic gas, low metallicities, long gas consumption times
and high gas mass fractions \citep[e.g.][]{fukui2010,schruba2017}.
Molecular structures (clumps and clouds) in dwarf galaxies may also be
quite different from those in the MW and Local Group galaxies, as they
are shaped by different galactic environments
\citep[e.g.][]{hughes2013,hughes2015,colombo2014,sun2018,sun2020}. So
far, detailed observations and analyses of molecular structures in
dwarf galaxies have been limited to a handful of nearby systems (e.g.\
IC~10, \citealt{leroy2006}; SMC, \citealt{muller2010}; NGC~6822,
\citealt{schruba2017}; Henize~2-10, \citealt{imara2019}; J1023+1952,
\citealt{querejeta2021}), as molecular tracers like CO and cold dust
are often too faint to detect at low metallicities
\citep[e.g.][]{leroy2011,elmegreen2013,cormier2017,madden2019,hunter2019}.

Dwarf galaxies may also serve as good analogues of the universe's
earliest galaxies \citep[e.g.][]{motinoflores2021a,
  motinoflores2021b}. It is well known that high-redshift star-forming
galaxies are significantly different from local spiral galaxies, the
former being gravitationally unstable (Toomre parameter $Q\le1$),
distinctly clumpy, gas rich and dynamically hot
\citep[e.g.][]{tacconi2012,forbes2014,genzel2014,swinbank2015,stott2016,tacconi2020,tadaki2018,rizzo2020}.
However, dwarf galaxies share many of these properties, e.g.\ young
ages, low metallicities, high gas-mass fractions and clumpy
morphologies \citep[e.g.][]{motinoflores2021a,
  motinoflores2021b}. Hence, studying dwarf galaxies may provide
unique insights into the evolution of the first galaxies.

Turbulence is a key factor regulating interstellar gas
\citep[e.g.][]{elmegreen2004} and star formation
\citep[e.g.][]{maclow2004,bournaud2010}, but there is an ongoing
debate over the source of the observed turbulence in dwarf galaxies. A
commonly discussed source is stellar feedback, including supernovae
and other stellar processes (e.g.\ winds and outflows). However, many
dwarf galaxies have very low star-formation rate (SFR) densities, and
stellar feedback cannot plausibly provide enough energy in these
systems \citep[e.g.][]{stilp2013}. The magneto-rotational and/or
thermal instabilities do not appear to be sufficient either
\citep[e.g.][]{kim2003,piontek2004,piontek2005,piontek2007,agertz2009}.
Possible drivers of turbulence in dwarf galaxies are thus large-scale
gravitational instabilities 
\citep[e.g.][]{elmegreen2010,elmegreen2011,goldbaum2015,krumholz2016},
that lead to a population of massive cold clumps undergoing mutual
gravitational interactions and merging (i.e.\ collisions). Such
clump-clump collisions can induce significant turbulent motions in the
gas, by extracting energy from the rotational energy of the galaxies
\citep[e.g.][]{agertz2009,tasker2009,bournaud2010,williamson2012,stilp2013,goldbaum2015,goldbaum2016,li2018}.

Clump-clump collisions can also create parsec-scale dense gas
structures \citep[e.g.][]{gammie2001,tasker2009}, regulate clump (or
cloud) properties \citep[e.g.][]{bournaud2010,dobbs2013,li2018} and
trigger star formation events
\citep[][]{hasegawa1994,fukui2021,maeda2021,sano2021}, particularly
high-mass star and star-cluster formation
\citep[e.g.][]{tan2000,myers2009,schneider2012,dobbs2014,kobayashi2018,
  wu2018,henshaw2019}.  Collisions between clumps have
  been clearly observed and identified in many galaxies, including
  high-speed ($\approx20$~km~s$^{-1}$) collisions between clouds in
  the barred galaxy NGC~1300 \citep{maeda2021} and extreme collisions
  ($>100$~km~s$^{-1}$) between clouds in the centre of the Milky Way
  \citep{sormani2019, henshaw2022} and the overlap region of the
  Antennae galaxies \citep{fukui2021}. However, not many analytical
  models have been developed to quantify the effects of these
  collisions. Dedicated numerical simulations have been performed
\citep[e.g.][]{tasker2009,wu2017a,wu2017b,li2018,wu2018}, but they
have mainly focused on MW-type gas discs.

In this paper, we perform statistical analyses of the multiple-scale
molecular structures of the dwarf lenticular galaxy NGC~404,
exploiting high-spatial resolution ($\approx0.86\times0.51$~pc$^2$)
Atacama Large Millimeter/sub-millimeter Array (ALMA) $^{12}$CO(2-1)
observations. The results are confronted with a new simple analytical
model of clump-clump collisions, and a good explanation of the
observational results in the molecular ring of NGC~404 is achieved.
We describe our target and ALMA data in
Section~\ref{sec:data_structural_decomposition}. Basic observational
results are presented in Section~\ref{sec:results}. Our analytic
model of clump-clump collisions and its comparison to observations are
described in Section~\ref{sec:clump-clump_collision_model}. In
Section~\ref{sec:implications_collision-induced_turbulence}, we
describe the implications of the clump-clump collision-induced
turbulence.  We provide a discussion in Section~\ref{sec:discussion}
and our conclusions in Section~\ref{sec:conclusions}.

\section{Data and Structural Decomposition}
\label{sec:data_structural_decomposition}

\subsection{Target}


NGC~404 is the nearest face-on S0 galaxy (distance
$D=3.06\pm0.37$~Mpc; \citealt{karachentsev2002}). It has a large
stellar disc (isophotal diameter $>20$~kpc at a $V$-band surface
brightness $\mu_V=31.5~{\rm mag~arcsec^{-2}}$;
\citealt{tikhonov2003,seth2010,nguyen2017}) that is known to be
dominated by old stellar populations (ages $>10$~Gyr;
\citealt{williams2010}). The centre of NGC~404, however, appears to be
dominated by young stellar populations (ages $\le1$~Gyr;
\citealt{maoz1998,boehle2018}). NGC~404 harbours a low-ionisation
nuclear emission region (LINER; \citealt{schmidt1990}). The dynamical
centre of the galaxy is revealed by a central radio continuum peak
that is spatially coincident with a hard X-ray source
\citep{taylor2015}. The galaxy exhibits bright extended H$\alpha$
emission \citep{nyland2017} and widespread shocks in the nucleus
\citep{boehle2018}. NGC~404 also appears to host an accreting massive
black hole, with a mass $M_{\rm BH}\approx5.7\times10^5$~M$_\odot$
constrained from both stellar and molecular gas kinematics
\citep{davis2020}.

NGC~404 contains an appreciable amount of atomic
($\approx1.5\times10^8$~M$_\odot$; \citealt{delrio2004}) and molecular
($\approx9.0\times10^6$~M$_\odot$; \citealt{taylor2015}) gas.
Approximately $75\%$ of the H{\small I} is located in a nearly face-on
doughnut-shaped distribution, with a central hole that nicely matches
the optical galaxy \citep{delrio2004}. The molecular gas, however, is
located on a much smaller spatial scale ($\approx140\times130$~pc$^2$)
at the centre of the optical galaxy and the H{\small I} hole, as
revealed by low-resolution ($7\farcs0\times7\farcs6$ synthesised beam)
Berkeley-Illinois-Maryland Association (BIMA) radio telescope array
$^{12}$CO(1-0) observations \citep{taylor2015}. The peak of CO(1-0)
emission is spatially-coincident with the dynamical centre of the
galaxy identified by the radio continuum and hard X-ray sources. The
total CO(1-0) flux detected by BIMA is $67.4$~Jy~km~s$^{-1}$
\citep{taylor2015}.
There is tentative evidence that NGC~404 acquired gas $0.5$ --
$1.0$~Gyr ago through a merger with a gas-rich dwarf irregular system
\citep{delrio2004,taylor2015,nguyen2017}.

\subsection{ALMA data}

NGC~404 was observed in the $^{12}$CO(2-1) line ($\approx230$~GHz)
using ALMA for a total of five tracks, three in extended
configurations and two with shorter baselines. This yielded a total
baseline range of $15$ -- $16,200$~m, a field of view (full-width at
half-maximum of the primary beam) of ${\rm FWHM}\approx24\farcs5$ that
extends far beyond the molecular gas disc, and a maximum recoverable
scale of $\approx10\farcs4$ that is much larger than than the largest
single molecular structure identified in this work. See
\citet{davis2020} for more details of the observations. The raw ALMA
data were calibrated using the standard ALMA pipeline and imaged using
Briggs weighting with a robust parameter of $0.5$, yielding a
synthesised beam of ${\rm FWHM}\approx0\farcs058\times0\farcs034$
($\approx0.86\times0.51$~pc$^2$) at a position angle of
$36^\circ$. Continuum emission was detected and subtracted in the $uv$
plane.
The data cube used in this paper is slightly different from that of
\citet{davis2020}, with a channel width of $2$~km~s$^{-1}$, a
root-mean-squared (RMS) noise $\sigma_{\rm rms}=1.13$~mJy~beam$^{-1}$
per channel ($13.1$~K converted to brightness temperature) and spaxels
of $0\farcs02\times0\farcs02$ (thus yielding $\approx5.6$ spaxels per
synthesised beam).

\begin{figure}
  \includegraphics[width=0.95\columnwidth]{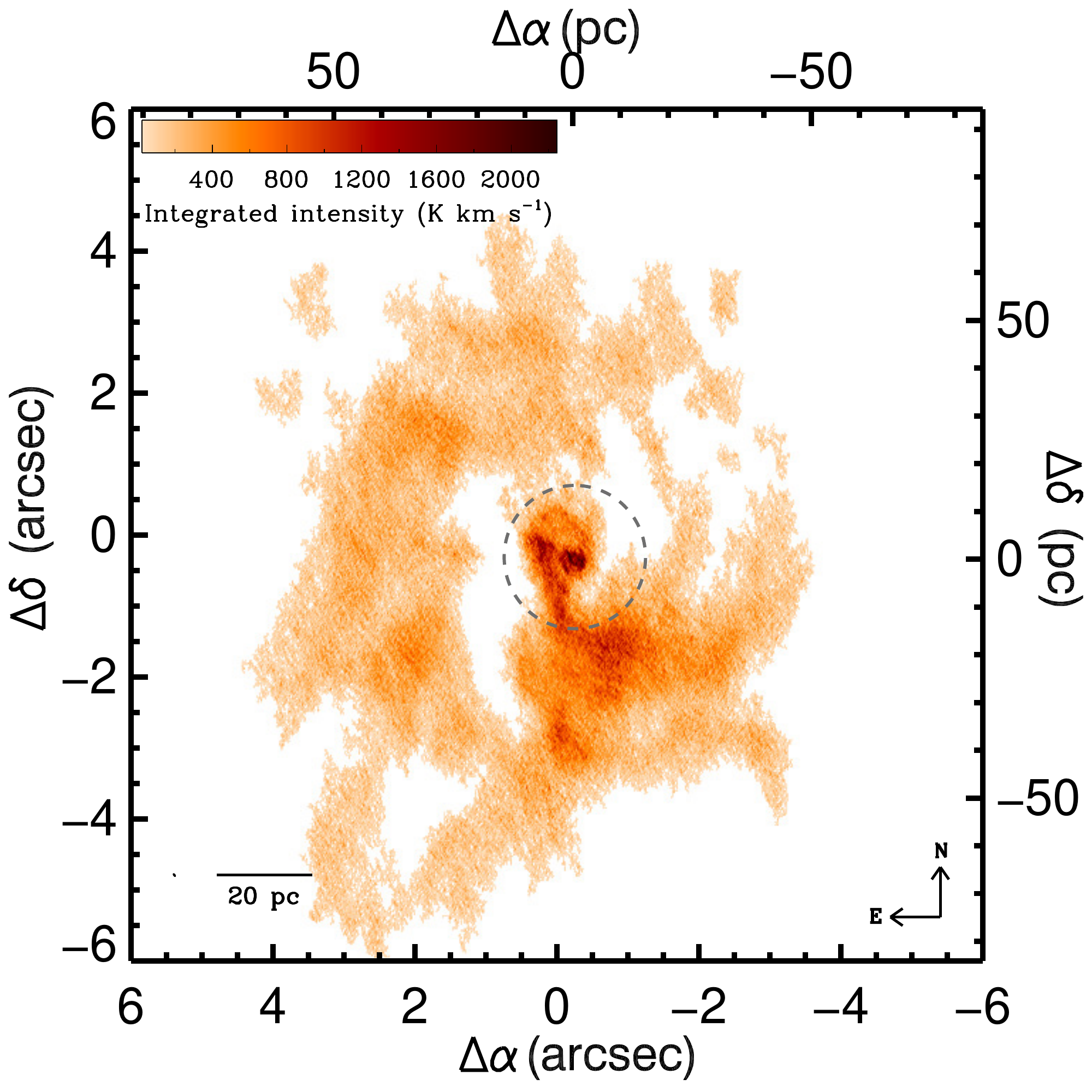}
  \caption{Moment-zero map of the $^{12}$CO(2-1) emission of NGC~404
    derived from our ALMA observations. The synthesised beam
    ($\approx0\farcs058\times0\farcs034$ or
    $\approx0.86\times0.51$~pc$^2$) is shown as a very small black
    ellipse to the left of the scale bar in the bottom-left
    corner. The grey dashed ellipse at a galactocentric distance of
    $15$~pc separates the central region and molecular ring discussed
    in the text.}
  \label{fig:mom0_image}
\end{figure}

Figure~\ref{fig:mom0_image} shows the zeroth-moment
(integrated-intensity) map of our adopted data cube, created using a
smooth-masking moment technique \citep[e.g.][]{dame2011}. To generate
the mask, the data cube was convolved spatially by a Gaussian of width
equal to that of the synthesised beam, Hanning-smoothed in velocity
and clipped at a fixed threshold ($0.3$~$\sigma_{\rm rms}$). The
integrated CO(2-1) intensity is $210$~Jy~km~s$^{-1}$. Comparing our
CO(2-1) integrated intensity to that in CO(1-0) by \citet{taylor2015},
we obtain a CO(2-1)/CO(1-0) line ratio of $\approx0.78$ in temperature
units, similar to the average MW ratio of $0.8$
\citep{carilli2013}. We thus estimate a total molecular gas mass of
$\approx9.4\times10^{6}$~M$_\odot$, derived using our integrated ALMA
CO(2-1) intensity and the CO(2-1)/CO(1-0) line ratio of $0.78$, and
assuming a standard Galactic CO-to-H$_2$ conversion factor
$X_{\rm CO}=2\times10^{20}$~cm$^{-2}$~(K~km~s$^{-1}$)$^{-1}$ (as
despite a low total stellar mass, NGC~404 has approximately solar
ionised gas metallicity; \citealt{bresolin2013,davis2020}).

The ALMA CO(2-1) observations reveal a complex molecular gas
morphology (see Fig.~\ref{fig:mom0_image}; \citealt{davis2020}),
including a central disc/ring within a radius of $\approx8$~pc (centre
${\rm R.A.\,(J2000)}=01^{\rm h}09^{\rm m}27\fs001$ and
${\rm Dec.\,(J2000)}=+35^{\circ}43\arcmin04\farcs942$, inclination
$i=37\fdg1$, position angle $PA=37\fdg2$), a single arm-like feature
with a radius of $\approx8$ -- $15$~pc, and an incomplete
(pseudo-)ring with a radius of $\approx15$ -- $50$~pc ($i=9\fdg3$,
$PA\approx1^\circ$). Hereafter we will refer to the central disc/ring
and arm-like feature as the central region (galactocentric distances
$R_{\rm gal}\le15$~pc), and to the outer incomplete (pseudo-)ring as
the molecular ring ($15<R_{\rm gal}\lesssim50$~pc). The grey dashed
ellipse in Fig.~\ref{fig:mom0_image} separates the two regions. The
central region's kinematic centre is spatially-coincident with the
continuum source detected with very long baseline interferometry
(VLBI; \citealt{nyland2017}), while the molecular ring contains
multiple spatially-resolved molecular structures corresponding to dust
features seen in absorption in {\it Hubble Space Telescope} ({\it
  HST}) images (see e.g.\ Fig.~3 in \citealt{davis2020}).

\subsection{Structural decomposition}
\label{sec:structural_decomposition}

\begin{figure*}
  \includegraphics[width=0.6\textwidth]{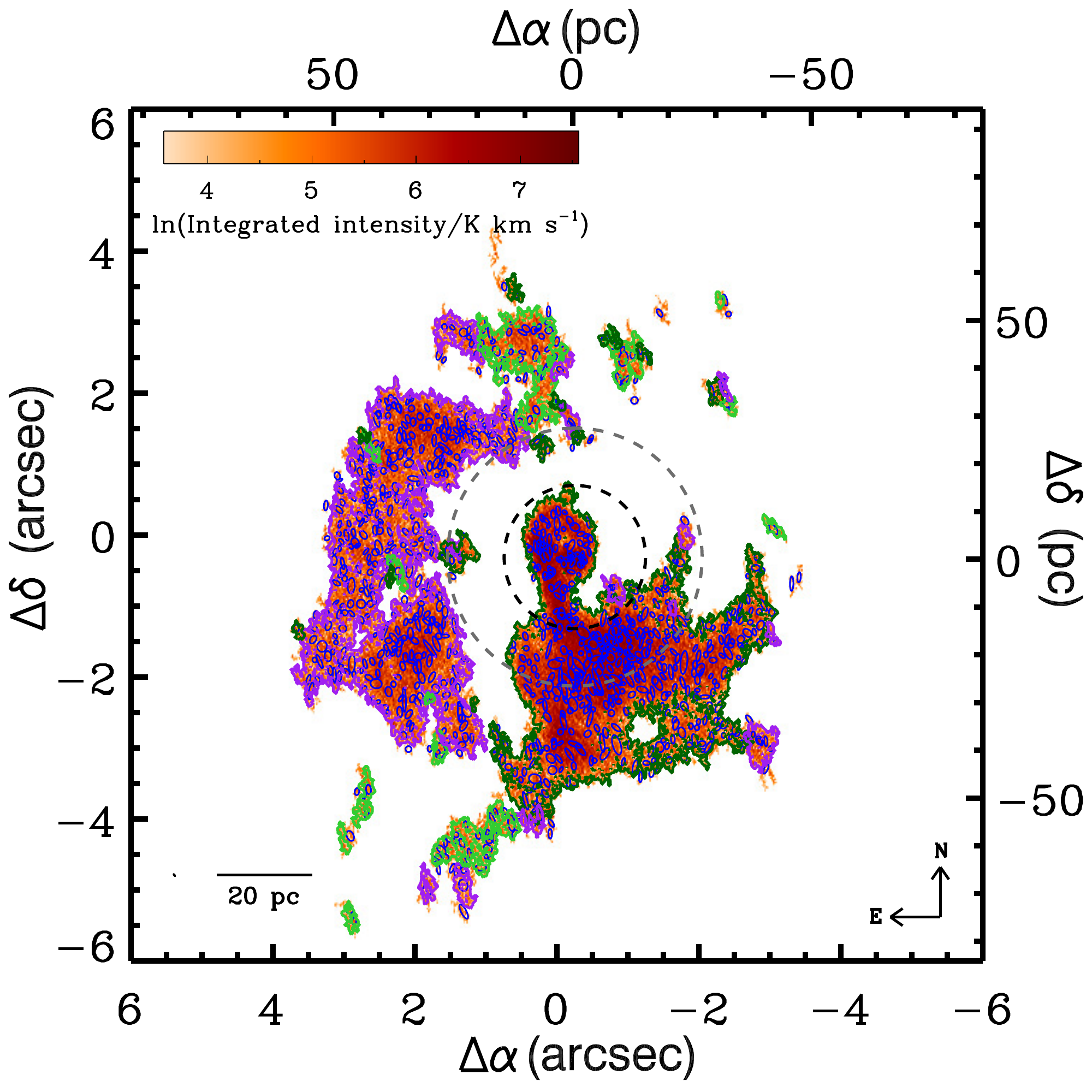}
  \caption{CO(2-1) integrated-intensity map of NGC~404 (colour scale),
    blanking out faint areas using the {\cpropstoo}-generated mask.
    The mask covers pixels with connected emission above
    $1.5$~$\sigma_{\rm rms}$ and at least one channel above
    $3$~$\sigma_{\rm rms}$, where $\sigma_{\rm rms}$ is the RMS noise
    of the data cube. Coloured contours show resolved trunks while
    blue ellipses show resolved leaves (i.e.\ clumps; extrapolated to
    the limit of perfect sensitivity but not corrected for the finite
    angular resolution). The black dashed ellipse at a galactocentric
    distance of $15$~pc separates the central region and molecular
    ring discussed in the text, while the grey dashed ellipse at a
    galactocentric distance of $27$~pc indicates the galactocentric
    distance beyond which the molecular gas disc is no longer
    gravitationally stable (i.e.\ the Toomre parameter $Q\le1$ at
    $R_{\rm gal}\ge27$~pc; see
    Section~\ref{sec:gravitational_instability_onset_collisions} for
    more details). The synthesised beam
    ($\approx0\farcs058\times0\farcs034$ or
    $\approx0.86\times0.51$~pc$^2$) is shown as a very small black
    ellipse to the left of the scale bar in the bottom-left corner.}
  \label{fig:gmc_figure}
\end{figure*}

We use the dendrogram (i.e.\ tree analysis) code {\astrodendro}
described by \citet{rosolowsky2008} to identify
molecular structures in NGC~404. A three-dimensional (3D) mask of
bright emission is initially created using the code {\cpropstoo}
\citep{rosolowsky2006,leroy2015}. All single pixels with brightness
temperatures $T_{\rm b}>3$~$\sigma_{\rm rms}$ ($\approx39$~K) are
first identified, and the pixel list is then expanded to include all
neighbouring pixels with $T_{\rm b}>1.5$~$\sigma_{\rm rms}$
($\approx20$~K). Each discrete region of signal in the mask (i.e.\
each \enquote{island}) is also required to have a minimum projected
diameter of $\approx3$~pc ($\approx100$~spaxels in area),
corresponding to the maximum Jeans length in the outer parts of the
molecular ring (see
Section~\ref{sec:gravitational_instability_onset_collisions}). 
 Fig.~\ref{fig:gmc_figure} shows the
zeroth-moment map of NGC~404 created using this mask. The integrated
CO(2-1) intensity within the masked region is
$\approx151$~Jy~km~s$^{-1}$, $\approx72\%$ of the total integrated
ALMA CO(2-1) intensity.

A hierarchy of molecular structures is then identified using the code
{\astrodendro}. The algorithm first sets a minimum significant
isophotal contour for every region; emission fainter than
$min\_value=1.5$~$\sigma_{\rm rms}$ ($\approx20$~K) is not
characterised. This step has almost no effect, however, as the
{\cpropstoo}-generated mask with a threshold of
$1.5$~$\sigma_{\rm rms}$ has already been applied to the data cube.
The algorithm then identifies local maxima of brightness temperatures
at least $min\_delta=1.5$~$\sigma_{\rm rms}$ ($\approx20$~K) above the
merger level with any other local maximum (i.e.\ above the highest
contour/isophotal level enclosing a pair of local maxima). The minimum
area ($min\_area$) and minimum number of pixels ($min\_npix$) that
each local maximum should span are also specified. Here we require
$min\_area=min\_npix=6$, so that each local maximum spans at least one
synthesised beam.

The isosurfaces surrounding local maxima are then categorised into
different types of structures: leaves, branches and trunks. Leaves are
the smallest structures and do not contain any sub-structure (i.e.\
they contain only one local maximum), trunks are the largest
contiguous structures (i.e.\ they have no parent structure), while
branches are intermediate in scale (and can have both sub-structures,
i.e.\ branches and leaves, and parent structures, i.e.\ branches and
trunks). Trunks therefore do not overlap any other trunks, and leaves
do not overlap any other leaves \citep{wong2019}, but both trunks and
branches can split up into branches and/or leaves, thus allowing
hierarchical structures to be adequately identified and represented
\citep{rosolowsky2008}. The algorithm identifies $3626$ molecular
structures in NGC~404: $50$ trunks, $1642$ branches and $1934$
leaves. The trunks in our catalogue recover almost all of the
{\cpropstoo}-masked CO(2-1) emission, while the leaves contain a total
of $\approx83$~Jy~km~s$^{-1}$ or $\approx55\%$ of this emission.

Once a hierarchy of structures has been identified by {\astrodendro},
we use our modified version of {\cpropstoo} \citep[see][]{liu2021} to
calculate the properties of the structures identified. The code
{\cpropstoo} is chosen because it attempts to correct the measured
quantities for the finite sensitivity of the observations,
extrapolating sizes, linewidths and luminosities to their infinite
signal-to-noise ratio ($S/N$) equivalents (i.e.\ to a brightness
temperature at the edge of the structures identified
$T_{\rm edge}=0$~K). {\cpropstoo} also \enquote{deconvolves} in two
dimensions the measured (and extrapolated) sizes, to account for the
finite size of the synthesised beam, thus roughly correcting for
possible resolution biases.

Table~\ref{tab:gmc_properties} lists the position and properties of
each identified structure: structure size (radius) $R_{\rm c}$,
one-dimensional (1D) observed linewidth (velocity dispersion)
$\sigma_{\rm obs,los}$, CO(2-1) luminosity $L_{\rm CO(2-1)}$, gaseous
mass $M_{\rm c}$ (assuming a CO(2-1)/CO(1-0) line ratio of 0.78 and a
standard Galactic CO-to-H$_2$ conversion factor
$X_{\rm CO}=2\times10^{20}$~${\rm cm^{-2}~(K~km~s^{-1})^{-1}}$;
referred to as $M_{\rm gas}$ in \citealt{liu2021}) and deprojected
galactocentric distance $R_{\rm gal}$. The uncertainties of the
measured properties are estimated via a bootstrapping technique. For a
detailed definition of each property and its uncertainty, refer to
\citet{liu2021}. About $50\%$ ($953/1934$) of the identified leaves,
$\approx97\%$ ($1592/1642$) of the identified branches and
$\approx94\%$ ($47/50$) of the identified trunks are resolved, i.e.\
have deconvolved diameters larger than or equal to the synthesised
beam in two dimentions and deconvolved velocity widths at least half
of one velocity channel \citep{donovanmeyer2013}.
Figure~\ref{fig:gmc_figure} shows the resolved leaves (blue ellipses)
and trunks (coloured contours) of NGC~404 overlaid on the
zeroth-moment map created with the {\cpropstoo}-generated mask.


\begin{table*}
  \centering
  \caption{Properties of the dendrogram-defined structures of NGC~404.}
  \label{tab:gmc_properties}
  \resizebox{\textwidth}{!}{%
    \begin{tabular}{rccccccccc}
      \hline\hline
      ID & R.A.\ (2000) & Dec.\ (2000) & $V_{\rm LSR}$ & $R_{\rm c}$ & $\sigma_{\rm obs,los}$ & $L_{\rm CO(2-1)}$ & $M_{\rm c}$ & $R_{\rm gal}$ & Structure \\
         & (h:m:s) & ($^\circ:^\prime:^{\prime \prime}$) & (${\rm km~s^{-1}}$) & (pc) & (${\rm km~s^{-1}}$) & ($10^4~{\rm K~km~s^{-1}~pc^2}$) & ($10^5~{\rm M_\odot}$) & (pc) & \\
      \hline\hline
      $1$ & 01:09:27.00 & 35:43:04.85 & $-111.1$ & $\dots$ & $\phantom{1}2.0\pm 0.9$ & $\phantom{1}\phantom{1}0.03\pm0.01$ & $\phantom{1}0.02\pm0.01$ & $\phantom{1}1.5$ & L \\
      $2$ & 01:09:26.98 & 35:43:03.64 & $-\phantom{1}58.0$ & $24.8\pm\phantom{1}3.3$ & $10.0\pm 1.2$ & $127.5\phantom{1}\pm1.9\phantom{1}$ & $70.1\phantom{1}\pm1.1\phantom{1}$ & $19.9$ & B  \\
      $3$ & 01:09:26.97 & 35:43:03.60 & $-\phantom{1}57.8$ & $26.7\pm\phantom{1}6.4$ & $10.4\pm 1.8$ & $139.4\phantom{1}\pm1.9\phantom{1}$ & $76.7\phantom{1}\pm1.0\phantom{1}$ & $20.8$ & B  \\
       $\dots$ & $\dots$ & $\dots$ & $\dots$ & $\dots$ & $\dots$ & $\dots$ & $\dots$ & $\dots$ & $\dots$ \\
      $3626$ & 01:09:27.01 & 35:43:04.97 & $\phantom{-11}7.2$ & $\dots$ & $\phantom{1}1.9\pm1.0$ & $\phantom{1}\phantom{1}0.03\pm0.02$ & $\phantom{1}0.01\pm0.01$ & $\phantom{1}0.9$ & L \\ 
      \hline\hline 
    \end{tabular}
  } \raggedright Notes.\ -- Measurements of $R_{\rm c}$ assume
  $\eta=1.91$. Measurements of $M_{\rm c}$ assume a CO(2-1)/CO(1-0)
  line ratio of $0.78$ in temperature units and a standard Galactic
  conversion factor
  $X_{\rm CO}=2\times10^{20}$~cm$^{-2}$~(K~km~s$^{-1}$)$^{-1}$
  (including the mass contribution from helium). Structure codes: L -
  leaf, B - branch, T - trunk. 
  Measurements of $R_{\rm gal}$ adopt a fixed 
 position angle ${\rm PA} = 1^\circ$ and inclination angle $i = 9\fdg3$.
  The uncertainty of the adopted
  distance $D$ to NGC~404 was not propagated through the tabulated
  uncertainties of the measured quantities. This is because an error
  on the distance to NGC~404 translates to a systematic (rather than
  random) scaling of some of the measured quantities (no effect on the
  others), i.e.\ $R_{\rm c}\propto D$, $L_{\rm CO(2-1)}\propto D^2$,
  $M_{\rm c}\propto D^2$ and $R_{\rm gal}\propto
  D$. Table~\ref{tab:gmc_properties} is available in its entirety in
  machine-readable form in the electronic edition.
\end{table*}

In this paper, we will refer to single centrally-concentrated
structures as \enquote{clumps}, and those objects with more complex
structures as \enquote{clouds}. We therefore treat dendrogram-defined
leaves as clumps, and both branches and trunks as clouds, because by
definition leaves can not contain any sub-structure while both
branches and trunks must harbour multiple sub-structures. Here,
\enquote{sub-structure} is understood to be emission within a contour
(in a 3D data cube) of $min\_delta=1.5$~$\sigma_{\rm rms}$ that has a
minimum area of one synthesised beam. Hereafter the subscript
\enquote{c} will denote both clumps and clouds, while the subscripts
\enquote{clump} and \enquote{cloud} will refer to only clumps and only
clouds, respectively.

\section{Results}
\label{sec:results}

\subsection{Mass function of clumps}
\label{sec:mass_function_clumps}

In this section, we analyse the mass distribution functions of
independent structures, i.e.\ clumps, in NGC~404. The distribution of
clumps by mass provides important information not only on the
mechanisms that influence clump formation, evolution and destruction
\citep[e.g.][]{rosolowsky2005,colombo2014,faesi2016}, but also on the
factors regulating star formation. Indeed, the SFR diversity across
galactic discs could simply originate from diverse clump populations
\citep{kobayashi2017,kobayashi2018}.

The differential clump mass distribution function can be characterised
by a power-law
\begin{equation}
  \label{eq:differential_mass_distribution_function}
  dN_{\rm clump}(M_{\rm clump})/dM_{\rm clump}\propto M_{\rm clump}^{\gamma_{\rm clump}}
\end{equation} 
(or equivalently
$dN_{\rm clump}(M_{\rm clump})/d\log M_{\rm clump}\propto M_{\rm
  clump}^{\gamma_{\rm clump}+1}$), where
$N_{\rm clump}(M_{\rm clump})$ is the number of clumps with masses
greater than $M_{\rm clump}$ and $\gamma_{\rm clump}$ is the power-law
index, usually compared to an index of $-2$. If
$\gamma_{\rm clump}<-2$, most of the molecular mass is found in
low-mass clumps, while if $\gamma_{\rm clump}>-2$, most of the
galaxy's molecular mass is located in high-mass
clumps.

\begin{figure*}
  \includegraphics[width=0.98\textwidth]{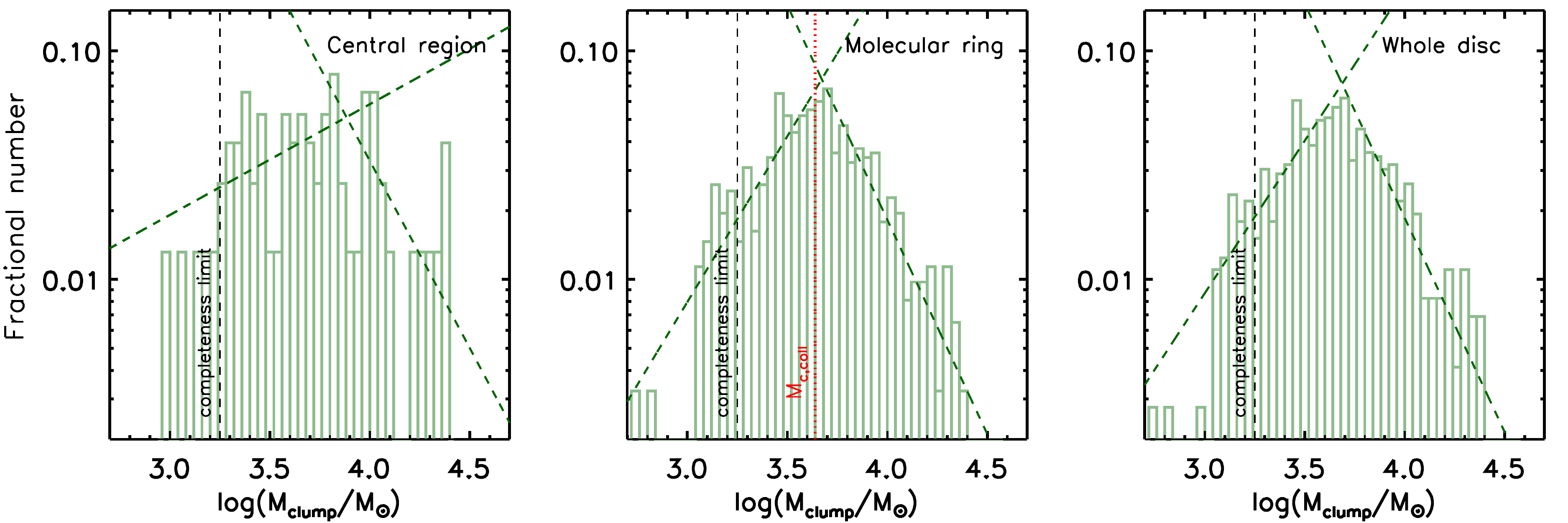}
  \caption{Normalised differential mass distribution function of the
    resolved clumps in the central region, molecular ring and whole
    disc of NGC~404, respectively. The two power-laws best fitting the
    differential mass distribution are overlaid as green dashed lines
    in each panel. Our mass completeness limit is indicated by a black
    vertical dashed line in each panel. The red dotted line in the
    middle panel indicates our model-predicted turn-over mass in the
    molecular ring ($M_{\rm c,coll}$; see
    Sections~\ref{sec:clump_mass_collision_model} and
    \ref{sec:clump_mass_function_collision_model} for more
    details).}
  \label{fig:clump_mass_distribution}
\end{figure*}

Figure~\ref{fig:clump_mass_distribution} shows the normalised
differential mass distribution functions of the resolved clumps in the
central region, molecular ring and whole disc of NGC~404. The black
vertical dashed line in each panel indicates the mass completeness
limit $M_{\rm comp}=1.8\times10^3$~M$_\odot$, estimated from the
minimum resolved clump (gaseous) mass
($M_{\rm min}\approx600$~M$_\odot$) and the observational sensitivity
$\delta_{\rm M}=120$~M$_\odot$, i.e.\
$M_{\rm comp}\equiv M_{\rm min}+10\delta_{\rm M}$. Here the
contribution to the mass due to noise $\delta_{\rm M}$ is estimated by
multiplying our RMS column density sensitivity limit of
$\approx250$~M$_\odot$~pc$^{-2}$ (derived from our RMS noise
$\sigma_{\rm rms}$) by the synthesised beam area of
$\approx0.49$~pc$^2$.

We find ``turn-overs'', that is break points in the single power-law
functions, in all the normalised differential mass distributions,
always at the same clump mass $M_{\rm clump}\approx4000$~M$_\odot$. As
this turn-over mass is much larger than our estimated mass
completeness limit, it is most likely real and thus informs on the
underlying formation and destruction of clumps. We therefore fit the
normalised differential mass distributions with two separate
power-laws, one for the high-mass regime
$M_{\rm clump}\gtrsim4000$~M$_\odot$, with a power-law index of
$\gamma_{\rm clump}^+$, and the other for the low-mass regime
$M_{\rm comp}\lesssim M_{\rm clump}\lesssim4000$~M$_\odot$, with a
power-law index of $\gamma_{\rm clump}^-$. All normalised differential
mass distribution functions can be well described by such broken
power-laws, and the best-fitting power-law indices of each region are
listed in Table~\ref{tab:results_summary}. We note that the power-law
index of the differential clump mass distribution function in the
high-mass regime $\gamma_{\rm clump}^{+}$ is consistent with the
best-fitting power-law indices of the cumulative clump mass
distribution function (see
Appendix~\ref{appendix:cumulative_mass_function} for the normalised
cumulative mass distribution functions of the resolved clumps in the
different regions of NGC404).

\begin{table*}
  \centering
  \caption{Summary of the observational results of NGC~404.}
  \label{tab:results_summary}
  \resizebox{0.9\textwidth}{!}{%
    \begin{tabular}{lccc}
      \hline\hline
      & Central region & Molecular ring & Whole disc \\
      \hline\hline 
      Mass functions of clumps & $\gamma^{-}_{\rm clump}=-0.52\pm0.49$ 
          & $\gamma^{-}_{\rm clump}=\phantom{-}0.37\pm0.08$
          & $\gamma^{-}_{\rm clump}=\phantom{-}0.25\pm0.19$ \\
          & $\gamma^{+}_{\rm clump}=-2.63\pm0.49$ 
          & $\gamma^{+}_{\rm clump}=-2.87\pm0.13$
          & $\gamma^{+}_{\rm clump}=-2.67\pm0.16$ \\
          \hline
      Size -- linewidth relation & $\sigma_{\rm obs,los}=(3.24\pm0.08)\,R_{\rm c}^{0.82\pm0.11}$
          & $\sigma_{\rm obs,los}=(2.95\pm0.13)\,R_{\rm c}^{0.30\pm0.03}$ 
          & $\sigma_{\rm obs,los}=(3.02\pm0.15)\,R_{\rm c}^{0.30\pm0.04}$ \\
          \hline
      Mass -- size relation & $\phantom{_{\rm lump}}D_{\rm m,c}=2.27\pm0.10$
          & $\phantom{_{\rm lump}}D_{\rm m,c}=2.12\pm0.01$ 
          & $\phantom{_{\rm lump}}D_{\rm m,c}=2.11\pm0.01$ \\
          & $D_{\rm m,clump}=2.07\pm0.16$
          & $D_{\rm m,clump}=1.63\pm0.04$ 
          & $D_{\rm m,clump}=1.65\pm0.04$ \\
          & $\phantom{.}D_{\rm m,cloud}=2.22\pm0.10$
          & $\phantom{.}D_{\rm m,cloud}=2.06\pm0.01$ 
          & $\phantom{.}D_{\rm m,cloud}=2.04\pm0.01$ \\
          \hline
      Virial parameter & $\phantom{_{\rm lump}}\langle\alpha_{\rm vir,c}\rangle=1.35\pm0.13$  
         & $\phantom{_{\rm lump}}\langle\alpha_{\rm vir,c}\rangle=0.67\pm0.10$ 
         & $\phantom{_{\rm lump}}\langle\alpha_{\rm vir,c}\rangle=0.73\pm0.10$ \\
         & $\langle\alpha_{\rm vir,clump}\rangle=1.52\pm0.11$  
         & $\langle\alpha_{\rm vir,clump}\rangle=1.82\pm0.07$ 
         & $\langle\alpha_{\rm vir,clump}\rangle=1.78\pm0.05$ \\
         & $\phantom{.}\langle\alpha_{\rm vir,cloud}\rangle=1.14\pm0.12$  
         & $\phantom{.}\langle\alpha_{\rm vir,cloud}\rangle=0.41\pm0.02$ 
         & $\phantom{.}\langle\alpha_{\rm vir,cloud}\rangle=0.41\pm0.02$ \\
         \hline
      $M_{\rm c} - \alpha_{\rm vir,c}$ relation  & --
         & $\phantom{_{\rm lump}}\alpha_{\rm vir,c}\propto M_{\rm c}^{-0.27\pm0.01}$
         & $\phantom{_{\rm lump}}\alpha_{\rm vir,c}\propto M_{\rm c}^{-0.27\pm0.01 }$ \\
         & --
         & $\alpha_{\rm vir,clump}\propto M_{\rm clump}^{-0.33\pm0.04}$
         & $\alpha_{\rm vir,clump}\propto M_{\rm clump}^{-0.33\pm0.03}$ \\
         & --
         & $\phantom{.}\alpha_{\rm vir,cloud}\propto M_{\rm cloud}^{-0.24\pm0.03}$
         & $\phantom{.}\alpha_{\rm vir,cloud}\propto M_{\rm cloud}^{-0.25\pm0.01}$ \\
         \hline\hline 
    \end{tabular}
  }\raggedright \newline Notes.\ -- The subscript ``c'' denotes both
  clumps and clouds, while the subscripts ``clump'' and ``cloud''
  refer to only clumps and only clouds, respectively. The power-law
  indices $\gamma_{\rm clump}^{+}$ and $\gamma_{\rm clump}^{-}$ are
  for the differential mass distribution functions of the clumps in
  the high-mass ($M_{\rm clump}\gtrsim4000$~M$_\odot$) and low-mass
  ($M_{\rm comp}\lesssim M_{\rm clump}\lesssim4000$~M$_\odot$) regime,
  respectively.
\end{table*}

The mass distribution functions of the NGC~404 clumps are unusually
steep in the high-mass regime ($\gamma_{\rm clump}^{+}\approx-2.8$ --
$-2.6$), much steeper (i.e.\ more negative) than those of MW clumps
and clouds ($\gamma_{\rm c}\approx-1.7$ -- $-1.5 $;
\citealt{simon2001,mckee2007,rice2016}) and Local Group galaxy clouds
($\gamma_{\rm cloud}=-2.0$ -- $-1.7$;
\citealt{fukui2001,leroy2006,wong2011,faesi2016}; except for M~33
clouds with $\gamma_{\rm cloud}\approx-2.5$;
\citealt{blitz2006}). They are also slightly steeper than those of the
clouds in the early-type galaxies (ETGs) NGC~4526 and NGC~4429
($\gamma_{\rm cloud}\approx-2.3$ -- $-2.1$; \citealt{utomo2015,
  liu2021}). Interestingly, the steep power-law indices
$\gamma_{\rm clump}^{+}$ of the NGC~404 clumps are similar to that of
the massive end of the observed stellar initial mass function
($\gamma\approx-2.4$; \citealt{salpeter1955,chabrier2003}).

The best-fitting power-law indices in the low-mass regime are much
larger ($\gamma_{\rm clump}^{-}\approx-0.5$ -- $0.4$).
The number of clumps in NGC~404 thus seems to increase with increasing
mass when $M_{\rm clump}$ is smaller than the turn-over mass
($\approx4000$~M$_\odot$), and to decrease sharply thereafter.

Overall, the clump mass distribution functions of NGC~404 seem to
favour the formation of clumps with a mass around that of the
turn-over mass ($\approx4000$~M$_\odot$), and we shall discuss the
implications of this observed turn-over mass in
Section~\ref{sec:clump_mass_function_collision_model}.

\subsection{Multiple-scale size -- linewidth relations}
\label{sec:multiple-scale_size-linewidth}

The empirical relation between size and linewidth has become the
yardstick against which to compare studies of clumps and clouds in the
MW and other galaxies \citep[e.g.][]{bolatto2008}. It is of
fundamental importance because it can be interpreted as a signature of
turbulent motions within molecular structures
\citep[e.g.][]{rosolowsky2005}.
As a dendrogram approach allows us to identify a full hierarchy of
molecular structures, we are able to investigate whether the size --
linewidth relation (and thus turbulence) is universal from the
large-scale ISM to its smallest and densest structures
\citep[e.g.][]{shetty2012}.

\begin{figure*}
  \includegraphics[width=0.98\textwidth]{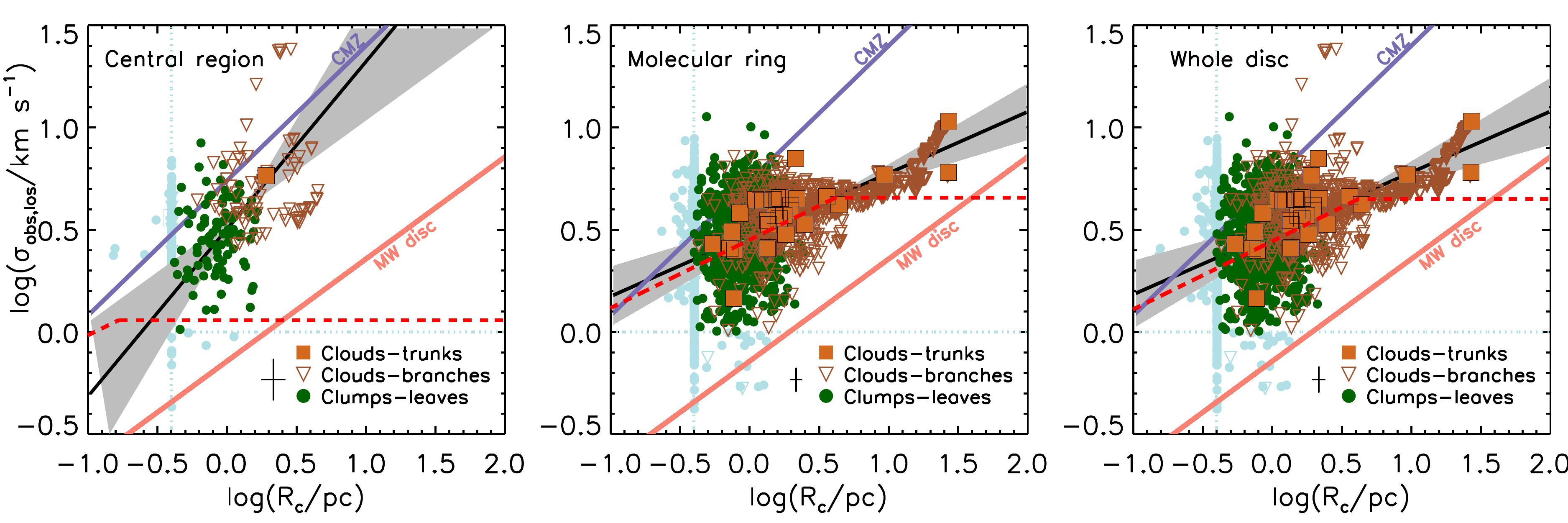}
  \caption{Size -- linewidth relation of the molecular structures in
    the central region, molecular ring and whole disc of NGC~404,
    respectively. Resolved clumps, i.e.\ leaves, are shown as filled
    green circled and resolved clouds, i.e.\ branches and trunks, are
    shown as open brown triangles and filled brown squares,
    respectively. Blue symbols denote unresolved structures. The
    horizontal and vertical blue dotted lines indicate our resolution
    limit of $1$~km~s$^{-1}$ (i.e.\ half the channel width) and
    $0.40$~pc (i.e.\ half the synthesised beam), respectively. In each
    panel, the black solid line with associated shading shows the
    best-fitting power-law relation of all resolved molecular
    structures with $1$~$\sigma$ confidence intervals, while the red
    dashed line shows the $R_{\rm c}$ -- $\sigma_{\rm obs,los}$
    relation predicted by our cloud-cloud collision model (see
    Section~\ref{sec:turbulence_collision_model}). The peach and
    purple solid lines show the size -- linewidth relation of the MW
    disc \citep{solomon1987} and MW central molecular zone
    (\citealt{kauffmann2017}), respectively. The typical uncertainty
    is shown as a black cross to the left of the legend in the
    bottom-right corner of each panel. }
  \label{fig:size_linewidth_relation}
\end{figure*}

Figure~\ref{fig:size_linewidth_relation} presents the size --
linewidth relations of the molecular structures in the central region,
molecular ring and whole disc of NGC~404. Clumps, i.e.\ leaves, are
shown as filled green circles. Clouds, i.e.\ branches and trunks, are
shown as open brown triangles and filled brown squares,
respectively. The molecular structures identified in NGC~404 span a
relatively large dynamic range, with sizes spanning about two orders
of magnitude and velocity dispersions spanning over one order of
magnitude. We note that the observed velocity dispersions
$\sigma_{\rm obs,los}$ shown in Fig.~\ref{fig:size_linewidth_relation}
should be almost exclusively comprised of turbulent motions. Following
the method of \citet{liu2021} (i.e.\ comparing $\sigma_{\rm obs,los}$
with the gradient-subtracted velocity dispersion measure
$\sigma_{\rm gs,los}$), we found no evidence of a significant
contribution of galactic rotation at the clump and cloud scales (i.e.\
$\sigma_{\rm obs,los}\approx\sigma_{\rm gs,los}$ for most clumps and
clouds). Besides, as the galaxy is nearly face-on ($i\approx9\degr$ at
$R_{\rm gal}\ge15$~pc; \citealt{davis2020}), contamination by
(orbital) motions driven by the background galaxy potential in the
disc plane ($\sigma_{\rm gal,r}$ radially and $\sigma_{\rm gal,t}$
azimuthally; see \citealt{liu2021}) should be negligible for most
molecular structures. The use of the gradient-subtracted velocity
dispersion $\sigma_{\rm gs,los}$ is therefore unnecessary in this
work.

The best-fitting size -- linewidth relation of each region is listed
in Table~\ref{tab:results_summary}. The molecular structures in the
central region exhibit an unusually steep size -- linewidth relation
with a slope of $0.82\pm0.11$, while those in the molecular ring have
a much shallower relation with a slope of $0.30\pm0.03$. This trend is
consistent with that in the MW, where the central molecular zone (CMZ)
clouds have a size -- linewidth relation steeper than that of clouds
elsewhere in the disc (slope $\approx0.7$ compared to $\approx0.5$;
e.g.\ \citealt{oka2001,kauffmann2017}). The molecular structures in
the central region of NGC~404 also generally have velocity dispersions
larger than those of similarly-sized structures in the molecular
ring. This is again similar to the behaviour observed in the MW, where
at a given size the clumps/clouds in the CMZ have velocity dispersions
larger than those of clumps/clouds in the disc
\citep[e.g.][]{oka2001,kauffmann2017}.

It is worth noting that the power-law index found for the molecular
ring ($0.30\pm0.03$) is similar to that of the \citet{kolmogorov1941}
law of incompressible turbulence, whereby
$\sigma_{\rm obs,los}\propto R_{\rm c}^{1/3}$ based on a constant
energy transmission rate in the turbulent cascade. However, the much
steeper slope of $0.82\pm0.11$ is found in the central region,
suggesting that the turbulence is highly compressible in this area,
and the energy transmission through decreasing spatial scales is no
longer conservative with kinetic energy also being expended to shock
and/or compress the gas
\cite[e.g.][]{maclow1999,federrath2013,cen2021}.

While a single power-law seems to fit well the size -- linewidth
relation observed in the central region, the same relation appears to
break down in the molecular ring at a scale of $R_{\rm c}\approx3$~pc,
where a flattening (i.e.\ a turn-over) is observed. The scatter around
the $R_{\rm c}$ -- $\sigma_{\rm obs,los}$ is also much larger at
scales below this turn-over scale (primarily clumps) than above
(primarily clouds). The flattening and scatter of the $R_{\rm c}$ --
$\sigma_{\rm obs,los}$ relation suggest a turbulence driving scale of
$\approx3$~pc in the molecular ring. Indeed, the velocity dispersion
is expected to remain constant at scales larger than that driving the
turbulence \citep[e.g.][]{blitz2006}. The flattening at
$R_{\rm c}\approx3$~pc is not observed in the central region, but
there are very few structures (all clouds) larger than this, and as we
will argue later (see Section~\ref{sec:turbulence_collision_model}) it
is likely that the mechanism responsible for driving the turbulence in
that region is different.


Overall, the molecular structures of NGC~404 exhibit complex size --
linewidth relations, that vary strongly between the central region and
molecular ring. It seems that molecular gas is more turbulent in the
central region, and thus different sources of turbulence may be
present in the two regions. These will be discussed in
Section~\ref{sec:turbulence_collision_model}.

\subsection{Multiple-scale mass -- size relations}
\label{sec:multiple-scale_mass-size}

A plot of $\log(M_{\rm c})$ versus $\log(R_{\rm c})$ often provides a
useful way to characterise the density structure
\citep[e.g.][]{kauffmann2010} and fractal nature
\citep[e.g.][]{mandelbrot1983,kritsuk2013} of the ISM. Assuming all
molecular structures have a power-law mass volume density radial
profile $\rho(r)\propto r^{-k}$, the total mass within a given radius
is then $M(r)\propto r^{3-k}$. Thus, the lower the slope of the mass
-- size relation, the more centrally condensed the structure. The
power-law index of the $M_{\rm c}$ -- $R_{\rm c}$ relation may also
reflect the fractal dimension $D_{\rm m}$ of the
sub-structures\footnote{The fractal dimension is an index
  characterising fractal patterns or sets, and quantifying their
  complexity as the ratio of the change in detail to the change in
  scale. In fractal geometry, a self-similar shape may be split up
  into $N$ parts, obtainable from the whole by a scaling factor $r$.
  The fractal dimension can then be defined mathematically as
  $D_{\rm m}\equiv-\log N/\log r$ \citep{mandelbrot1983}, and it need
  not be an integer.}  \citep{mandelbrot1983}. Indeed, as our
identified molecular structures can be treated as sub-structures of a
fractal ISM (to zeroth-order approximation; \citealt{romanduval2010}),
it is useful to adopt the fractal dimension $D_{\rm m}$ to describe
how fully the molecular structures fill the space of the underlying
ISM,
i.e.\ the degree of porosity of the ISM.

\begin{figure*}
  \includegraphics[width=0.98\textwidth]{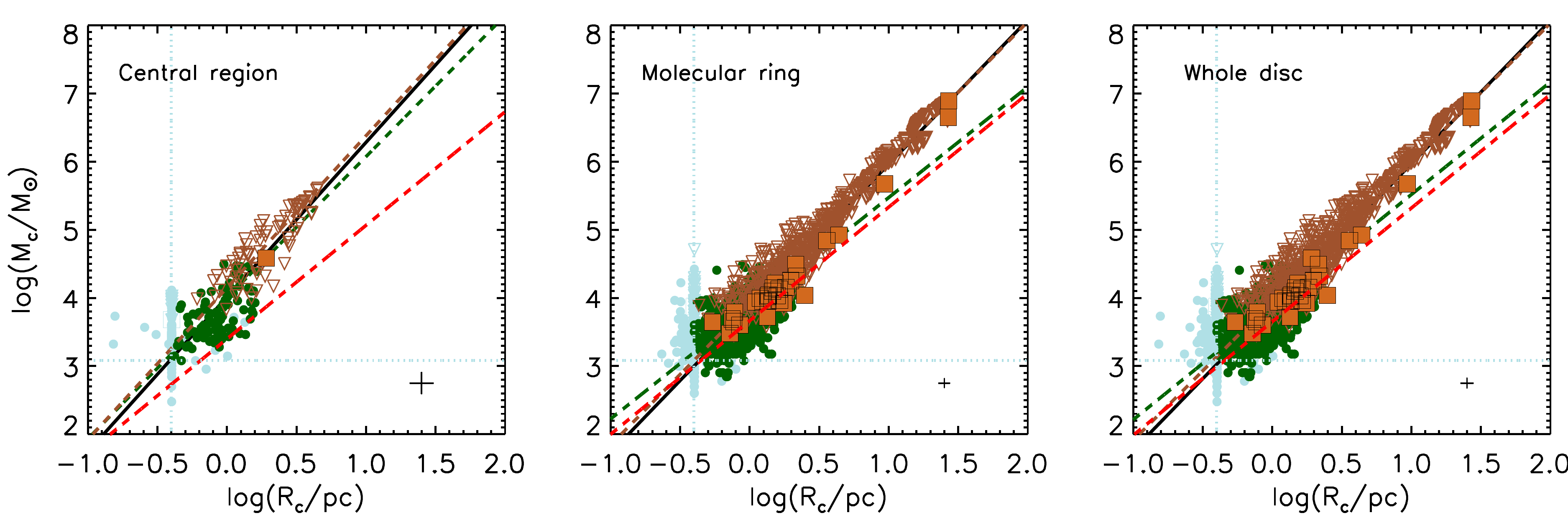}
  \caption{Mass -- size relation of the molecular structures in the
    central region, molecular ring and whole disc of NGC~404,
    respectively. Symbols are as for
    Fig.~\ref{fig:size_linewidth_relation}. The vertical blue dotted
    line indicates our spatial resolution limit of $0.40$~pc (i.e.\
    half the synthesised beam), while the horizontal blue dotted line
    indicates our mass detection limit
    ($\log(10\delta_{\rm M}/{\rm M}_\odot)=3.08$; see
    Section~\ref{sec:mass_function_clumps}). In each panel, the solid
    black, green and brown lines show the best-fitting power-law
    relation of all resolved molecular structures, only resolved
    clumps (i.e.\ leaves) and only resolved clouds (i.e.\ branches and
    trunks), respectively, while the red dashed line shows the
    $M_{\rm c}$ -- $R_{\rm c}$ relation of clumps predicted by our
    cloud-cloud collision model (see Eq.~\ref{eq:predicted_mass_size}
    and Section~\ref{sec:implications_mass-size}).}
  \label{fig:mass_size_relation}
\end{figure*}

Figure~\ref{fig:mass_size_relation} shows the mass -- size relations
of the molecular structures of the central region, molecular ring and
whole disc of NGC~404, all very tight. The best-fitting mass -- size
relations of each region are listed in
Table~\ref{tab:results_summary}. Overall, the power-law index for all
molecular structures in the whole disc of NGC~404,
$D_{\rm m}=2.11\pm0.01$, is similar to (although statistically
inconsistent with) that derived for MW clouds
($D_{\rm m}=2.39\pm 0.01$; \citealt{romanduval2010}) and is within the
range inferred from observations of nearby galaxies
($1.6\le D_{\rm m}\le3.0$;
\citealt{kauffmann2010,lombardi2010,urquhart2013,zhang2017,li2020}).
We note that $D_{\rm m}\approx2.1$ implies a molecular gas mass volume
density $\rho_{\rm gas}\propto R_{\rm c}^{-0.9}$ and mass surface
density $\Sigma_{\rm gas}\propto R_{\rm c}^{0.1}$. The latter is the
usual nearly-constant mass surface density, but the former indicates
inhomogeneous mass volume densities for the resolved molecular
structures of NGC~404.

Interestingly, while clumps and clouds show similar power-law indices
in the central region ($D_{\rm m,clump}=2.07\pm0.16$ versus
$D_{\rm m,cloud}=2.22\pm0.10$), clumps exhibit power-law indices much
shallower than those of clouds in the molecular ring
($D_{\rm m,clump}=1.65\pm0.04$ versus $D_{\rm m,cloud}=2.04\pm0.01$).
The power-law index $D_{\rm m,clump}$ of clumps in the molecular ring
(and thus the whole disc) is smaller than $2$, suggesting that in that
region the larger the clumps, the smaller their mass surface
densities. The discrepancy between the power-law indices of clumps and
clouds in the molecular ring is similar to that observed in the MW,
where $M_{\rm c}\propto R_{\rm c}^{1.6}$ for clumps of sizes
$0.01\le R_{\rm c}\le1$~pc
\citep{kauffmann2010,lombardi2010,zhang2017}, while
$M_{\rm c}\propto R_{\rm c}^{2.2-2.3}$ for clouds of sizes
$3\le R_{\rm c}\le100$~pc
\citep{romanduval2010,mivilledeschenes2016}. These trends thus suggest
that smaller structures are generally more centrally concentrated
(i.e.\ denser), and that the clumps and clouds in the molecular ring
may constitute different populations of molecular structures within
the same region. We will discuss the physical origins of the observed
mass -- size relations of NGC~404 in
Section~\ref{sec:implications_mass-size}.

\subsection{Multiple-scale virial analyses}
\label{sec:virial_analysis_multiple-scale}

A useful tool to quantify the dynamical state of a molecular structure
is the virial theorem. The virial parameter
\begin{equation}
  \label{eq:virial_parameter}
  \alpha_{\rm vir,c}\equiv M_{\rm vir,c}/M_{\rm c}~,
\end{equation}
that compares the virial mass
\begin{equation}
  \label{eq:virial_mass}
  M_{\rm vir,c}\equiv\frac{5\sigma_{\rm obs,los}^2R_{\rm c}}{G}
\end{equation}
of the molecular structure to its gaseous mass $M_{\rm c}$, provides a
useful measure of its degree of gravitational binding. If
$\alpha_{\rm vir,c}\ll1$, the gravitational binding energy is more
important than the kinetic energy, and the structure is
gravitationally bound.
If $\alpha_{\rm vir,c}\approx1$, the kinetic energy is roughly half
the gravitational potential energy, and the structure is
gravitationally bound and in (gravitational) virial equilibrium.
If $\alpha_{\rm vir,c}\gg1$, the kinetic energy is more important than
the gravitational energy, and the structure is gravitationally unbound
(and thus transient).
A virial parameter $\alpha_{\rm vir,crit}\approx2$ is often considered
the threshold betwen gravitationally-bound and unbound objects
\citep[e.g.][]{kauffmann2013,kauffmann2017}.

\begin{figure*}
  \includegraphics[width=0.98\textwidth]{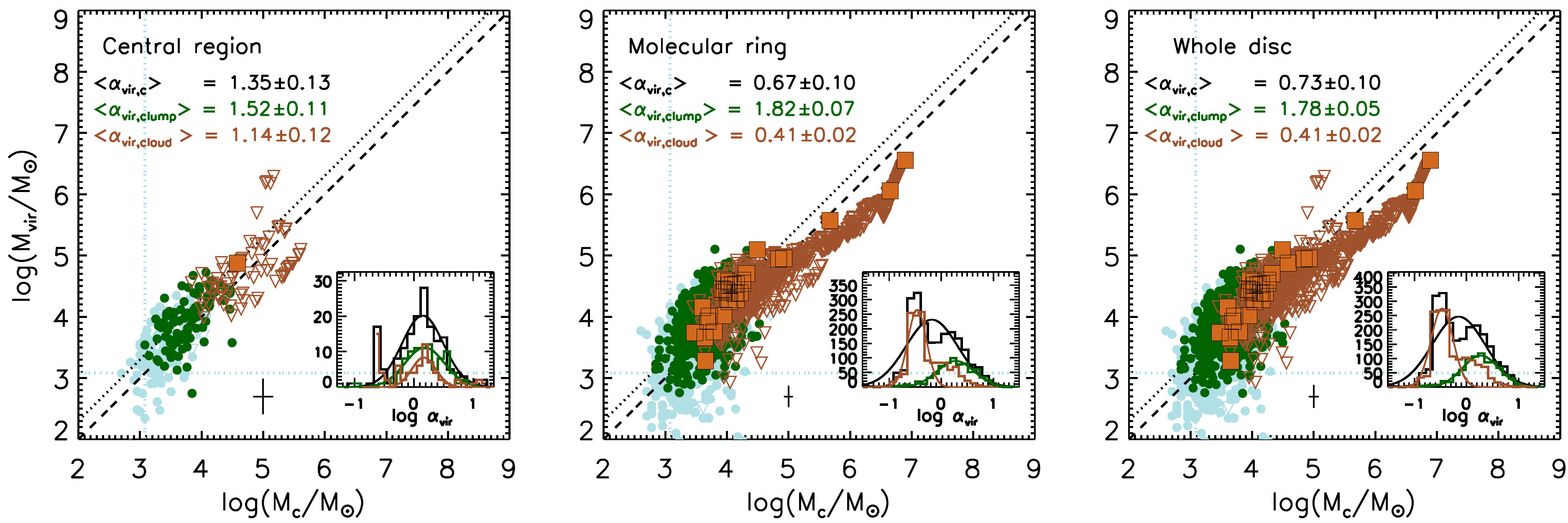}
  \caption{Virial mass -- gaseous mass relation of molecular
    structures in the central region, molecular ring and whole disc of
    NGC~404, respectively. Symbols are as for
    Fig.~\ref{fig:size_linewidth_relation}. The black dashed lines
    show the $1:1$ relations (i.e.\ $\alpha_{\rm vir,c}=1$) and the
    black dotted lines the $2:1$ relations (i.e.\
    $\alpha_{\rm vir,c}=2$). The vertical and horizontal blue dotted
    lines indicate our mass detection limit
    ($\log(10\delta_{\rm M}/{\rm M}_\odot)=3.08$; see
    Section~\ref{sec:mass_function_clumps}). In the inset of each
    panel, the distributions of $\log\alpha_{\rm vir,c}$ of all
    resolved molecular structures (black), clumps (green) and clouds
    (brown) are shown, with their best log-normal fits overlaid in
    matching colours. The corresponding mean virial parameters are
    listed in the legend.}
  \label{fig:alpha}
\end{figure*}

Figure~\ref{fig:alpha} compares the virial and gaseous masses of
molecular structures in the central region, molecular ring and whole
disc of NGC~404. The distributions of the resulting virial parameters
of all resolved molecular structures (black line), only resolved
clumps (green line) and only clouds (brown line) are also shown in an
inset in each panel. The corresponding mean virial parameters are
listed in the legend of each panel and
Table~\ref{tab:results_summary}.

Overall, the molecular structures identified in NGC~404 tend to be
gravitationally bound, as their mean virial parameter is smaller than
unity ($\langle\alpha_{\rm vir,c}\rangle=0.73\pm0.10$ for all the
molecular structures of the whole disc). However, the distributions of
virial parameters vary significantly across the different regions.
First, the molecular structures in the central region tend to have
much larger virial parameters than those in the molecular ring
($\langle\alpha_{\rm vir,c}\rangle=1.35\pm0.13$ versus
$\langle\alpha_{\rm vir,c}\rangle=0.67\pm0.10$). This is consistent
with the fact that gas seems to be more turbulent in the central
region (see Section\ \ref{sec:multiple-scale_size-linewidth}).

Second, while clumps and clouds have similar virial parameters in the
central region ($\langle\alpha_{\rm vir,clump}\rangle=1.52\pm0.11$
versus $\langle\alpha_{\rm vir,cloud}\rangle=1.14\pm0.12$), clumps
have much larger virial parameters than clouds in the molecular ring
($\langle\alpha_{\rm vir,clump}\rangle=1.82\pm0.07$ versus
$\langle\alpha_{\rm vir,cloud}\rangle=0.41\pm0.02$). Indeed, clumps
have a $\log(\alpha_{\rm vir,c})$ distribution that is clearly
distinct from that of clouds in the molecular ring (see middle panel
of Fig.~\ref{fig:alpha}). It thus seems that the dynamical states of
clumps and clouds are regulated by different physical mechanisms in
that region. This appears to support the suggestion that clumps and
clouds in the molecular ring may constitute different populations of
molecular structures (see Section~\ref{sec:multiple-scale_mass-size}).
Possible reasons for the different dynamical states of clumps and
clouds in the molecular ring will be discussed in
Section~\ref{sec:stability_multiple-scale}.

It is also worth noting that the clouds in the molecular ring clearly
have a double-peaked, double-Gaussian $\log(\alpha_{\rm vir,c})$
distribution, implying two distinct cloud populations. Indeed,
``low-mass'' clouds clearly have virial parameters larger than those
of ``high-mass'' clouds there
($\langle\alpha_{\rm vir,cloud}\rangle=1.10\pm0.07$ for
$M_{\rm cloud}\leq10^5$~M$_\odot$ and
$\langle\alpha_{\rm vir,cloud}\rangle=0.21\pm0.0$ for
$M_{\rm cloud}>10^5$~M$_\odot$).

\begin{figure*}
  \includegraphics[width=0.98\textwidth]{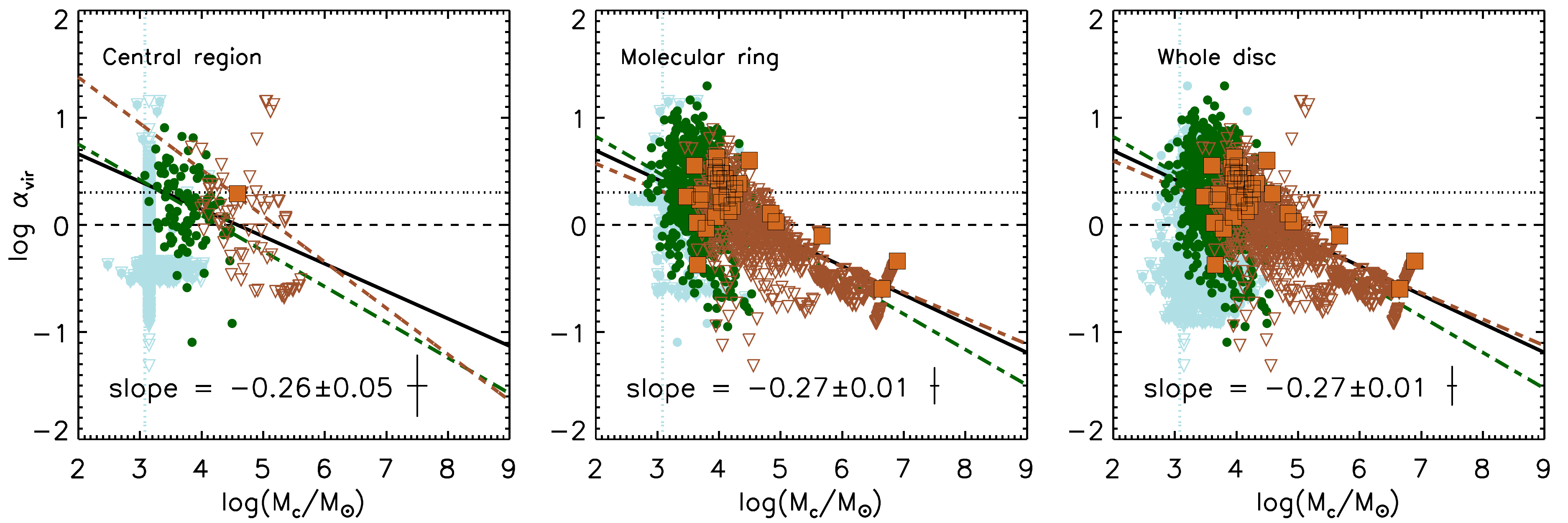}
  \caption{Virial parameter -- gaseous mass relation of molecular
    structures in the central region, molecular ring and whole disc of
    NGC~404, respectively. Symbols are as for
    Fig.~\ref{fig:size_linewidth_relation}. The black dashed lines
    indicate $\alpha_{\rm vir,c}=1$ and the black dotted lines
    $\alpha_{\rm vir,c}=2$, above which structures are unbound in the
    absence of other confining mechanisms. In each panel, the black,
    green and brown lines show the best-fitting power-law relation of
    all resolved molecular structures, only resolved clumps (i.e.\
    leaves) and only resolved clouds (i.e.\ branches and trunks),
    respectively. The slope of the best-fitting power-law relation of
    all resolved molecular structures is listed in the legend. The
    vertical blue dotted lines indicate our mass detection limit
    ($\log(10\delta_{\rm M}/{\rm M}_\odot)=3.08$; see
    Section~\ref{sec:mass_function_clumps}). }
  \label{fig:alpha_vs_mgas}
\end{figure*}

Figure~\ref{fig:alpha_vs_mgas} shows the dependences of
$\alpha_{\rm vir,c}$ on $M_{\rm c}$ for the central region, molecular
ring and whole disc of NGC~404. The corresponding power-law indices of
the $\alpha_{\rm vir,c}$ -- $M_{\rm c}$ relations are listed in
Table~\ref{tab:results_summary}. There is no correlation between
$\alpha_{\rm vir,c}$ and $M_{\rm c}$ in the central region (left panel
of Fig.~\ref{fig:alpha_vs_mgas}; Spearman rank correlation coefficient
$-0.20\pm0.07$), but there is a clear trend in the molecular ring
(middle panel of Fig.~\ref{fig:alpha_vs_mgas}; Spearman rank
correlation coefficient $-0.75\pm0.01$), where the best-fitting
power-law is $\alpha_{\rm vir,c}\propto M_{\rm c}^{-0.27\pm0.01}$
(black solid line in the middle panel of
Fig.~\ref{fig:alpha_vs_mgas}). It thus seems that the higher-mass
molecular structures tend to be more bound than the lower-mass ones in
the molecular ring (albeit with much scatter). An anti-correlation of
$\alpha_{\rm vir,c}$ and $M_{\rm c}$ has also been observed in the MW,
where $\alpha_{\rm vir,c}\proptosim M_{\rm c}^{-0.6}$ for Galactic
clumps ($R_{\rm c}\le5$~pc; \citealt{zhang2016,veltchev2018}) and
$\alpha_{\rm vir,c}\propto M_{\rm c}^{-0.53\pm0.30}$ for $\approx8000$
molecular clouds across the entire Galactic plane
\citep{mivilledeschenes2017}.

\begin{figure}
  \includegraphics[width=0.95\columnwidth]{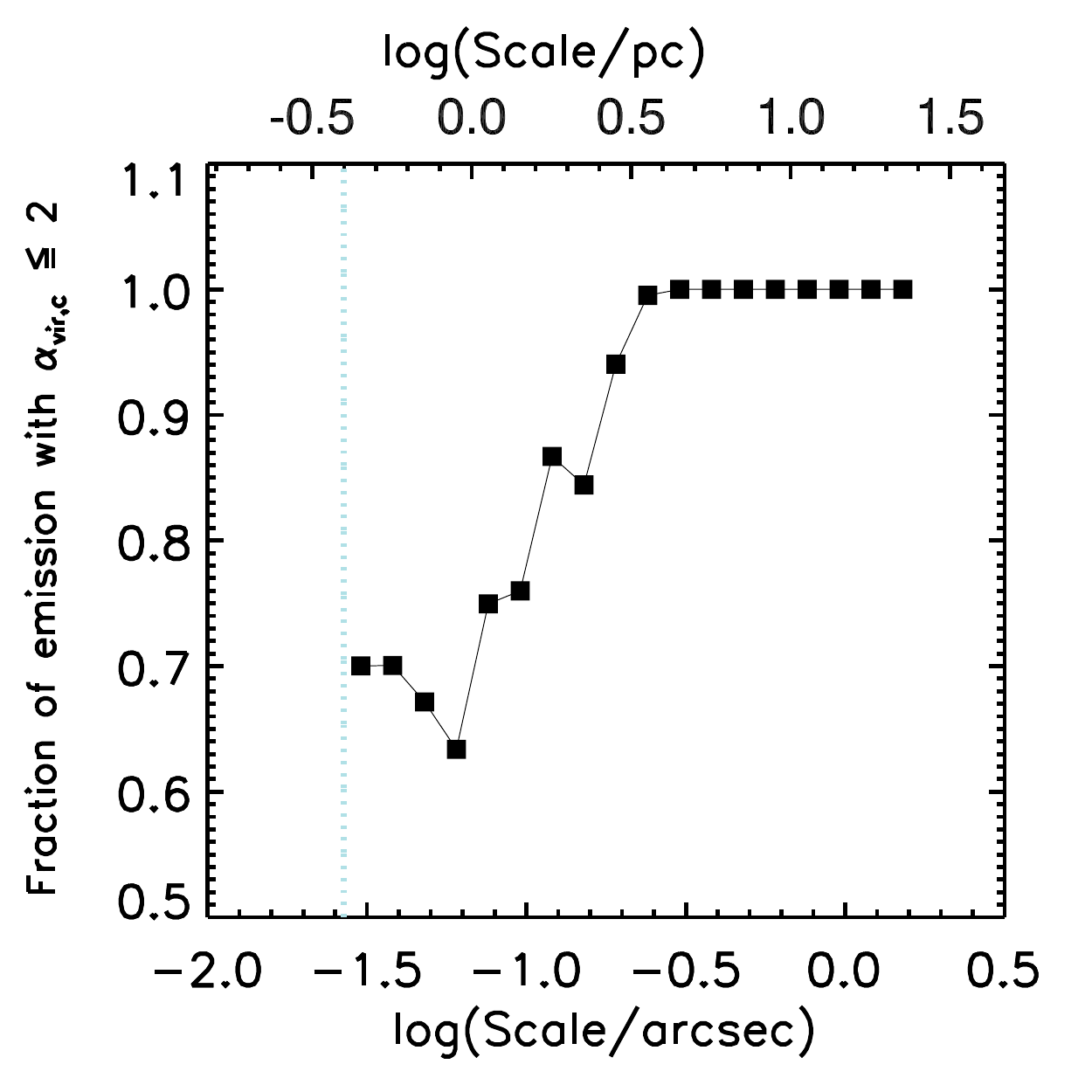}
  \caption{Fraction of emission originating from gravitationally-bound
    (i.e.\ self-gravitating) structures
    ($\alpha_{\rm vir,c}\le\alpha_{\rm vir,crit}=2$) as a function of
    the structures' spatial scales in NGC~404. The fraction grows with
    scale and saturates at $1$ at scales $\gtrsim3$~pc. The blue
    vertical dotted line indicates our spatial resolution limit of
    $0\farcs027$ or $0.40$~pc (i.e.\ half the synthesised beam).}
  \label{fig:selffrac_vs_r_obs}
\end{figure}

We also calculate the fraction of emission originating from
gravitationally-bound (i.e.\
$\alpha_{\rm vir,c} \le \alpha_{\rm vir,crit}=2$) structures, and plot
this fraction as a function of the structures' spatial scales in
Fig.~\ref{fig:selffrac_vs_r_obs}. To calculate this, we first measure
the virial parameter of the emission contained within each resolved
isosurface in the data cube, and define the emission enclosed by an
isosurface as self-gravitating if its
$\alpha_{\rm vir,c}\le\alpha_{\rm vir,crit}=2$. The fraction of
self-gravitating gas is then defined as the ratio of the total
emission of the structures with
$\alpha_{\rm vir,c}\le\alpha_{\rm vir,crit}=2$ to the total emission
of all structures (of any $\alpha_{\rm vir,c}$), this within a small
range of spatial scales (i.e.\ a spatial scale ``bin''; see Eq.~6 of
\citealt{rosolowsky2008}). Similarly to the aforementioned trend, we
find the fraction of self-gravitating gas to be about unity at spatial
scales $\gtrsim3$~pc, implying that all these large-scale structures
(that are exclusively clouds) are gravitationally bound. However, the
fraction decreases for smaller spatial scales and it drops below
$\approx0.7$ at spatial scales $\lesssim1$~pc. A similar trend has
been observed in the multiple-scale structures of the L1448 molecular
cloud in the MW, where only a small fraction of small-scale objects
appear to be self-gravitating, but the fraction of
gravitationally-bound gas grows to unity at larger spatial scales
\citep{rosolowsky2008,goodman2009}. We will discuss this scale (and
mass) dependence of the gravitational boundedness in more detail in
Section~\ref{sec:stability_multiple-scale}.

\section{Clump-Clump Collision Model}
\label{sec:clump-clump_collision_model}

Cloud-cloud collisions have been proposed in the past as a potentially
important mechanism for giant molecular cloud (GMC) formation
\citep[e.g.][]{kwan1979,cowie1980} and star formation
\citep[e.g.][]{dobbs2008,tasker2009,dobbs2011,fukui2021,maeda2021,sano2021},
but they were rejected because of their supposedly long timescale
($\sim100$~Myr; e.g.\
\citealt{blitz1980,das1996,mckee2007,dobbs2008,hirota2011}). However,
recent theory \citep{gammie1991,tan2000} and high-resolution
hydrodynamic simulations \citep{tasker2009,li2018,wu2018} suggested
that cloud-cloud collisions can be efficient in a
differentially-rotating disc, where collisions between clouds are
driven by galactic shear.
In such a shear-driven collision scenario, the collision timescale is
much shorter than traditional estimates (a small fraction,
$\approx1/5$, of the orbital time rather than hundreds of Myr;
\citealt{tasker2009}). A short collision timescale has several
important implications: (1) cloud-cloud collisions can be crucial to
regulate cloud properties such as mass and size
\citep{tasker2009,li2018}; (2) cloud-cloud collisions can be efficient
to disturb the ISM and induce turbulence
\citep{li2018,wu2017a,wu2017b,wu2018}.

In this work, instead of cloud-cloud collisions, we will hereafter use
the term \enquote{clump-clump collisions}, to better reflect the fact
that most of the molecular structures we have identified in NGC~404
have sizes much smaller than those of molecular clouds (i.e.\ $<10$~pc
rather than tens of parsecs; \citealt{solomon1987}). More importantly,
as we will see in Section~\ref{sec:clump_mass_collision_model}, in
NGC~404 the collisions between clouds are much less important than
those between clumps. In this section, we will thus develop and
explore a new simple analytical model connecting clump-clump
collisions to the clump properties and gas turbulence. We will also
demonstrate that a key ingredient of our collision model is
gravitational instabilities (i.e.\ a Toomre parameter $Q\lesssim1$).

\subsection{Collision timescale}
\label{sec:collision_timescale}

We consider an idealised model of a galactic disc that ignores the
effects of supernovae, stellar winds, ram pressure and magnetic
fields, i.e.\ all the physical processes other than galactic
rotational shear, gravitational instabilities and clump-clump
collisions. Molecular clumps are assumed to be uniformly distributed
over a small (coarse-grained) region of an infinitely thin,
two-dimensional disc,
and to populate perfectly circular orbits determined by the
gravitational potential of the galaxy. We take clump-clump collisions
to be any mutual gravitational interaction and merging of clumps. In
other words, collisions between clumps are assumed to be completely
inelastic, and to lead to the coalescence of the clumps
\citep{fleck1987, gammie1991}.
Finally, we assume that collisions can only occur between (two) clumps
of equal mass; i.e.\ collisions can be described as ``major mergers'',
as hydrodynamic simulations have shown that the mass ratio
distribution of colliding objects peaks at $\approx1$ \citep{li2018}.

Following \citet{tan2000}, we set the velocity of the
collision between two clumps to be the shear velocity
\begin{equation}
  \label{eq:vshear}
  v_{\rm shear}\equiv v_{\rm shear}(R)=2Ab
\end{equation}
and the radius of the collision cross section to be the tidal radius
\begin{equation}
  \label{eq:radius_tidal}
  R_{\rm t}\equiv R_{\rm t}(R)\approx\left(\frac{G}{2A^2}\right)^{1/3}M_{\rm c}^{1/3}~,
\end{equation}
where $A\equiv A(R)=-\frac{R}{2}\frac{d\Omega}{dR}$ is Oort's constant
$A$ evaluated at the galactocentric distance $R$ of the clump
($R_{\rm gal}$ in Table~\ref{tab:gmc_properties}),
$\Omega\equiv \Omega(R)=V_{\rm circ}/R$ is the angular velocity of
orbital circular rotation, $V_{\rm circ}\equiv V_{\rm circ}(R)$ is the
circular velocity of the galaxy, and $b$ is the radial distance
between the orbits of the two colliding clumps.  See
Appendix~\ref{appendix:collision_timescale} for a full derivation of
the shear velocity and tidal radius. The tidal radius $R_{\rm t}$
rather than the actual clump radius $R_{\rm clump}$ is adopted for the
collision cross section, because two clumps will gravitationally
attract (i.e.\ collide with) each other despite the effects of the
galaxy gravity (i.e.\ the effects of shear and tidal forces) as long
as their relative distance is smaller than their tidal radius
$R_{\rm t}$. We note that our adopted tidal radius $R_{\rm t}$ is
derived by assuming a spherical galaxy mass distribution, and is
similar to the Roche limit defined in the literature
\citep[e.g.][]{stark1978}.

The clump-clump collision timescale can then be shown to be
\begin{equation}
  \label{eq:collision_timescale}
  t_{\rm coll}\equiv t_{\rm coll}(R)\approx\frac{1}{2AN_{\rm A}R_{\rm t}^2}\approx\frac{A^{1/3}M_{\rm c}^{1/3}}{2^{1/3}G^{2/3}\Sigma_{\rm gas,disc}}
\end{equation}
(see, again, Appendix~\ref{appendix:collision_timescale}), where
$N_{\rm A}\equiv N_{\rm A}(R)=\Sigma_{\rm gas,disc}/M_{\rm c}$ is the
number surface density of clumps and
$\Sigma_{\rm gas,disc}\equiv \Sigma_{\rm gas,disc}(R)$ is the
coarse-grained gaseous mass surface density of the disc.

\subsection{Clump mass}
\label{sec:clump_mass_collision_model}

We assume here that clumps (i.e.\ connected, locally-peaked
structures) are formed from the collisional agglomeration of smaller
clumps (i.e.\ a ``bottom-up'' scenario of clump formation;
\citealt{mckee2007}).
We first define an accumulation length as the length scale at which
uniformly distributed gas can coalesces into a single centrally-peaked
clump via clump-clump collisions.
The accumulation length can thus be approximated as the average
distance between neighbouring clumps:
\begin{equation}
  \label{eq:accumulation_length}
  L_{\rm acc,clump}\equiv L_{\rm acc,clump}(R)=(M_{\rm clump}/\Sigma_{\rm gas,disc})^{1/2}=1/N_{\rm A}^{1/2}~.
\end{equation}
This accumulation length naturally increases during the process of
clump-clump collisions, as successive collisions constantly increase
the clump mass and size (thereby decreasing the number surface density
of clumps).

We then propose that the clumps regulated by collisions should have
their accumulation length approximately equal to their tidal radius,
i.e.\ $L_{\rm acc,clump}\approx R_{\rm t,clump}$. This is intuitively
easy to understand. If $L_{\rm acc,clump}<R_{\rm t,clump}$, the
distance between neighbouring clumps is smaller than the tidal radius,
and thus the clumps will continue to coalesce with (i.e.\
gravitationally attract) each other, forming more massive clumps. If
$L_{\rm acc,clump}>R_{\rm t,clump}$, the distance between neighbouring
clumps is larger than the tidal radius, and thus the clumps will be
pulled apart/sheared away from each other due to the effects of
external gravity. Therefore, clump growth via clump-clump collisions
naturally ceases when $L_{\rm acc,clump}\approx R_{\rm t,clump}$.

By posing
$L_{\rm acc,clump}=R_{\rm
  t,clump}\approx\left(\frac{G}{2A^2}\right)^{1/3}M_{\rm c}^{1/3}$
(assuming a spherical galaxy mass distribution, see
Eq.~\ref{eq:radius_tidal}), we can rewrite the clump tidal radius and
accumulation length as
\begin{equation}
  \label{eq:lambda_coll}
 R_{\rm t,clump}\approx L_{\rm acc,clump}\approx G\Sigma_{\rm gas,disc}/2A^2\equiv\lambda_{\rm coll}(R)~.
\end{equation}
This naturally defines a critical collision length
$\lambda_{\rm coll}\equiv\lambda_{\rm coll}(R)=G\Sigma_{\rm
  gas,disc}/2A^2$, that is a key length scale of our collision
model. When $R_{\rm t,clump}=\lambda_{\rm coll}$ or equivalently
$L_{\rm acc,clump}=\lambda_{\rm coll}$, then necessarily
$R_{\rm t,clump}=L_{\rm acc,clump}$.

Another way to understand this limitation is to compare the collision
timescale of clumps $t_{\rm coll,clump}$
(Eq.~\ref{eq:collision_timescale}) with the shear timescale
\begin{equation}
  \label{eq:shear_time}
  t_{\rm shear}\equiv t_{\rm shear}(R)=1/2A~,
\end{equation}
i.e.\ the timescale for gas instabilities to develop and grow before
the clumps are sheared apart \citep[see][]{kruijssen2014,liu2021}.
If $t_{\rm coll,clump}< t_{\rm shear}$, collisions dominate over
shear, and clumps can collide and merge into more massive clumps. If
$t_{\rm coll,clump}>t_{\rm shear}$, shear dominates over collisions,
and collisions between clumps are disrupted by shear. Hence, clumps
regulated by clump-clump collisions should have
$t_{\rm coll}\approx t_{\rm shear}=1/2A$, that is equivalent to
$L_{\rm acc,clump}\approx R_{\rm t,clump}$.

A simple scenario of clump formation therefore emerges, whereby
small-scale clumps initially form and then collide and coalesce into
more massive clumps, until the clumps' tidal radii reach
$\lambda_{\rm coll}$. The resulting clump mass ($M_{\rm c,coll}$) is
thus directly obtained:
\begin{equation}
  \label{eq:mc_crit}
  \begin{split}
   M_{\rm clump} & \equiv M_{\rm clump}(R)=\Sigma_{\rm gas,disc}\,\lambda_{\rm coll}^2 \\
   & \approx(2A^2/G)\,\lambda_{\rm coll}^3 \approx G^2\Sigma_{\rm gas,disc}^3/4A^4\equiv M_{\rm c,coll}(R)
  \end{split}
\end{equation}
(see Section~\ref{sec:clump_mass_function_collision_model} for a more
detailed discussion). The typical mass of clumps formed via
clump-clump collisions is thus determined only by galaxy properties
($\Sigma_{\rm gas,disc}$ and $A$).

\subsection{First comparison to NGC~404}
\label{sec:application_ngc404}

\begin{figure*}
  \includegraphics[width=0.9\textwidth]{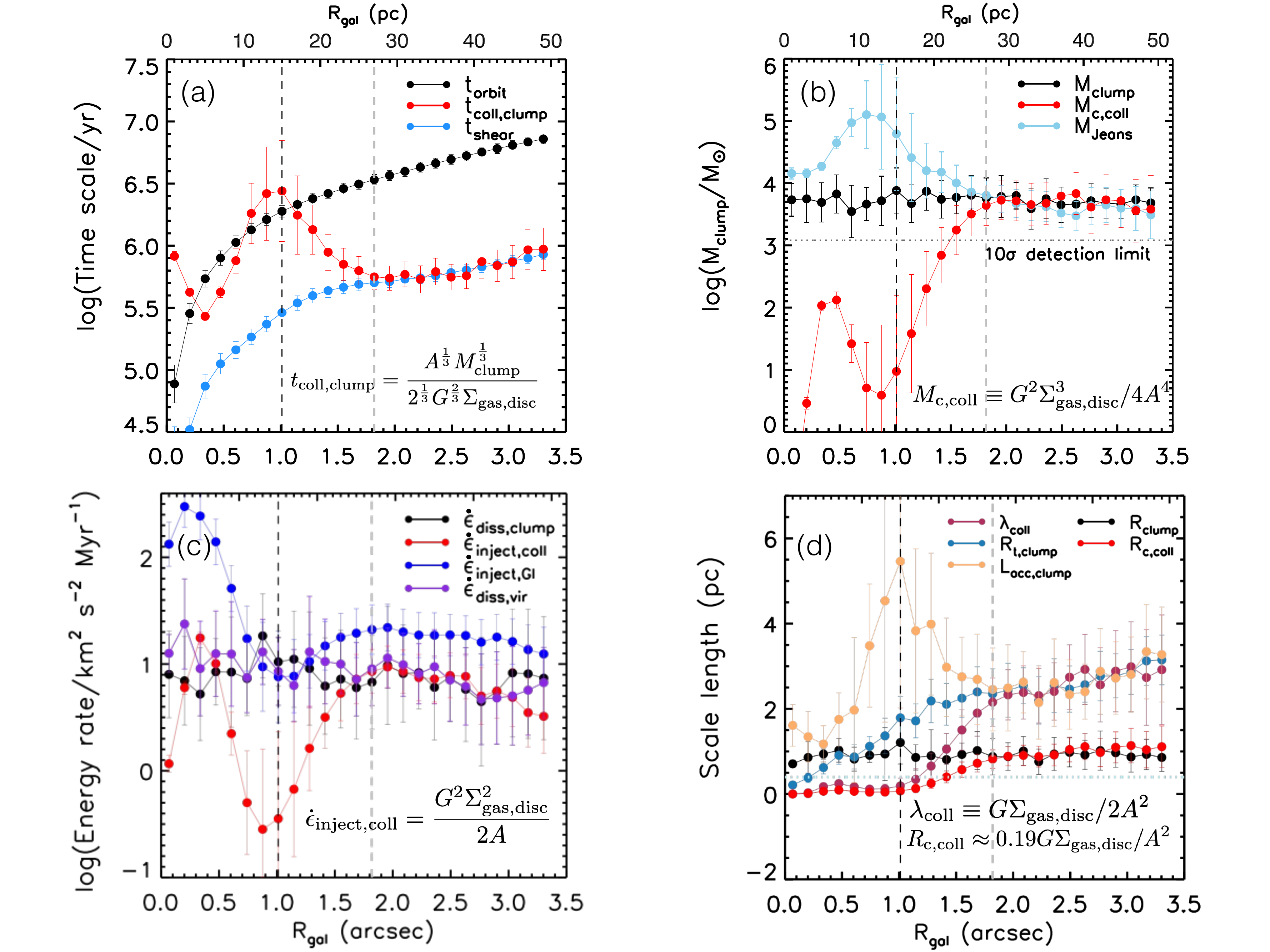}
  \caption{Radial variation of the predicted collision timescale
    ($t_{\rm coll,clump}$), clump mass ($M_{\rm c,coll}$), turbulence
    energy injection rate ($\dot{\epsilon}_{\rm inject,coll}$) and
    clump size ($R_{\rm c,coll}$), respectively, of our clump-clump
    collision model, averaged over galactocentric radius bins of
    $2$~pc in NGC~404. In each panel, the black vertical dashed line
    indicates the boundary ($R_{\rm gal}=15$~pc) between the central
    region and molecular ring, while the grey vertical dashed line
    indicates the galactocentric distance ($R_{\rm gal}=27$~pc) beyond
    which the molecular gas disc is no longer gravitationally stable
    (i.e.\ Toomre parameter's $Q\le1$ at $R_{\rm gal}\ge27$~pc; see
    Section~\ref{sec:gravitational_instability_onset_collisions}).
    The error bars of each quantity indicate the $1~\sigma$ scatter of
    the values within each radial bin (not the uncertainty on the mean
    within each bin, that is much smaller). {\bf Panel~(a):}
    comparisons of the orbital timescale $t_{\rm orbit}$
    (Eq.~\ref{eq:orbital_timescale}), our model-predicted clump-clump
    collision timescale $t_{\rm coll,clump}$
    (Eq.~\ref{eq:collision_timescale}) and the shear timescale
    $t_{\rm shear}$ (Eq.~\ref{eq:shear_time}) at each galactocentric
    radius. The clump-clump collision timescale $t_{\rm coll,clump}$
    is much smaller than the orbital timescale $t_{\rm orbit}$ but is
    in good agreement with the shear timescale $t_{\rm shear}$ in the
    molecular ring (especially at $R_{\rm gal}\ge27$~pc where $Q\le1$;
    see Section~\ref{sec:clump_mass_collision_model}). {\bf
      Panel~(b):} comparisons of the observed median clump mass
    $M_{\rm clump}$, that also happens to be the most common clump
    mass, our model-predicted clump mass $M_{\rm c,coll}$
    (Eq.~\ref{eq:mc_crit}) and the Jeans mass $M_{\rm Jeans}$
    (Eq.~\ref{eq:Jeans_mass}) at each galactocentric radius. The black
    horizontal dotted line indicates our mass detection limit
    ($\log(10\delta_{\rm M}/{\rm M}_\odot)=3.08$; see
    Section~\ref{sec:mass_function_clumps}). The median/most common
    clump mass $M_{\rm clump}$ is in good agreement with the predicted
    clump mass $M_{\rm c,coll}$ (and the Jeans mass $M_{\rm Jeans}$)
    in the molecular ring (especially at $R_{\rm gal}\ge 27$~pc where
    $Q\le1$; see Section~\ref{sec:clump_mass_collision_model}). {\bf
      Panel~(c):} comparisons of the observed median energy
    dissipation rate (per unit mass) of clumps
    $\dot{\epsilon}_{\rm diss,clump}$
    (Eq.~\ref{eq:energy_dissipation_rate}), our model-predicted energy
    injection rate (per unit mass) due to clump-clump collisions
    $\dot{\epsilon}_{\rm inject,coll}$
    (Eq.~\ref{eq:energy_injection_rate}), the predicted energy
    injection rate (per unit mass) from gravitational instabilities
    $\dot{\epsilon}_{\rm inject,GI}$ (Eq.~\ref{eq:e_gi}) and the
    median virial energy dissipation rate (per unit mass) of clumps
    $\dot{\epsilon}_{\rm diss,vir}$
    (Eq.~\ref{eq:virial_energy_dissipation_rate}) at each
    galactocentric radius. The observed median energy dissipation rate
    $\dot{\epsilon}_{\rm diss,clump}$ is consistent with the median
    virial energy dissipation rate $\dot{\epsilon}_{\rm diss,vir}$ at
    all radii, but it agrees well with our predicted energy injection
    rate $\dot{\epsilon}_{\rm inject,coll}$ and is within one order of
    magnitude of $\dot{\epsilon}_{\rm inject,GI}$ in the molecular
    ring only (especially at $R_{\rm gal}\ge27$~pc where $Q\le1$; see
    Section~\ref{sec:turbulence_collision_model}). {\bf Panel~(d):}
    comparison of the observed median clump size $R_{\rm clump}$, that
    also happens to be the most common clump size, and our
    model-predicted clump size $R_{\rm c,coll}$
    (Eq.~\ref{eq:predicted_clump_size}), and comparisons of the
    observed clump tidal radius $R_{\rm t,clump}$
    (Eq.~\ref{eq:tidal_radius2_app}), accumulation length
    $L_{\rm acc,clump}$ (Eq.~\ref{eq:accumulation_length}) and
    collision critical length $\lambda_{\rm coll}$
    (Eq.~\ref{eq:lambda_coll}) at each galactocentric radius. The blue
    horizontal dotted line indicates our spatial resolution limit of
    $0\farcs027$ or $0.40$~pc (i.e.\ half the synthesised beam). The
    clump tidal radius $R_{\rm t,clump}$, accumulation length
    $L_{\rm acc,clump}$ and collision critical length
    $\lambda_{\rm coll}$ are consistent with each other (see
    Section~\ref{sec:clump_mass_collision_model}) in the molecular
    ring (especially at $R_{\rm gal}\ge27$~pc where $Q\le1$; see
    Section~\ref{sec:implications_mass-size}), similarly for the
    observed clump size $R_{\rm clump}$ and predicted clump size
    $R_{\rm c,coll}$. The clump collision timescale
    $t_{\rm coll,clump}$, tidal radius $R_{\rm t,clump}$ and
    accumulation length $L_{\rm acc,clump}$ are calculated using the
    median clump mass $M_{\rm clump}$ at each galactocentric radius
    (see panel~(b)). The quantities $t_{\rm coll,clump}$,
    $M_{\rm clump}$, $\dot{\epsilon}_{\rm diss,clump}$,
    $\dot{\epsilon}_{\rm diss,vir}$, $R_{\rm clump}$,
    $R_{\rm t,clump}$ and $L_{\rm acc,clump}$ are calculated for
    resolved clumps only. }
  \label{fig:ccc_model}
\end{figure*}

We now apply our clump-clump collision model to NGC~404. We first
compare the clump-clump collision timescale $t_{\rm coll,clump}$
(Eq.~\ref{eq:collision_timescale}) with the shear timescale
$t_{\rm shear}$ (Eq.~\ref{eq:shear_time}) at each galactocentric
radius (see panel~(a) of Fig.~\ref{fig:ccc_model}). To achieve this,
we adopt radially-varying $\Omega$ and $A$ (see
Fig.~\ref{fig:a_vs_r}), $\Sigma_{\rm gas,disc}$ (see
Fig.~\ref{fig:Sigma_gas_vs_r}) and median clump mass $M_{\rm clump}$
(that also happens to be the most common clump mass; see panel~(b) of
Fig.~\ref{fig:ccc_model}). In the central region, the collision
timescale is much longer than the shear timescale (i.e.\
$t_{\rm coll,clump}>t_{\rm shear}$), implying that clump-clump
collisions are not important in this region. In the molecular ring,
however, the collision timescale decreases significantly with
increasing galactocentric distance, and it becomes comparable to the
shear timescale at $R_{\rm gal}\ge27$~pc, where most ($\approx87\%$)
of the molecular gas in the molecular ring is located (see the grey
vertical dashed line in each panel of Fig.~\ref{fig:ccc_model}). This
implies clump-clump collisions must be important in the molecular
ring.

Panel~(a) of Fig.~\ref{fig:ccc_model} also shows the orbital timescale
of clumps
\begin{equation}
  \label{eq:orbital_timescale}
  t_{\rm orbit}\equiv t_{\rm orbit}(R)=2\pi/\Omega
\end{equation}
as a function of the galactocentric distance $R$ (i.e.\ $R_{\rm gal}$)
in NGC~404. The orbital timescales of clumps are generally longer than
their collision timescales. This is particularly the case in the
molecular ring, where at $R_{\rm gal}\ge27$~pc the clump-clump
collision timescales are only $0.1$ -- $0.2$ of the orbital
timescales. Our observations are thus consistent with simulation
results ($t_{\rm coll}\approx0.2\,t_{\rm orbit}$;
\citealt{tasker2009,dobbs2015,li2018}), suggesting clump-clump
collisions can indeed be frequent in galactic discs with strong shear.
We also note that, for a flat circular velocity curve ($\Omega=2A$),
the shear timescale $t_{\rm shear}=1/2A=t_{\rm orbit}/2\pi$, and thus
in the molecular ring (where the rotation curve is almost flat)
\begin{equation}
  t_{\rm coll,clump}\approx t_{\rm shear}\approx t_{\rm orbit}/2\pi~,
\end{equation}
that naturally accounts for our finding that
$t_{\rm coll,clump}/t_{\rm orbit}\approx0.1$ -- $0.2$ in the molecular ring.

The observed clump accumulation length $L_{\rm acc,clump}$
(Eq.~\ref{eq:accumulation_length}) and tidal radius $R_{\rm t,clump}$
(Eq.~\ref{eq:radius_tidal})
are also compared in panel~(d) of Fig.~\ref{fig:ccc_model} (calculated
again using the median clump mass $M_{\rm clump}$ at each
galactocentric radius; see panel~(b) of
Fig.~\ref{fig:ccc_model}). Again, the clump accumulation length is
much larger than the clump tidal radius in the central region, but the
two are comparable in the molecular ring (especially at
$R_{\rm gal}\ge27$~pc). This again suggests that clump-clump
collisions are important in the molecular ring but not in the central
region.

We also compare our predicted clump mass $M_{\rm c,coll}$
(Eq.~\ref{eq:mc_crit}) with the observed (median) clump masses
$M_{\rm clump}$, and our predicted clump tidal radius
$\lambda_{\rm coll}$ (Eq.~\ref{eq:lambda_coll}) with the observed
(median) clump tidal radii $R_{\rm t,clump}$ at each galactocentric
distance. As shown in panels~(b) and (d) of Fig.~\ref{fig:ccc_model},
our model successfully accounts for the measurements in the molecular
ring, especially at $R_{\rm gal}\ge27$~pc, where
$M_{\rm clump}\approx M_{\rm c,coll}$ and
$R_{\rm t,clump}\approx L_{\rm acc,clump}\approx\lambda_{\rm
  coll}$. The good match between the predicted and observed clump
masses in the molecular ring can also be seen in the middle panel of
Fig.~\ref{fig:clump_mass_distribution}, where our predicted clump mass
$M_{\rm c,coll}$ for the molecular ring (the red vertical dashed line)
agrees well with the observed most common clump mass (i.e.\ the
turn-over mass at $M_{\rm clump}\approx4000$~M$_\odot$).

Having said that, our collision model leads to significant
underestimates of the clump masses in the central region.
This significant mismatch between the model $M_{\rm c,coll}$ (or
$\lambda_{\rm coll}$) and the observed (median) $M_{\rm clump}$ (or
$R_{\rm t,clump}$) in the central region is unlikely to be due to the
limited spatial resolution and/or sensitivity of our data, as the
clumps in the central region have deconvolved sizes larger than our
spatial resolution limit (see the blue horizontal dotted line in
panel~(d) of Fig.~\ref{fig:ccc_model}) and masses well above our
detection limit ($\log(10\delta_{\rm M}/{\rm M}_\odot)=3.08$; see the
grey horizontal dotted line in panel~(b) of Fig.~\ref{fig:ccc_model}).

Overall, the good match between our predictions and the observations
in the molecular ring suggests that clump-clump collisions are an
important mechanism regulating clumps in regions where
$t_{\rm coll,clump}\le t_{\rm shear}$. Other physical mechanisms are
required to explain the formation of massive clumps in the central
region, where $t_{\rm coll,clump}>t_{\rm shear}$. It is interesting to
note that the clumps in the central region nevertheless have masses
and sizes similar to those in the molecular ring (see panels~(b) and
(d) of Fig.~\ref{fig:ccc_model}). We will discuss this fact in
Section~\ref{sec:clump_migration}.
 
So far, we have only considered the importance of collisions between
clumps, but are collisions between clouds also likely to be important?
The answer is probably no. This is because
$t_{\rm coll}\propto M_{\rm c}^{1/3}$ (see
Eq.~\ref{eq:collision_timescale}), so clouds tend to have much longer
collision timescales than clumps. We have also seen that most clumps
have already reached a critical state whereby
$t_{\rm coll,clump}\approx t_{\rm shear}$ in the molecular ring (and
$t_{\rm coll,clump}>t_{\rm shear}$ in the central region). Clouds,
that by definition contains several clumps, should therefore have
$t_{\rm coll,cloud}>t_{\rm shear}$ everywhere. In other words, it is
likely that clouds in NGC~404 will be pulled away from each other (by
tidal and shear forces) before they have a chance to collide and
coalesce with each other. Cloud-cloud collisions are therefore likely
to be much less important than clump-clump collisions in NGC~404.
Hereafter, we will thus ignore cloud-cloud collisions, and only
discuss clump-clump collisions.

\subsection{Turbulence driven by clump-clump collisions}
\label{sec:turbulence_collision_model}

High-resolution hydrodynamic simulations suggest that frequent
clump-clump collisions can be an important source of turbulence in
galaxies
\citep[e.g.][]{agertz2009,namekata2011,tan2013,li2017,li2018,wu2018}.
According to these simulations, clump-clump collisions alone can
provide sufficient energy to maintain the observed level of turbulence
in the ISM \citep[e.g.][]{aumer2010,li2017,wu2017a,wu2017b,wu2018}.
In this section, we thus explore the impact of clump-clump collisions
on turbulence, by relating the turbulent energy injected via
clump-clump collisions to the kinetic energy of ordered
differential-rotation motions in the disc.
In the process, we will show that our model of collision-induced
turbulence matches well the observed turbulence in the molecular ring
of NGC~404.

\subsubsection{Energy injection rate by clump-clump collisions}

We have implicitly assumed that the collisions between clumps are
inelastic. This assumption is reasonable, as theoretical and
simulation works suggest that clumps in a differentially rotating
galactic disc can have inelastic collisions, that give rise to
viscosity and lead to energy dissipation and the transport of angular
momentum
\citep[e.g.][]{jog1988,ozernoy1998,vollmer2001,williamson2012,li2017}.
Given inelastic collisions, all the available kinetic energy is
dissipated into turbulence, i.e.\ the kinetic energy extracted from
galactic differential rotation by clump-clump collisions is completely
converted to turbulence. The average rate of \enquote{turbulent
  energy} injection due to clump-clump collision (per unit mass),
$\dot{\epsilon}_{\rm inject,coll}$, can then be expressed as
 \begin{equation}
\label{eq:energy_injection_rate}
 \begin{split}
    \dot{\epsilon}_{\rm inject,coll}(R) & =\frac{1}{t_{\rm coll}}\,\frac{2\,\int^{R_{\rm t,clump}}_0\left(\frac{1}{2}M_{\rm clump}v_{\rm shear}^2(b)\right)\,z_{\rm coll}(b)\,db}{2\,\int^{R_{\rm t,clump}}_0M_{\rm clump}\,z_{\rm coll}(b)\,db}\\
    & \approx\frac{1}{t_{\rm coll}}\,\frac{(2A)^3N_{\rm A}\,\int^{R_{\rm t,clump}}_0b^{3}\,db}{Z_{\rm coll}}\\
    & \approx2A^3\,N_{\rm A}\,R_{\rm t,clump}^4 \\ 
    & \approx2A^3N_{\rm A}\lambda_{\rm coll}^4\approx(G\Sigma_{\rm gas,disc})^2\big/2A~,\\
  \end{split}
\end{equation}
where $z_{\rm coll}(b)=N_{\rm A}v_{\rm shear}(b)=N_{\rm A}(2Ab)$ is
the collision rate per unit length and $N_{\rm A}$ is approximately
constant over a region of radius $R_{\rm t,clump}$ (see
Eqs.~\ref{eq:collision_rate_tmp_app} --
\ref{eq:collision_timescale_app}), and we have used
$R_{\rm t,clump}\approx\lambda_{\rm coll}$ as demonstrated in
Section~\ref{sec:clump_mass_collision_model} (see
Eq.~\ref{eq:lambda_coll}). The factor of $2$ in both numerator and
denominator of Eq.~\ref{eq:energy_injection_rate} accounts for clumps
either catching up with other clumps at larger $R_{\rm gal}$ or being
caught up by other clumps at smaller $R_{\rm gal}$. As
$N_{\rm A}\propto M_{\rm clump}^{-1}$ and
$R_{\rm t}\propto M_{\rm clump}^{1/3}$,
$\dot{\epsilon}_{\rm inject,coll}\propto M_{\rm clump}^{1/3}$,
suggesting that collisions between more massive clumps are more
effective at injecting turbulent energy. We note that our derived
$\dot{\epsilon}_{\rm inject,coll}$ again comprises galaxy properties
only ($\Sigma_{\rm gas,disc}$ and $A$) and therefore does not depend
on the clump properties.


The question is then whether the energy injected by clump-clump
collisions in NGC~404 is sufficient to maintain the observed level of
turbulence. An important parameter to quantify the properties of
turbulence is the rate at which energy is dissipated in the turbulent
cascade (per unit mass), or equivalently the rate at which energy is
transferred from large to small scales (per unit mass),
$\dot{\epsilon}_{\rm diss}$ \citep{mivilledeschenes2017}. This can be
estimated for each clump as the total kinetic energy (per unit mass)
divided by the dynamical time
($t_{\rm dyn}=2R_{\rm c}/\sigma_{\rm obs,los}$;
\citealt{maclow1998,maclow1999}):
\begin{equation}
  \label{eq:energy_dissipation_rate}
  \dot{\epsilon}_{\rm diss,clump}=\frac{1}{M_{\rm clump}}\,\frac{\frac{1}{2}M_{\rm clump}\sigma_{\rm obs,los}^2}{2R_{\rm clump}/\sigma_{\rm obs,los}}=\frac{1}{4}\frac{\sigma_{\rm obs,los}^3}{R_{\rm clump}}~.
\end{equation}
If turbulence is maintained by clump-clump collisions, we should expect
\begin{equation}
  \dot{\epsilon}_{\rm diss,clump}\approx\dot{\epsilon}_{\rm inject,coll}~.
\end{equation}

{\bf Comparison to NGC~404.} Panel~(c) of Fig.~\ref{fig:ccc_model}
compares the energy injection rate from clump-clump collisions
$\dot{\epsilon}_{\rm inject,coll}$ to the clumps' turbulent energy
dissipation rate $\dot{\epsilon}_{\rm diss,clump}$ in NGC~404.  The
latter is derived by utilising the observed median clump size
$R_{\rm clump}$ and median clump velocity dispersion
$\sigma_{\rm obs,los}$ at each radius. We find the energy injection
rates of collisions are sufficient to balance the turbulent energy
dissipation rates of the clumps in the molecular ring and the inner
parts ($R_{\rm gal}\lesssim8$~pc) of the central region, but not in
the outer parts ($8\lesssim R_{\rm gal}\lesssim15$~pc) of the central
region.

\subsubsection{Size -- linewidth relation regulated by clump-clump
  collisions}

To check whether the gas turbulence is indeed triggered by clump-clump
collisions, one must also compare the observed size -- linewidth
relation with that predicted from the collision model. This is because
the size -- linewidth relation is often interpreted as a signature of
turbulent motions, representing the length scales of velocity
correlations within a turbulent flow
\citep[e.g.][]{falgarone1991,elmegreen1996,romanduval2010}. Energy is
generally considered to be fed into the turbulent medium on its
largest spatial scale (the turbulence driving scale $L_{\rm D}$), and
to be transmitted through the so-called \enquote{turbulent cascade} to
smaller and smaller spatial scales $l$, until the energy reaches a
(very small) scale at which it is dissipated into heat
\citep[e.g.][]{bodenheimer2011}. If turbulence is maintained by
clump-clump collisions, the turbulence driving scale should be equal
to the critical collision length, i.e.\
\begin{equation}
  \label{eq:turbulence_driving_scale}
  L_{\rm D,coll}\approx\lambda_{\rm coll}=G\Sigma_{\rm gas,disc}/2A^2~.
\end{equation}
This is because $\lambda_{\rm coll}$ is the maximum distance within
which clumps can collide with each other and thus trigger turbulence.

As turbulence has self-similar properties only on spatial scales below
the driving scale, the size -- linewidth relation should exhibit a
power law on spatial scales $l<L_{\rm D}$, and then turn over and
flatten at $l\ge L_{\rm D}$ \citep[e.g.][]{blitz2006}. If we assume
the turbulent energy due to clump-clump collisions
$\dot{\epsilon}_{\rm inject,coll}$ cascades down to small spatial
scales with a constant energy dissipation rate (per unit mass), i.e.\
if we assume a Kolmogorov spectrum of turbulence
\citep[e.g.][]{kolmogorov1941,lighthill1955,kritsuk2007,mivilledeschenes2017},
we can pose
\begin{equation} 
  \label{eq:predicted_ediss}
  \dot{\epsilon}_{\rm diss}\equiv\frac{1}{4}\frac{\sigma_{\rm obs,los}^3(l)}{l}={\rm constant}=\dot{\epsilon}_{\rm inject,coll}~~~~~\mbox{for $l\le L_{\rm D}\approx\lambda_{\rm coll}$}~
\end{equation}
(see Eq.~\ref{eq:energy_dissipation_rate}). Combining this with
Eqs.~\ref{eq:energy_injection_rate} and
\ref{eq:turbulence_driving_scale}, we can infer the velocity
dispersion of clumps (and clouds) due to clump-clump (or cloud-cloud)
collisions $\sigma_{\rm c,coll}$ at all spatial scales:
\begin{equation}
  \label{eq:velocity_dispersion_pred_total} 
  \sigma_{\rm c,coll}=\left\{
    \begin{array}{lr}
      \big(2G^2\Sigma_{\rm gas,disc}^2/A\big)^{1/3}R_{\rm c}^{1/3} & \mbox{if $R_{\rm c}\lesssim L_{\rm D}\approx\lambda_{\rm coll}$}~;\\
      G\Sigma_{\rm gas,disc}/A\equiv\sigma_{\rm max,coll} & \mbox{if $R_{\rm c}\gtrsim L_{\rm D}\approx\lambda_{\rm coll}$}~.
    \end{array}
  \right.
\end{equation}
Hence, if clump-clump collisions constitute a major driver of
turbulence, the size -- linewidth relation is linked to the galaxy
properties $\Sigma_{\rm gas,disc}$ and $A$ only.

{\bf Comparison to NGC~404.} The red dashed line shown in each panel
of Fig.~\ref{fig:size_linewidth_relation} shows our predicted size --
linewidth relation using Eq.~\ref{eq:velocity_dispersion_pred_total},
with a slope of $1/3$, a flattening at a clump/cloud radius
$L_{\rm D}\approx\lambda_{\rm coll}=G\Sigma_{\rm gas,disc}/2A^2$
(Eq.~\ref{eq:turbulence_driving_scale}) and a maximum clump/cloud
velocity dispersion $\sigma_{\rm max,coll}=G\Sigma_{\rm gas,disc}/A$,
where the molecular gas mass surface density $\Sigma_{\rm gas,disc}$
and Oort's constant $A$ were spatially averaged within each region
(i.e.\ the central region, molecular ring and whole disc,
respectively). Our predicted size -- linewidth relation (with no free
parameter!) strongly resembles the observed $R_{\rm c}$ --
$\sigma_{\rm obs,los}$ trend in the molecular ring (see the middle
panel of Fig.~\ref{fig:size_linewidth_relation}), including its slope,
normalisation, turn-over scale and plateau. This lends strong support
to our conjecture that the observed turbulence in the molecular ring
of NGC~404 is maintained by clump-clump collisions.

In the central region, although the energy injection rate due to
collisions appears to be sufficient to balance the turbulent energy
dissipation rate of the clumps in the very inner parts
($R_{\rm gal}\le8$~pc; see panel~(c) of Fig.~\ref{fig:ccc_model}), the
observed velocity dispersions are much larger than those predicted by
our collision model (and a turn-over at the scale predicted is not
observed). Mechanisms other than collisions are therefore required to
explain the higher level of turbulence (and larger turbulence driving
scale) in that region. We will compare clump-clump collisions with
stellar feedback in
Section~\ref{sec:effects_stellar_feedback_turbulence}, and discuss the
possibility of clump migration as a source of turbulence in the
central region in Section~\ref{sec:clump_migration}.

\subsubsection{Vertical support from collision-induced turbulence}
\label{section:vertical_support_collision_turbulence}

If the disc velocity dispersion $\sigma_{\rm gas,disc}$ is driven by
clump-clump collisions and maintains the gas disc in vertical
equilibrium, it should be approximately equal to (or smaller than) the
maximum velocity dispersion possibly maintained by clump-clump
collisions ($\sigma_{\rm max,coll}$; see
Eq.~\ref{eq:velocity_dispersion_pred_total}), i.e.\
\begin{equation}
  \label{eq:disc_velocity_dispersion}
  \sigma_{\rm gas,disc}\approx\sigma_{\rm max,coll}\approx G\Sigma_{\rm gas,disc}/A ~.
\end{equation}
In addition, as the gas disc scale height $h_{\rm gas,disc}$ should be
comparable to the spatial scale on which the turbulence is driven
\citep[e.g.][]{blitz2006,swinbank2011,hughes2013,kim2015}, we also
expect
\begin{equation}
  \label{eq:disc_scale_height}
  h_{\rm gas,disc}\approx L_{\rm D}\approx\lambda_{\rm coll}=G\Sigma_{\rm gas,disc}/2A^2~
\end{equation}  
(see Eq.~\ref{eq:turbulence_driving_scale}). Therefore, if the
turbulent kinetic energy that supports the gas disc in vertical
equilibrium is maintained by clump-clump collisions, the disc velocity
dispersion $\sigma_{\rm gas,disc}$ and vertical scale height
$h_{\rm gas,disc}$ are again linked only to the galaxy properties
$\Sigma_{\rm gas,disc}$ and $A$.
 
\begin{figure*}
  \includegraphics[width=0.98\textwidth]{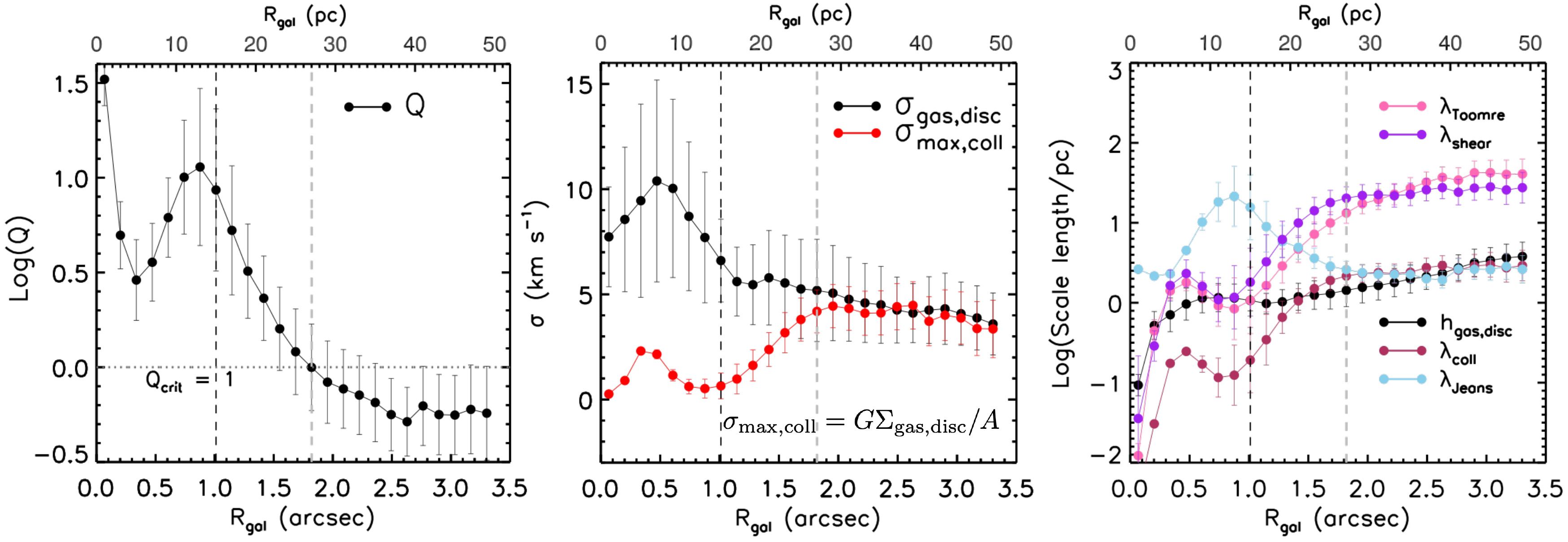}
  \caption{Radial variations of molecular gas disc properties of
    NGC~404, averaged over galactocentric radius bins of $2$~pc. In
    each panel, the black vertical dashed line indicates the boundary
    ($R_{\rm gal}=15$~pc) between the central region and molecular
    ring, while the grey vertical dashed line indicates the
    galactocentric distance ($R_{\rm gal}=27$~pc) beyond which the
    molecular gas disc is no longer gravitationally stable (i.e.\
    $Q\le1$ at $R_{\rm gal}\ge27$~pc). The error bars of each quantity
    indicate the $1~\sigma$ scatter of the values within each radial
    bin (not the uncertainty on the mean within each bin, that is much
    smaller). {\bf Left:} Toomre parameter $Q$ (Eq.~\ref{eq:Q}) as a
    function of galactocentric radius. The horizontal black dotted
    line indicates $Q_{\rm crit}=1$. The Toomre parameter $Q\gg1$ in
    the central region but decreases significantly in the molecular
    ring, and drops below unity at $R_{\rm gal}\ge27$~pc. {\bf
      Centre:} Comparison of the observed coarse-grained velocity
    dispersion of the molecular gas disc $\sigma_{\rm gas,disc}$ and
    our predicted maximum velocity dispersion sustained by clump-clump
    collisions $\sigma_{\rm max,coll}$
    (Eq.~\ref{eq:velocity_dispersion_pred_total}) at each
    galactocentric radius. We note that
    $\sigma_{\rm gas,disc}\gg\sigma_{\rm max,coll}$ in the central
    region, while $\sigma_{\rm gas,disc}\approx\sigma_{\rm max,coll}$
    in the molecular ring (especially at $R_{\rm gal}\ge27$~pc where
    $Q\le1$). {\bf Right:} Comparisons of the molecular gas disc scale
    height $h_{\rm gas,disc}$ (Eq.~\ref{eq:disc_scale height}),
    collision critical length $\lambda_{\rm coll}$
    (Eq.~\ref{eq:lambda_coll}) and Jeans length $\lambda_{\rm Jeans}$
    (Eq.~\ref{eq:Jeans_length}), and comparison of the Toomre length
    $\lambda_{\rm Toomre}$ (Eq.~\ref{eq:Toomre_length}) and shear
    length $\lambda_{\rm shear}$
    (Eq.~\ref{eq:max_tidally_stable_size}) at each galactocentric
    radius.  The scale height
    $h_{\rm gas,disc}\approx\lambda_{\rm coll}\approx\lambda_{\rm
      Jeans}$ and $\lambda_{\rm Toomre}\approx\lambda_{\rm Shear}$ in
    the molecular ring of NGC~404 (especially at $R_{\rm gal}\ge27$~pc
    where $Q\le1$; see
    Sections~\ref{sec:gravitational_instability_onset_collisions} and
    \ref{sec:stability_multiple-scale}, respectively).}
  \label{fig:galaxy_instability}
\end{figure*}

{\bf Comparison to NGC~404.} In NGC~404, we do find that the observed
one-dimensional (coarse-grained) gas disc velocity dispersions
$\sigma_{\rm gas,disc}$ agree well with our predicted maximum velocity
dispersions $\sigma_{\rm max,coll}$ in the molecular ring (especially
at $R_{\rm gal}\ge27$~pc; see the middle panel of
Fig.~\ref{fig:galaxy_instability}), where $\sigma_{\rm gas,disc}(R)$
was obtained again by averaging measurements within galactocentric
annuli of increasing $R_{\rm gal}$ (with a radial bin size of $2$~pc
and a fixed position angle $PA=1^\circ$ and inclination angle
$i=9\fdg3$). A very good agreement between our estimated scale height
$h_{\rm gas,disc}$ and the turbulence driving scale
$L_{\rm D}\approx\lambda_{\rm coll}$ is also found in the molecular
ring (see the right panel of Fig.~\ref{fig:galaxy_instability}). We
note that for this purpose we estimated the disc scale height as
\begin{equation}
  \label{eq:disc_scale height}
  h_{\rm gas,disc}\approx\sigma_{\rm gas,disc}/\sqrt{4\pi G\rho_\ast}~,
\end{equation}
that assumes a disc in vertical equilibrium
\citep[e.g.][]{koyama2009}, where $\rho_\ast\equiv \rho_\ast(R)$ is
the stellar mass volume density at each galactocentric radius, here
taken from the stellar mass model presented in \citet{nguyen2017} and
\citet{davis2020}.

In the central region, however, the observed disc velocity dispersions
$\sigma_{\rm gas,disc}$ and disc scale heights $h_{\rm gas,disc}$ are
significantly larger than the model predictions,
suggesting other sources of turbulence are required in that region.

\subsection{Gravitational instabilities as the onset of clump-clump collisions}
\label{sec:gravitational_instability_onset_collisions}

The above sections have shown that clump-clump collisions play an
important role to regulate clump properties and gas turbulence in the
molecular ring of NGC~404. In the central region, however, the clump
masses (see panel~(b) of Fig.~\ref{fig:ccc_model}) and velocity
dispersions (see left panel of Fig.~\ref{fig:size_linewidth_relation})
are significantly underestimated by our collision model. A question
then arises: under what circumstance does our clump-clump collision
model hold? Numerical simulations have shown that gravitational
instabilities coupled with galactic shear is probably the only
mechanism able to generate a population of clumps that undergo mutual
gravitational interactions and merging
\citep{kim2007,agertz2009,agertz2015,dekel2009b,elmegreen2010,aumer2010,goldbaum2015}.
It has been further shown that gravitational instabilities are of
great importance to generate turbulent motions from ordered circular
motions
\citep[e.g.][]{agertz2009,dekel2009,goldbaum2016,krumholz2016}.
Indeed, gas would remain on approximately circular orbits without the
onset of gravitational instabilities \citep{agertz2009}.

According to the standard analysis, a thin rotating gaseous disc
becomes unstable if the Toomre parameter $Q$ is smaller than a
critical value $Q_{\rm crit}$ that is approximately unity
\citep{toomre1964,lin1987,binney2008}:
\begin{equation}
  \label{eq:Q}
  Q\equiv Q(R)=\frac{\sigma_{\rm gas,disc}\kappa}{\pi G\Sigma_{\rm gas,disc}}<Q_{\rm crit}\approx1~,
\end{equation}
where
$\kappa\equiv\kappa(R)=\left(R\frac{d\Omega^2(R)}{dR}+4\Omega^2(R)\right)^\frac{1}{2}$
is the epicyclic frequency.
Figure~\ref{fig:a_vs_r} shows the dependence of $\kappa$ on the
galactocentric distance $R$ (i.e.\ $R_{\rm gal}$) in NGC~404.

As shown in the left panel of Fig.~\ref{fig:galaxy_instability}, the
central region of NGC~404 ($R_{\rm gal}<15$~pc) has $Q$ significantly
larger than unity at all radii ($Q=3$ -- $30$), while the molecular
ring ($R_{\rm gal}\ge15$~pc) has $Q$ smaller than unity at
$R_{\rm gal}\ge27$~pc, where most ($\approx87\%$) of the molecular gas
in the molecular ring is located. This suggests the central region is
strongly gravitationally stable, while the molecular ring is
gravitationally unstable or only marginally stable. The observed
trends of the Toomre parameter $Q$ in NGC~404 seem to suggest that
gravitational instabilities can indeed trigger clump-clump collisions,
as already pointed out by many numerical simulations
\citep[e.g.][]{dekel2009b,elmegreen2010,goldbaum2015,agertz2009,agertz2015}.
Indeed, we find a much better match between our model predictions and
the observations at $R_{\rm gal}\ge27$~pc where $Q\le1$ (see e.g.\ all
the panels of Fig.~\ref{fig:ccc_model} and the middle and right panels
of Fig.~\ref{fig:galaxy_instability}).

More importantly, using
$\sigma_{\rm gas,disc}\approx\sigma_{\rm max,coll}$ (see
Eq.~\ref{eq:disc_velocity_dispersion}) and assuming a flat circular
velocity curve ($\kappa=\sqrt{2}\Omega=2\sqrt{2}A$), as is
approximately the case in the molecular ring of NGC~404 (see
Fig.~\ref{fig:a_vs_r}), we can rewrite the Toomre parameter
(Eq.~\ref{eq:Q}) as
\begin{equation}
  \label{eq:predicted_Q}
  \begin{split}
    Q_{\rm coll}\equiv Q_{\rm coll}(R) & \approx\frac{\sigma_{\rm max,coll}\,\kappa}{\pi G\Sigma_{\rm gas,disc}}\\
    & \approx\frac{(G\Sigma_{\rm gas,disc}/A)(2\sqrt{2}A)}{\pi G\Sigma_{\rm gas,disc}}=\frac{2\sqrt{2}}{\pi}\approx1 ~.
  \end{split}
\end{equation}
It thus seems that, if the velocity dispersions supporting a gas disc
are generated by clump-clump collisions, this disc will necessarily
self-regulate and remain marginally gravitationally stable (i.e.\
$Q\approx1$). Conversely, for a gas disc with $Q\gg1$, the disc
velocity dispersions $\sigma_{\rm gas,disc}$ are expected to be
significantly larger than the maximum velocity dispersions generated
by clump-clump collisions $\sigma_{\rm max,coll}$, as is indeed the
case in the central region of NGC~404 (see middle panel of
Fig.~\ref{fig:galaxy_instability}). Thus, other mechanisms are
required to stir the turbulence in such a disc (e.g.\ 
  clump migration, stellar feedback and gas inflows/outflows;
\citealt{krumholz2015}).

A simple criterion to ascertain whether clump-clump collisions are
efficient was given in Section~\ref{sec:clump_mass_collision_model}, a
comparison of the collision timescale $t_{\rm coll,clump}$
(Eq.~\ref{eq:collision_timescale}) and shear timescale $t_{\rm shear}$
(Eq.~\ref{eq:shear_time}), i.e.\
$t_{\rm coll,clump}\le t_{\rm shear}$. But how does this criterion
imply a gas disc that is gravitationally unstable or marginally stable
(i.e.\ a gas disc with $Q\lesssim1$)?
For a thin gas disc, the length scale at which gravitational
fragmentation occurs is approximately the two-dimensional Jeans length
\citep{bournaud2010,swinbank2011,dobbs2013,swinbank2015}
\begin{equation}
  \label{eq:Jeans_length}
  \lambda_{\rm Jeans}\equiv\lambda_{\rm Jeans}(R)\approx\pi\sigma^2_{\rm gas,disc}/8G\Sigma_{\rm gas,disc}~.
\end{equation}
Thus, the mass of the clumps formed via large-scale gravitational
fragmentation should be approximately equal to
\begin{equation}
  \label{eq:Jeans_mass}
  M_{\rm clump} \approx M_{\rm Jeans}\equiv M_{\rm Jeans}(R)=\Sigma_{\rm gas,disc}\lambda_{\rm Jeans}^2=\left(\frac{\pi}{8G}\right)^2\frac{\sigma_{\rm gas,disc}^4}{\Sigma_{\rm gas,disc}}
\end{equation}
\citep{kim2001}. We can then obtain the collision timescale that
depends on the Jeans mass:
\begin{equation}
  \begin{split}
  \label{eq:tcoll_tshear_relation}
    t_{\rm coll,clump} & \approx \frac{A^{1/3}M_{\rm Jeans}^{1/3}}{2^{1/3}G^{2/3}\Sigma_{\rm gas,disc}}=\frac{\pi^2A^{1/3}Q^{4/3}}{2^{7/3}\kappa^{4/3}}\\
      & \approx \frac{\pi^2}{2^{10/3}}\left(\frac{1}{2A}\right)Q^{4/3}\approx t_{\rm shear}\,Q^{4/3}~,
  \end{split}
\end{equation}
where for the last line we have assumed a flat circular velocity curve
(i.e.\ $\kappa=2\sqrt{2}A$). Equation~\ref{eq:tcoll_tshear_relation}
implies that the timescale ratio $t_{\rm coll,clump}/t_{\rm shear}$ is
directly related to the Toomre parameter $Q$. If $Q\gg1$,
$t_{\rm coll,clump}\gg t_{\rm shear}$ and clump-clump collisions will
not be relevant in the disc. Only when $Q\le1$ does
$t_{\rm coll,clump}\le t_{\rm shear}$ and clump-clump collisions
become important.

It thus seems that gravitational instabilities are the ultimate
sources of the turbulent energy. As a sanity check, we now compare the
energy injection rate from gravitational instabilities
$\dot{\epsilon}_{\rm inject,GI}$ to that from clump-clump collisions
$\dot{\epsilon}_{\rm inject,coll}$
(Eq.~\ref{eq:energy_injection_rate}). The energy injection rate (per
volume) from gravitational instabilities is of the order of
\begin{equation}
  \dot{e}_{\rm inject,GI}\approx G(\Sigma_{\rm gas,disc}/h_{\rm gas,disc})^2L_{\rm D}^2\,\Omega\approx G\Sigma_{\rm gas,disc}^2\Omega
\end{equation}
(see Eq.~46 of \citealt{maclow2004}),
where we have used $h_{\rm gas,disc}\approx L_{\rm D}$ as the scale
height of the gas disc should be approximately equal to the turbulence
driving scale for gravitationally-unstable gas discs
\citep{blitz2006,swinbank2011,hughes2013,kim2015}, and as is indeed
the case for the molecular ring of NGC~404 (see the right panel of
Fig.~\ref{fig:galaxy_instability}). Thus, the energy injection rate
(per unit mass) from gravitational instabilities
$\dot{\epsilon}_{\rm inject,GI}$ is
\begin{equation}
  \label{eq:e_gi}
  \begin{split}
    \dot{\epsilon}_{\rm inject,GI} & =\dot{e}_{\rm inject,GI}\,/\rho_{\rm gas,disc} \\
    & \approx G\Sigma_{\rm gas,disc}^2\Omega/(\Sigma_{\rm gas,disc}/2h_{\rm gas,disc}) \\
    & \approx 2G\Sigma_{\rm gas,disc}\,\Omega\,h_{\rm gas,disc}~,
  \end{split}
\end{equation} 
where we have used
$\rho_{\rm gas,disc}=\Sigma_{\rm gas,disc}/2h_{\rm gas,disc}$.
Panel~(c) in Fig.~\ref{fig:ccc_model} shows the derived energy
injection rate (per unit mass) from gravitational instabilities
$\dot{\epsilon}_{\rm inject,GI}$ as a function of the galactocentric
radius  $R_{\rm gal}$ in NGC~404. In the molecular ring,
these energy injection rates (per unit mass) from gravitational
instabilities $\dot{\epsilon}_{\rm inject,GI}$ are indeed within an
order of magnitude of those from clump-clump collisions
$\dot{\epsilon}_{\rm inject,coll}$
($\dot{\epsilon}_{\rm inject,GI}\approx4\,\dot{\epsilon}_{\rm
  inject,coll}$). This provides further evidence for gravitational
instabilities as the onset of clump-clump collisions and
turbulence. We also note that $\dot{\epsilon}_{\rm inject,GI}$ should
in fact always be $\approx4$ times larger than
$\dot{\epsilon}_{\rm inject,coll}$ for a flat rotation curve (i.e.\
$\Omega=2A$), as
$h_{\rm gas,disc}\approx\lambda_{\rm coll}=G\Sigma_{\rm
  gas,disc}/2A^2$ (see Eqs.~\ref{eq:disc_scale_height} and
\ref{eq:energy_injection_rate}).

\section{Implications of Collision-Induced Turbulence}
\label{sec:implications_collision-induced_turbulence}

In Section~\ref{sec:turbulence_collision_model}, we characterised the
driving scale and energy injection rate of turbulence from clump-clump
collisions. The results have profound implications for the dynamics
and density structures of molecular gas, as interstellar turbulence is
one of the main agents opposing gravity
\citep[e.g.][]{hennebelle2012,federrath2012,padoan2014} and shaping
the molecular ISM \citep[e.g.][]{kritsuk2011,orkisz2017}.

In this section, we will discuss how clump-clump collision-driven
turbulence affects the dynamical states (virial parameters
$\alpha_{\rm vir,c}$) and density structures (mass -- size relations)
of clumps and clouds. We will tackle the following questions with
regard to NGC~404: (1) why are clumps in rough virial equilibria
(i.e.\ $\alpha_{\rm vir,clump}\approx2$) while clouds are strongly
self-gravitating (i.e.\ $\alpha_{\rm vir,cloud}<1$) in the molecular
ring (see Section~\ref{sec:virial_analysis_multiple-scale}); (2) why
do clumps have a mass -- size relation ($D_{\rm m,clump}\approx1.7$)
different from that of clouds ($D_{\rm m,cloud}\approx2.1$) in the
molecular ring (see Section~\ref{sec:multiple-scale_mass-size}); and
(3) based on our estimates of the dynamical states and mass -- size
relation of clumps, can we predict the clump sizes in the molecular
ring?

\subsection{Stability of multiple-scale molecular structures}
\label{sec:stability_multiple-scale}

Neglecting magnetic fields, the stability and dynamics of molecular
gas structures are generally governed by self-gravity and interstellar
turbulence. In a differentially-rotating gas disc, shear (and tides,
i.e.\ external/galactic gravity) is an additional factor. This leads
to a complex interplay between shear and turbulence, both opposing
self-gravity.
The key question is then in which ranges of size and mass do shear,
turbulence and self-gravity individually dominate the gas dynamics.

In a differentially-rotating gas disc, there are two critical
lengths. One is the Toomre length
\begin{equation}
  \label{eq:Toomre_length}
  \lambda_{\rm Toomre}\equiv\lambda_{\rm Toomre}(R)=2\pi^2G\Sigma_{\rm gas,disc}/\kappa^2~,
\end{equation}
where the numerical coefficient $2\pi^2$ applies to infinitely thin
gas discs \citep{tasker2009}. The other is the Jeans length
$\lambda_{\rm Jeans}\approx\pi\sigma^2_{\rm gas,disc}/8G\Sigma_{\rm
  gas,disc}$ (Eq.~\ref{eq:Jeans_length}). Molecular structures with
$L_{\rm acc}\ge\lambda_{\rm Toomre}$ (where $L_{\rm acc}$ is the
accumulation length; Eq.~\ref{eq:accumulation_length}) are supported
against gravity by shear motions and thus can not collapse, while
molecular structures with $L_{\rm acc}\le\lambda_{\rm Toomre}$ are
supported against gravity by turbulent motions. Only molecular
structures with $\lambda_{\rm Jeans}<L_{\rm acc}<\lambda_{\rm Toomre}$
are dominated by gravity, and thus can collapse gravitationally. The
two critical length scales also correspond to two critical masses: the
Toomre mass
\begin{equation}
  \label{eq:Toomre_mass}
  M_{\rm Toomre}\equiv M_{\rm Toomre}(R)=\Sigma_{\rm gas,disc}\lambda^2_{\rm Toomre}=\left(\frac{2\pi^2G}{\kappa^2}\right)^2\Sigma_{\rm gas,disc}^3
\end{equation}
and the Jeans mass
$M_{\rm Jeans}\equiv\Sigma_{\rm gas,disc}\lambda^2_{\rm
  Jeans}=\left(\frac{\pi}{8G}\right)^2\frac{\sigma_{\rm
    gas,disc}^4}{\Sigma_{\rm gas,disc}}$ (Eq.~\ref{eq:Jeans_mass}).

\begin{figure}
  \includegraphics[width=0.95\columnwidth]{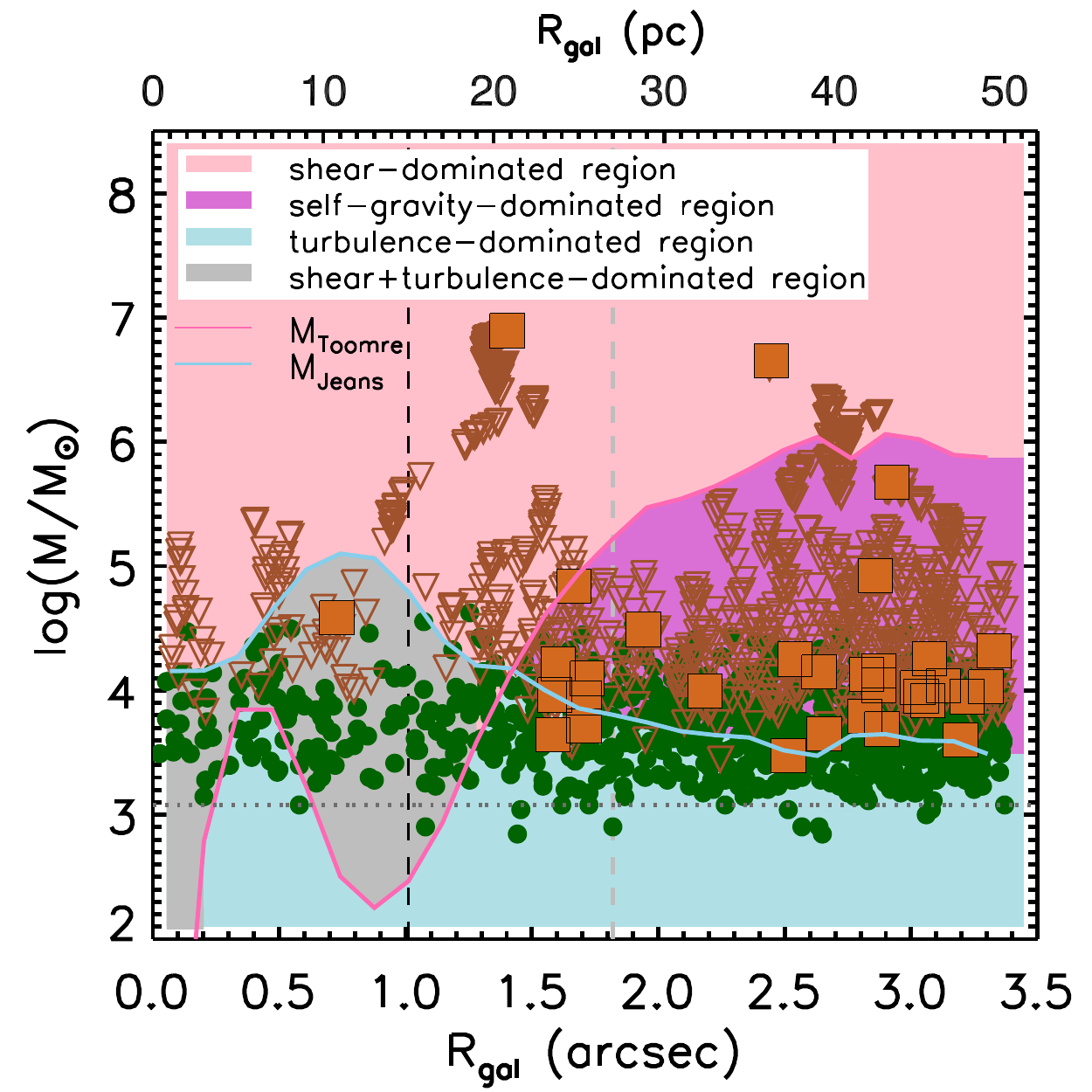}
  \caption{Various masses as a function of galactocentric distance in
    NGC~404. The Toomre mass ($M_{\rm Toomre}$) is shown as a pink
    line, the Jeans mass ($M_{\rm Jeans}$) as a blue line, and
    individual clumps and clouds as data points. Symbols are as for
    Fig.~\ref{fig:size_linewidth_relation}. In the pink-shaded region
    ($M_{\rm c}>M_{\rm Jeans}$ and $M_{\rm c}>M_{\rm Toomre}$),
    molecular gas structures are supported against gravity by
    shear. In the blue-shaded region ($M_{\rm c}\le M_{\rm Toomre}$
    and $M_{\rm c}\le M_{\rm Jeans}$), molecular gas structures are
    supported against gravity by turbulence. In the grey-shaded region
    ($M_{\rm Toomre}<M_{\rm c}\le M_{\rm Jeans}$), molecular gas
    structures are supported against gravity by both shear and
    turbulence. In the magenta-shaded region
    ($M_{\rm Jeans}<M_{\rm c}\le M_{\rm Toomre}$), molecular gas
    structures are gravitational unstable due to dominant
    self-gravity.
    The black vertical dashed line indicates the boundary
    ($R_{\rm gal}=15$~pc) between the central region and molecular
    ring, while the grey vertical dashed line indicates the
    galactocentric distance ($R_{\rm gal}=27$~pc) beyond which the
    molecular gas disc is no longer gravitationally stable (i.e.\
    Toomre parameter's $Q\le1$ at $R_{\rm gal}\ge27$~pc; see
    Section~\ref{sec:gravitational_instability_onset_collisions}).
    The black horizontal dotted line indicates our mass detection
    limit ($\log(10\delta_{\rm M}/{\rm M}_\odot)=3.08$; see
    Section~\ref{sec:mass_function_clumps}).  }
  \label{fig:balance_dynamics}
\end{figure}

Figure~\ref{fig:balance_dynamics} shows the Toomre and Jeans masses as
a function of the galactocentric radius $R$ (i.e.\ $R_{\rm gal}$), as
well as the masses of all clumps and clouds at their respective
galactocentric radii. There are four regions in
Fig.~\ref{fig:balance_dynamics}: (1) $M_{\rm c}>M_{\rm Jeans}$ and
$M_{\rm c}>M_{\rm Toomre}$ (pink-shaded region), where molecular
structures are supported against gravity or even disrupted by shear
(i.e.\ $\alpha_{\rm vir,c}>\alpha_{\rm vir,crit}$); (2)
$M_{\rm c}\le M_{\rm Toomre}$ and $M_{\rm c}\le M_{\rm Jeans}$
(blue-shaded region), where molecular structures are supported against
gravity by turbulence (i.e.\
$\alpha_{\rm vir,c}\ge\alpha_{\rm vir,crit}$); (3)
$M_{\rm Toomre}<M_{\rm c}\le M_{\rm Jeans}$ (grey-shaded region),
where molecular structures are supported against gravity or even
disrupted by both shear and turbulence (i.e.\
$\alpha_{\rm vir,c}\ge\alpha_{\rm vir,crit}$); and (4)
$M_{\rm Jeans}<M_{\rm c}\le M_{\rm Toomre}$ (magenta-shaded region),
where molecular structures are gravitationally unstable due to
dominant self-gravity (i.e.\
$\alpha_{\rm vir,c}<\alpha_{\rm vir,crit}$). We note that
$M_{\rm Jeans}>M_{\rm Toomre}$ (or equivalently
$\lambda_{\rm Jeans}>\lambda_{\rm Toomre}$) in the central region,
while $M_{\rm Jeans}<M_{\rm Toomre}$ (or equivalently
$\lambda_{\rm Jeans}<\lambda_{\rm Toomre}$) at
$R_{\rm gal}\gtrsim20$~pc in the molecular ring (see also the right
panel of Fig.~\ref{fig:galaxy_instability}). This is consistent with
our earlier conclusion that the central region is gravitationally
stable while the molecular ring is gravitationally unstable or only
marginally stable
(Section~\ref{sec:gravitational_instability_onset_collisions}).

\subsubsection{Gas dynamics in the molecular ring}
\label{sec:dynamics_molecular_ring}

We first assess the dynamical states of the molecular structures in
the molecular ring of NGC~404, where our collision model holds. We
find that the clumps in the molecular ring have their Jeans masses
$M_{\rm Jeans}$ (or Jeans lengths $\lambda_{\rm Jeans}$) approximately
equal to the collision critical masses $M_{\rm c,coll}$ (or the
collision critical lengths $\lambda_{\rm coll}$), i.e.\
$M_{\rm Jeans}\approx M_{\rm c,coll}$ (or
$\lambda_{\rm Jeans}\approx\lambda_{\rm coll}$), as shown in
Fig.~\ref{fig:balance_dynamics} (or the right panel of
Fig.~\ref{fig:galaxy_instability}). In fact, for a
gravitationally-unstable disc where the disc velocity dispersion
$\sigma_{\rm gas,disc}$ is set by the maximum velocity dispersion
sustained by clump-clump collisions
$\sigma_{\rm max,coll}=G\Sigma_{\rm gas,disc}/A$ (see
Eq.~\ref{eq:disc_velocity_dispersion} and
Section~\ref{section:vertical_support_collision_turbulence}), one
naturally expects
\begin{equation}
  \begin{split}
    \lambda_{\rm Jeans} & =\pi\sigma_{\rm gas,disc}^2/8G\Sigma_{\rm gas,disc} \\
                      & \approx(\pi/8G\Sigma_{\rm gas,disc})(G\Sigma_{\rm gas,disc}/A)^2 \\
                      &  =(\pi/4)\lambda_{\rm coll}\approx\lambda_{\rm coll}
  \end{split}~,
\end{equation}
and thus $M_{\rm Jeans}\approx M_{\rm c,coll}$. Our findings that
$\lambda_{\rm Jeans}\approx\lambda_{\rm coll}$ and
$h_{\rm gas,disc} \approx \lambda_{\rm coll}$ (see
Eq.~\ref{eq:disc_scale_height}) are consistent with calculations and
simulations showing that the average separation of clumps and the disc
thickness are primarily regulated by gravitational instabilities and
are about the Jeans length
\citep[e.g.][]{elmegreen2001,dutta2009,bournaud2010}.

{\bf Clumps are supported by turbulence.} The fact that
$\lambda_{\rm Jeans}\approx\lambda_{\rm coll}$ (or
$M_{\rm Jeans}\approx M_{\rm c,coll}$) has important implications for
the dynamical states of clumps and clouds in our collision model.
First, clumps should have their masses
$M_{\rm clump}\approx M_{\rm c,coll}\approx M_{\rm Jeans}$ (see
Fig.~\ref{fig:balance_dynamics}), and thus should be supported against
gravity by collision-induced turbulence (i.e.\
$\alpha_{\rm vir,clump}\approx\alpha_{\rm vir,crit}$). Indeed, clumps
are in rough virial equilibrium (mean virial parameter
$\langle\alpha_{\rm vir,clump}\rangle=1.82\pm0.07$; see
Section~\ref{sec:virial_analysis_multiple-scale}) in the molecular
ring. This is consistent with numerical simulations that show the
kinetic energy injected by clump-clump collisions is enough to
counterbalance the self-gravity of clumps in rough virial equilibrium
\citep[e.g.][]{tasker2009,wu2018,li2018}.

We note that turbulent energy injected by clump-clump collisions is
only sufficient to support clumps against self-gravity, but is not
enough to drive them out of virial equilibrium (i.e.\ disrupt them),
i.e.\ clumps should always remain in rough virial equilibrium with
$\alpha_{\rm vir,clump}\approx\alpha_{\rm vir,crit}$. This is because
clump-clump collisions can only happen effectively between
gravitationally-bound objects
\citep[e.g.][]{tan2000,gammie2001,tasker2009,goldbaum2016,takahira2018}.
If the turbulent velocity dispersion is abnormally high temporarily,
such that clumps become unbound
($\alpha_{\rm vir,clump}>\alpha_{\rm vir,crit}$), the collisions
between clumps become less frequent (as the number of bound clumps
decreases), driving $\alpha_{\rm vir,clump}$ downward again.
Thus, clumps will be self-regulated by their collisions to have
$\alpha_{\rm vir,c}\approx\alpha_{\rm vir,crit}$.

{\bf Clouds are dominated by self-gravity.} On the other hand, clouds
(that by definition contain more than one clump) should have their
masses $M_{\rm cloud}>M_{\rm c,coll}\approx M_{\rm Jeans}$ (see
Fig.~\ref{fig:balance_dynamics}), and thus should be dominated by
self-gravity (i.e.\ $\alpha_{\rm vir,cloud}<\alpha_{\rm vir,crit}$),
i.e.\ collision-induced turbulence can not support clouds against
gravity in virial equilibrium. Indeed, clouds have a mean virial
parameter much smaller than unity
($\langle\alpha_{\rm vir,cloud}\rangle=0.41\pm0.02$; see
Section~\ref{sec:virial_analysis_multiple-scale}) in the molecular
ring.

{\bf Most massive clouds are dominated by shear.} The most massive
clouds, however, may not be gravitationally bound as their dynamics
are likely dominated by shear rather than turbulence and
self-gravity. As one can see from Fig.~\ref{fig:balance_dynamics}, in
the molecular ring of NGC~404, the two most massive trunks (labelled
by the two largest contours in Fig.~\ref{fig:gmc_figure}) and some of
their largest branches (with $M_{\rm cloud}\gtrsim10^6$~M$_\odot$)
have masses larger than the Toomre masses (pink line in
Fig.~\ref{fig:balance_dynamics}), suggesting that these structures may
not be gravitationally bound.

In fact, \citet{liu2021} introduced a maximum size (diameter) for a
cloud to stay gravitationally bound in the presence of shear motions,
 \begin{equation}
  \label{eq:max_tidally_stable_size}
  \lambda_{\rm shear}\equiv 2R_{\rm shear}\approx\frac{3\pi G\Sigma_{\rm gas,disc}}{2A^2}
\end{equation}
(see Eq.~54 in \citealt{liu2021}), assuming that the cloud has a
spherical homogeneous density distribution.  An effective virial
parameter $\alpha_{\rm eff,vir}$, that provides a straightforward
measurable diagnostic of cloud boundedness in the presence of a
non-negligeable external potential, was also defined in
\citet{liu2021} (see their Eq.~34). A molecular structure with a size
(diameter) larger than $\lambda_{\rm shear}$ should have
$\alpha_{\rm eff,vir}>\alpha_{\rm vir,crit}$ \citep{liu2021}.

Our estimated maximum cloud size $\lambda_{\rm shear}$ is in good
agreement with the Toomre length $\lambda_{\rm Toomre}$
(Eq.~\ref{eq:Toomre_length}) in the molecular ring, with the ratio of
$\lambda_{\rm Toomre}$ and $\lambda_{\rm shear}$ between $0.5$ and
$1.5$ (see the right panel of Fig.~\ref{fig:galaxy_instability}).
This is expected, since the Toomre length also sets a natural spatial
scale over which molecular gas structures can not be bound by gravity
due to shear. The fact that
$\lambda_{\rm Toomre}\approx\lambda_{\rm shear}$ implies
$\alpha_{\rm eff,vir}>\alpha_{\rm vir,crit}$ for structures with
$M_{\rm cloud}>M_{\rm Toomre}$. Indeed, we find structures with
$M_{\rm cloud}>M_{\rm Toomre}$ to have estimated effective virial
parameters $\alpha_{\rm eff,vir}$ ranging from $2$ to $5$ in the
molecular ring.
As such, these massive clouds are not gravitationally bound. They are
likely to be disrupted or even torn apart by galactic shear.

\subsubsection{Gas dynamics in the central region}
\label{sec:dynamics_central_region}

The central region ($R_{\rm gal}\le15$~pc) of NGC~404 is found to be
gravitationally stable, with $Q=3$ -- $30$ (see
Fig.~\ref{fig:galaxy_instability}). Our collision model is therefore
not expected to hold there (see
Section~\ref{sec:gravitational_instability_onset_collisions}), and
indeed $M_{\rm Jeans}$ is significantly larger than $M_{\rm Toomre}$
throughout that region (see Fig.~\ref{fig:balance_dynamics}). This
implies that molecular structures in the central region should not
collapse gravitationally. As sources of turbulence other than
clump-clump collisions are likely present in the central region (see
Sections~\ref{sec:turbulence_collision_model} and
\ref{sec:gravitational_instability_onset_collisions}), not only clumps
but even clouds there appear to be supported against gravity by
turbulent motions ($\langle\alpha_{\rm vir,clump}\rangle=1.52\pm0.11$
and $\langle\alpha_{\rm vir,cloud}\rangle=1.14\pm0.12$ in the central
region).

A few of the most massive molecular structures in the very centre
($R_{\rm gal}\lesssim10$~pc) region have masses $M_{\rm cloud}$ much
larger than $M_{\rm Toomre}$, suggesting shear motions may dominate
their dynamics and tear them apart (i.e.\
$\alpha_{\rm eff,vir}>\alpha_{\rm vir,crit}$).

\subsection{Implications for the mass -- size relation}
\label{sec:implications_mass-size}

As shown in Section~\ref{sec:stability_multiple-scale}, clumps are
primarily dominated by turbulence and clouds by self-gravity (except
for the most massive clouds that are dominated by shear) in the
molecular ring. As the two main drivers of density structures in
molecular gas (thus the mass -- size relation) are turbulence and
self-gravity \citep{field2008,field2011,kritsuk2011b,gouliermis2018},
could it be that the different mass -- size relations for the clumps
and clouds in the molecular ring ($D_{\rm m,clump}=1.63\pm0.04$ versus
$D_{\rm m,cloud}=2.06\pm0.01$; see
Section~\ref{sec:multiple-scale_mass-size}) also originate from these
two different physical mechanisms (respectively turbulence and
self-gravity)? To answer this question, we predict the mass -- size
trend for both the turbulence-dominated and the self-gravity-dominated
regime below, and compare our predictions with observations.

\subsubsection{Mass -- size relation regulated by turbulence}
\label{sec:implications_mass-size_turbulence}

We first assess the mass -- size relation regulated by
collision-induced turbulence, i.e.\ the mass -- size relation of
clumps in the molecular ring. In our clump-clump collision scenario,
the turbulent energy injected by the collisions cascades down to small
scales and counterbalances self-gravity to produce a rough virial
equilibrium (i.e.\ the clumps are able to maintain virialisation; see
Section~\ref{sec:stability_multiple-scale}). These virialised clumps
should always have their energy dissipation rates (per unit mass)
$\dot{\epsilon}_{\rm diss,clump}=\frac{1}{4}\frac{\sigma_{\rm
    obs,los}^3}{R_{\rm clump}}$ (Eq.~\ref{eq:energy_dissipation_rate})
match their virial energy dissipation rates (per unit mass)
\begin{equation}
  \label{eq:virial_energy_dissipation_rate}
  \begin{split}
    \dot{\epsilon}_{\rm diss,vir} & =\frac{1}{4}\left(\alpha_{\rm vir,crit}\,GM_{\rm clump}/5\right)^{3/2}R_{\rm clump}^{-5/2}\\
  \end{split}
\end{equation} 
\citet{li2017}, i.e.\
$\dot{\epsilon}_{\rm diss,clump}\approx\dot{\epsilon}_{\rm diss,vir}$,
where as before $\alpha_{\rm vir,crit}\approx2$ is the boundary
between gravitationally-bound and unbound objects
\citep{kauffmann2013,kauffmann2017}. The virial energy dissipation
rate (per unit mass) $\dot{\epsilon}_{\rm diss,vir}$ is the energy
dissipation rate (per unit mass) a molecular gas structure would have
if it were virialised.
We therefore expect
\begin{equation}
  \dot{\epsilon}_{\rm diss,vir}\approx\dot{\epsilon}_{\rm diss,clump}\approx\dot{\epsilon}_{\rm inject,coll}
\end{equation} 
for the clumps in the molecular ring of NGC~404, where we have used
$\dot{\epsilon}_{\rm diss,clump}\approx\dot{\epsilon}_{\rm
  inject,coll}$, i.e.\ the energy dissipation rates (per unit mass) of
clumps are approximately equal to the energy dissipation rates (per
unit mass) from clump-clump collisions in the molecular ring (see
Section~\ref{sec:turbulence_collision_model}). We indeed find very
good agreements between the estimated $\dot{\epsilon}_{\rm diss,vir}$,
the measured $\dot{\epsilon}_{\rm diss,clump}$ and our predicted
$\dot{\epsilon}_{\rm inject,coll}$ in the molecular ring (see
panel~(c) of Fig.~\ref{fig:ccc_model}), providing more evidence that
the turbulence induced by clump-clump collisions can support and
maintain molecular ring clumps in virial equilibrium.

The facts that
$\dot{\epsilon}_{\rm diss,vir}=\frac{1}{4}\left(\alpha_{\rm
    vir,crit}\,GM_{\rm clump}/5\right)^{3/2}R_{\rm
  clump}^{-5/2}\approx\dot{\epsilon}_{\rm inject,coll}$, and that
$\dot{\epsilon}_{\rm inject,coll}$ is independent of clump properties
(it depends only on $\Sigma_{\rm gas,disc}$ and $A$; see
Eq.~\ref{eq:energy_injection_rate}), imply
$M_{\rm clump}^{3/2}R_{\rm clump}^{-5/2}\approx constant$ at any given
location in the disc (i.e.\ for any given $\Sigma_{\rm gas,disc}$ and
$A$). Thus, the mass -- size relation should have the form
$M_{\rm clump}\propto R_{\rm clump}^{5/3}$, as suggested by
\citet{li2017b}. Specifically, by equating the virial energy
dissipation rate $\dot{\epsilon}_{\rm diss,vir}$
(Eq.~\ref{eq:virial_energy_dissipation_rate}) to the energy injection
rate due to clump-clump collisions $\dot{\epsilon}_{\rm inject,coll}$
(Eq.~\ref{eq:energy_injection_rate}), we obtain
\begin{gather}
  \begin{split}
    \label{eq:predicted_mass_size}
    M_{\rm clump} & =\big(5/\alpha_{\rm vir,crit}G\big)\big(4\dot{\epsilon}_{\rm inject,coll}\big)^{2/3}R_{\rm clump}^{5/3}\\
    & =\big(5/\alpha_{\rm vir,crit}G\big)\big(2G^2\Sigma_{\rm gas,disc}^2/A\big)^{2/3}R_{\rm clump}^{5/3}
  \end{split}
\end{gather}
for virialised clumps supported by collision-induced turbulence. The
mass -- size relation of clumps should thus depend only on the
galactic properties (i.e.\ $\Sigma_{\rm gas,disc}$ and $A$). The red
dashed line in the middle panel of Fig.~\ref{fig:mass_size_relation}
shows our predicted mass -- size relation using
Eq.~\ref{eq:predicted_mass_size}, that is in very good agreement (both
slope {\em and} normalisation) with the observed trend of the clumps
in the molecular ring (green dashed line),
$M_{\rm clump}\propto R_{\rm clump}^{1.63\pm0.04}$.
In Eq.~\ref{eq:predicted_mass_size}, we have therefore predicted an
accurate mass -- size relation regulated by collision-induced
turbulence only, with no free parameter. We also note that the
molecular structures lying above the predicted mass -- size relation
in Fig.~\ref{fig:mass_size_relation} should have
$\alpha_{\rm vir,c}<\alpha_{\rm vir,crit}$, while those lying below
the relation should have $\alpha_{\rm vir,c}>\alpha_{\rm vir,crit}$.

Equation~\ref{eq:predicted_mass_size} also predicts the sizes of
clumps, i.e.\ the clump size $R_{\rm c,coll}$ corresponding to the
clump mass $M_{\rm c,coll}$:
\begin{gather}
  \begin{split}
    \label{eq:predicted_clump_size}
    R_{\rm clump}  & =\big(4\,\dot{\epsilon}_{\rm inject,coll}\big)^{-2/5}\left(\alpha_{\rm vir,crit}\,GM_{\rm clump}/5\right)^{3/5}\\
    & =\big(4\,\dot{\epsilon}_{\rm inject,coll}\big)^{-2/5}\left(\alpha_{\rm vir,crit}\,GM_{\rm c,coll}/5\right)^{3/5}\\
    & =\frac{1}{5^{3/5}}\,\frac{G\Sigma_{\rm gas,disc}}{2A^2}\\
    & \approx0.19\,G\Sigma_{\rm gas,disc}/A^2=0.38\,\lambda_{\rm coll}\equiv R_{\rm c,coll}~,\\
  \end{split}
\end{gather}
where we have used
$\dot{\epsilon}_{\rm inject,coll}=G^2\Sigma_{\rm gas,disc}^2/2A$
(Eq.~\ref{eq:energy_injection_rate}),
$M_{\rm c,coll}=G^2\Sigma_{\rm gas,disc}^3/4A^4$
(Eq.~\ref{eq:mc_crit}) and $\alpha_{\rm vir,crit}=2$. It therefore
seems that the typical clump size regulated by clump-clump collisions
$ R_{\rm c,coll}$ is determined solely by galactic properties
($\Sigma_{\rm gas,disc}$ and $A$). A comparison between the observed
most common clump size, that also happens to be the median clump size
$R_{\rm clump}$, and our predicted $R_{\rm c,coll}$ at each
galactocentric radius is shown in panel~(d) of
Fig.~\ref{fig:ccc_model}. Again, a very good agreement is found in the
molecular ring of NGC~404.
Our results thus support numerical simulations suggesting that
clump-clump collisions can be an important mechanism to determine both
the masses and the sizes of clumps
\citep[e.g.][]{tasker2009,li2018}. In the central region, clumps have
sizes much larger than our predictions, implying that in that region
other physical mechanisms are required to explain their
formation. Clumps in the central region nevertheless have sizes
similar to those of clumps in the molecular ring, a feature we will
discuss in Section~\ref{sec:clump_migration}.

It is also worth mentioning here that, for a flat rotation curve
($\Omega=2A$), we have
$\lambda_{\rm coll}=G\Sigma_{\rm gas,disc}/2A^2=\sigma_{\rm
  max,coll}/\Omega\approx\sigma_{\rm gas,disc}/\Omega$ (see
Eqs.~\ref{eq:lambda_coll} and \ref{eq:velocity_dispersion_pred_total},
Section~\ref{sec:gravitational_instability_onset_collisions} and the
middle panel of Fig.~\ref{fig:galaxy_instability}) and therefore
\begin{equation}
  \begin{split}
    R_{\rm clump}/R_{\rm gal} & =R_{\rm c,coll}/R_{\rm gal}\approx0.38\,\lambda_{\rm coll}/R_{\rm gal}=0.38\,\lambda_{\rm coll}\,\Omega/V_{\rm circ}\\
    & \approx0.38\,\sigma_{\rm gas,disc}/V_{\rm circ}
  \end{split}
\end{equation}
(see Eq.~\ref{eq:predicted_clump_size}), where as usual $R_{\rm gal}$
is the galactocentric distance of the clump. This relation is a
perfect match to the best-fitting relation of the clumps in a sample
of local analogues to high-redshift galaxies in DYnamics of
Newly-Assembled Massive Objects (DYNAMO)-{\it HST} survey
\citep{fisher2017b}:
$R_{\rm clump}/R=(0.38\pm0.02)\,\sigma_{\rm gas,disc}/V_{\rm
  circ}$. \citet{fisher2017b} explained their observed relation by
instabilities in a self-gravitating disc, but only predicted a range
of $(R_{\rm clump}/R)/(\sigma_{\rm obs,los}/V_{\rm circ})$ ratios of
$1/3$ to $\sqrt{2}/3\approx0.47$ (see their Eq.~3). A precise match
between our model ($0.38$) and DYNAMO-{\it HST} observations
($0.38\pm0.02$) does not only suggest that the molecular ring of
NGC~404 is akin to high-$z$ star-forming disc galaxies, but also that
clump-clump collisions are an important mechanism to regulate clump
sizes across redshits.

\subsubsection{Mass -- size relation regulated by self-gravity}
\label{sec:implications_mass-size_gravity}

We now analyse the mass -- size relation regulated by self-gravity,
i.e.\ the mass -- size relation of clouds (rather than clumps) in the
molecular ring of NGC~404. As clouds are gravitationally unstable and
have $\alpha_{\rm vir,cloud}\ll\alpha_{\rm vir,crit}$, they should lie
above the mass -- size relation regulated by turbulence in the middle
panel of Fig.~\ref{fig:mass_size_relation}
(Eq.~\ref{eq:predicted_mass_size}; i.e.\ the red dashed line). Indeed,
the clouds in the molecular ring are located above both the predicted
(red dashed line) and measured (green dashed line) mass -- size
relations of clumps there.

In the molecular ring, the mass -- size relation of clouds also
exhibits a steeper power-law
($M_{\rm cloud}\propto R_{\rm cloud}^{2.06\pm0.01}$, brown dashed
line) than that of clumps
($M_{\rm clump}\propto R_{\rm clump}^{1.63\pm0.04}$, green dashed
line; see Section~\ref{sec:multiple-scale_mass-size}). This again
appears to be consistent with theoretical results suggesting that
structures subject to gravitational collapse should have fractal
dimensions $D_{\rm m}$ larger than the critical value
$D_{\rm m,crit}=2$ for gravitational instabilities
(\citealt{perdan1990,kritsuk2011b}).

\subsubsection{Clumps versus clouds}
\label{sec:clumps_clouds}

In summary, turbulence triggered by clump-clump collisions has
important implications for the structure and dynamics of molecular
gas. In a gravitationally-unstable or only marginally stable gas disc
where clump-clump collisions are efficient, such as the molecular ring
of NGC~404, the transition from the turbulence- to the
self-gravity-dominated regime seems to occur at a mass
$M_{\rm c, coll}=\Sigma_{\rm gas,disc}\lambda_{\rm coll}^2\approx
M_{\rm Jeans}$. Molecular structures with masses
$\approx M_{\rm c,coll}$ (or $\approx M_{\rm Jeans}$), that are
typically clumps, have their internal structures and dynamics
dominated by collision-induced turbulence (see
Section~\ref{sec:implications_mass-size_turbulence}), that is not only
able to support them against gravity in a rough virial equilibrium,
but also shapes them to yield a mass -- size relation with a power-law
index $D_{\rm m}$ smaller than $2$:
$M_{\rm clump}\propto R_{\rm clump}^{5/3}$. Molecular structures with
masses much larger than $M_{\rm c,coll}$ (or $M_{\rm Jeans}$), that
are primarily clouds, have their internal structures and dynamics
increasingly governed by self-gravity and thus appear gravitationally
unstable, as the turbulent kinetic energy injected by clump-clump
collisions is unable to support them against gravitational
collapse. The dominant self-gravity breaks the self-similarity of gas
structures established by collision-induced turbulence, leading to a
mass -- size relation with a steeper slope ($D_{\rm m}\gtrsim2$).

\section{Discussion}
\label{sec:discussion}

\subsection{Clump mass function due to clump-clump collisions}
\label{sec:clump_mass_function_collision_model}

If clumps are formed via collisions and mergers of (smaller) clumps
(i.e.\ clump-clump collisions), the clumps' lifetimes
$t_{\rm life,clump}$ should be approximately equal to their collision
timescales $t_{\rm coll,clump}$, as the clumps are destroyed by
becoming part of larger clumps; \citealt{tasker2009}. If we make the
zeroth-order approximation that the number density of clumps is
proportional to their lifetimes (i.e.\ their collision times), we
obtain
\begin{equation} 
  dN_{\rm clump}(M_{\rm clump})/dM_{\rm clump}\propto t_{\rm life,clump}\approx t_{\rm coll,clump}\propto M_{\rm clump}^{1/3}
\end{equation}
(see Eq.~\ref{eq:collision_timescale}), where as before
$N_{\rm clump}(M_{\rm clump})$ is the number of clumps with masses
$\ge M_{\rm clump}$ (see
Eq.~\ref{eq:differential_mass_distribution_function}). Thus, for those
clumps regulated by collisions (i.e.\ those clumps with
$M_{\rm clump}\le M_{\rm c,coll}$), the power-law index of the
differential mass distribution function should be
$\gamma_{\rm coll,clump}\approx1/3$.

Having said that, clumps with $M_{\rm clump}>M_{\rm c,coll}$, can not
form efficiently via clump-clump collisions, as they are disrupted by
shear before they can collide and merge with other clumps (see
Section~\ref{sec:clump_mass_collision_model}). The number density of
clumps with $M_{\rm clump}>M_{\rm c,coll}$ should thus decrease
sharply with mass, and the differential mass distribution function in
the high-mass regime ($M_{\rm clump}>M_{\rm c,coll}$) should have
$\gamma_{\rm coll,clump}<-2$. Counter-intuitively, this is also true
of the overall mass distribution function, that is primarily sensitive
to the mass distribution at high masses. This is consistent with the
result of hydrodynamic simulations by \citet{li2018}, who found the
cloud population regulated by cloud-cloud collisions to have a
cumulative mass distribution function with a power-law index
$\gamma\approx-2.4$. Hence, our collision model predicts the
differential mass distribution function of clumps to exhibit {\em two}
separate power laws: $\gamma_{\rm coll,clump}\approx1/3$ in the
low-mass regime ($M_{\rm clump}\le M_{\rm c,coll}$) and
$\gamma_{\rm coll,clump}<-2$ in the high-mass regime
($M_{\rm clump}>M_{\rm c,coll}$). The observed clump mass distribution
function in the molecular ring of NGC~404 indeed closely follows this
prediction (see Table~\ref{tab:results_summary} and
Section~\ref{sec:mass_function_clumps}), further suggesting that the
clump population in the molecular ring is indeed primarily regulated
by clump-clump collisions.

\subsection{Effect of stellar feedback on turbulence}
\label{sec:effects_stellar_feedback_turbulence}

As argued in Sections~\ref{sec:turbulence_collision_model} and
\ref{sec:gravitational_instability_onset_collisions}, clump-clump
collisions are likely to be an important mechanism driving the
turbulence in (the molecular ring of) NGC~404. Another important
source of turbulence in galaxies is stellar feedback, including
supernova explosions, stellar winds, ionising radiation and
protostellar outflows
\citep[e.g.][]{maclow2004,dib2006,agertz2009,krumholz2016}.
The question then arises: can stellar feedback also play an important
role to drive the turbulence in NGC~404, that is a dwarf galaxy?

One way to assess whether stellar feedback is important to drive
turbulence is to analyse whether clump-clump collisions alone are
sufficient to maintain the observed velocity
dispersions. High-resolution 3D numerical simulations of ISM
turbulence suggest that clump-clump collisions serve as a baseline
level for turbulence in galaxies, and any excess random motions
observed must be caused by additional sources such as stellar
feedback, AGN feedback and/or the magneto-rotational instability
\citep[e.g.][]{agertz2009,goldbaum2016}. If this is right, then
stellar feedback should not be necessary in the molecular ring of
NGC~404, as clump-clump collisions are sufficient to maintain the
observed velocity dispersions (see panel~(c) of
Fig.~\ref{fig:ccc_model} and
Section~\ref{sec:turbulence_collision_model}). In the central region,
however, the observed velocity dispersions are much larger than our
collision model predictions (see
Eq.~\ref{eq:velocity_dispersion_pred_total} and
Fig.~\ref{fig:size_linewidth_relation}), and additional energy sources
are required to disturb the gas there. Stellar feedback is one such
potential energy source.

\begin{figure}
  \includegraphics[width=0.95\columnwidth]{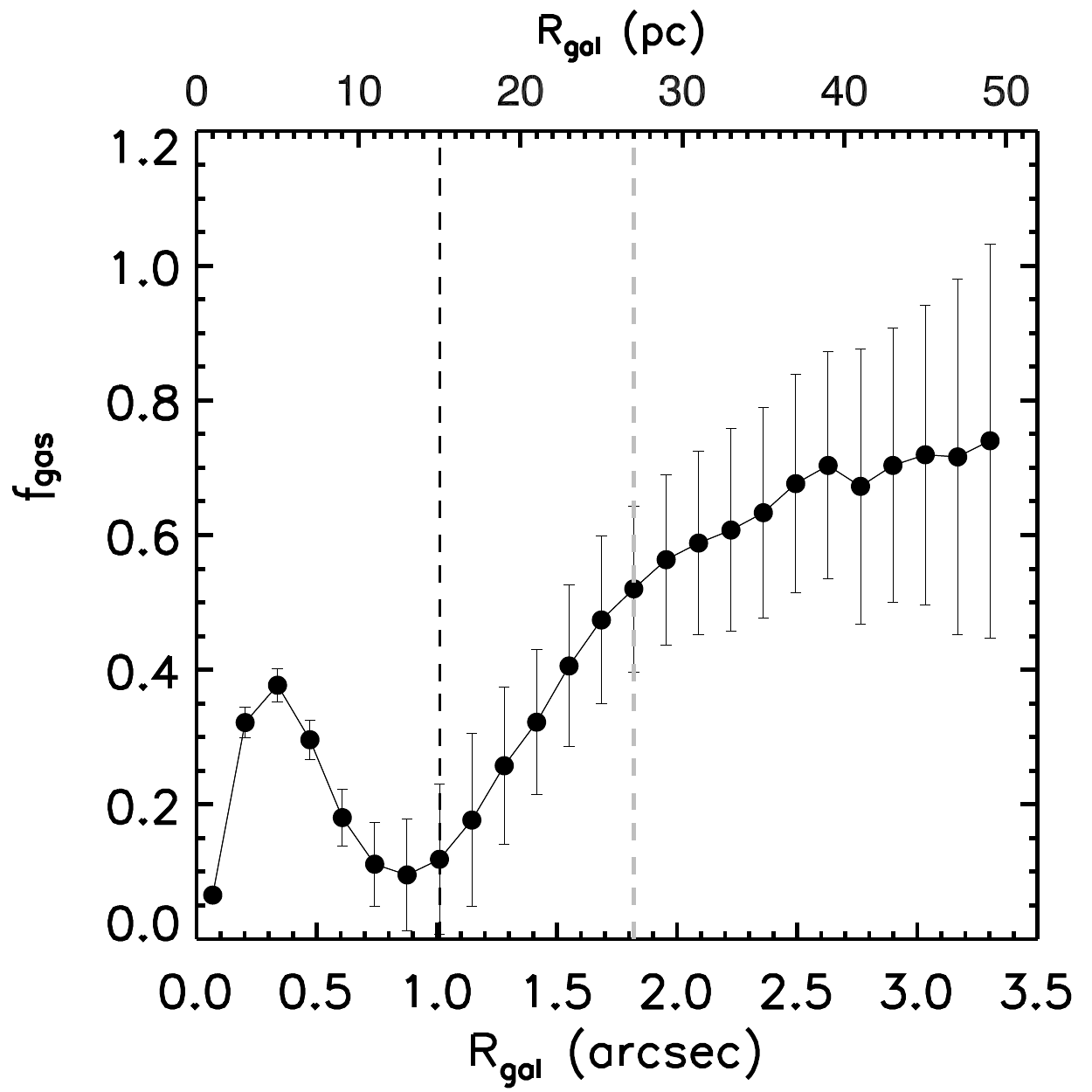}
  \caption{Galactocentric distance dependence of the gas fraction
    $f_{\rm gas}$ of NGC~404. The black vertical dashed line indicates
    the boundary ($R_{\rm gal}=15$~pc) between the central region and
    molecular ring, while the grey vertical dashed line indicates the
    galactocentric distance ($R_{\rm gal}=27$~pc) beyond which the
    molecular gas disc is no longer gravitationally stable (i.e.\
    Toomre parameter's $Q\le1$ at $R_{\rm gal}\ge27$~pc; see
    Section~\ref{sec:gravitational_instability_onset_collisions}).
    The error bars indicate the $1\sigma$ scatter of the different
    fractions within each radial bin (not the uncertainty on the mean
    within each bin, that is much smaller).}
  \label{fig:fraction_gas_vs_r}
\end{figure}

To explore the origin of disc turbulence using observational tests,
\citet{krumholz2016} constructed two simple analytical models
describing collision-driven turbulence and feedback-driven turbulence,
respectively. The models predict
${\rm SFR}\propto\sigma_{\rm obs,los}^2$ for feedback-driven
turbulence and ${\rm SFR}\propto\sigma_{\rm obs,los}\,f_{\rm gas}^2$
for collision-driven turbulence (i.e.\ turbulence triggered by
gravitational instabilities), where
$f_{\rm gas}\equiv\frac{\Sigma_{\rm gas,disc}}{\Sigma_{\rm
    gas,disc}+\Sigma_{\rm \ast,disc}}$ is the galaxy gas fraction and
$\Sigma_{\rm \ast,disc}$ is the (coarse-grained) stellar mass
surface density of the galactic disc.
Figure~\ref{fig:fraction_gas_vs_r} presents the galactocentric
distance dependence of the gas fraction $f_{\rm gas}$ of NGC~404,
assuming only the molecular gas observed here for the gas. The stellar
mass surface density is taken from the stellar mass model presented in
\citet{nguyen2017} and \citet{davis2020}. One can see that the gas
fraction $f_{\rm gas}$ is very high across the molecular ring, with a
median of $\approx0.5$, similar to those found in high-redshift
star-forming gas discs
\citep[e.g.][]{swinbank2015,kanekar2020,tacconi2020}.

To properly test the predictions of these two models in NGC~404,
spatially-resolved SFR measurements are required. High-resolution
($\approx0\farcs28$) and high-sensitivity
($\mu_{\rm 33GHz}\approx3.1$~$\mu$Jy~beam$^{-1}$) observations of the
$33$~GHz radio continuum emission (tracing free-free emission from
H{\scriptsize II} regions) have been carried out in with Very Large
Array (VLA; Liu et al.\ in prep). However, except for a very bright
central point source possibly associated with the AGN, no extended
emission is detected, suggesting an upper limit to the SFR surface
density of
\begin{equation}
  \begin{split}
    \left(\frac{\Sigma_{\rm SFR}}{\rm M_\odot~yr^{-1}~pc^{-2}}\right) & <4.6\times10^{-28}\left(\frac{T_{\rm e}}{10^4~{\rm K}}\right)\left(\frac{\nu}{{\rm GHz}}\right)^{0.1} \\
    ~~~~~ & \left(\frac{L_{v,{\rm beam}}}{\rm erg~s^{-1}~Hz^{-1}}\right)\left(\frac{\rm beam}{\rm pc^2}\right)^{-1} \\
    \left(\frac{L_{v,{\rm beam}}}{\rm erg~s^{-1}~Hz^{-1}}\right) & =4\pi\left(\frac{D}{\rm pc}\right)^2\times10^{-23}\times(3.0857\times10^{18})^2\left(\frac{\mu_{\rm 33GHz}}{\rm Jy~beam^{-1}}\right)
  \end{split}
\end{equation}
(see Eq.~11 in \citealt{murphy2011}), where $T_{\rm e}\approx11,000$~K
is the electron temperature, $\nu=33$~GHz the frequency,
$L_{v,{\rm beam}}$ the free-free spectral luminosity per beam,
$D=3.06\times10^6$~pc the distance of NGC~404,
$\mu_{\rm 33GHz}\approx3.1$~$\mu$Jy~beam$^{-1}$ the sensitivity of our
VLA observations and ${\rm beam}\approx20$~pc$^{2}$ the VLA
synthesised beam size. Thus, we derive a maximum SFR surface density
of
$\Sigma_{\rm SFR}\approx1\times10^{-6}$~M$_\odot$~yr$^{-1}$~pc$^{-2}$
for NGC~404, or an upper limit to the SFR volume density of
$\rho_{\rm SFR}\equiv\Sigma_{\rm SFR}/2h_{\rm
  gas,disc}\approx1.6\times10^{-7}$~M$_\odot$~yr$^{-1}$~pc$^{-3}$
(using $h_{\rm gas,disc}\approx3$~pc; see panel~(d) in
Fig.~\ref{fig:ccc_model}).
  
The energy injection rate (per volume) from supernova (SN) explosions
is given by
\begin{equation}
  \dot{e}_{\rm SN}=\xi_{\rm SN}E_{\rm SN}(f_{\rm SN}\rho_{\rm SFR})
\end{equation}
(see Eq.~4 of \citealt{kawakatu2020} and Eq.~55 of
\citealt{maclow2004}), where $\xi_{\rm SN}=0.1$ is the efficiency with
which SN energy is transferred to the gas
\citep[e.g.][]{thornton1998}, $E_{\rm SN}=10^{51}$~ergs (or
$5\times10^7$~M$_\odot$~km$^2$~s$^{-2}$) is the total energy injected
by one SN, and $f_{\rm SN}=10^{-2}$~M$_\odot^{-1}$ is the fraction of
supernovae per solar mass of star formation
\citep[e.g.][]{thompson2005}. The energy injection rate (per unit
mass) from supernova explosions is therefore
\begin{equation}
  \begin{split}
    \dot{\epsilon}_{\rm SN} & =\dot{e}_{\rm SN}/\rho_{\rm gas}=\xi_{\rm SN}E_{\rm SN}(f_{\rm SN}\rho_{\rm SFR})/\rho_{\rm gas} \\
    & \le2.5~{\rm km}^2~{\rm s}^{-2}~{\rm Myr}^{-1}~,
  \end{split}
\end{equation}
where we have used the $\rho_{\rm SFR}$ upper limit above and the
global gas volume density of
$\rho_{\rm gas} = \Sigma_{\rm gas,disc}/2h_{\rm
  gas,disc}\approx10^{3.5}$~M$_\odot$~pc$^{-3}$ as measured in NGC~404
(see Fig.\ \ref{fig:Sigma_gas_vs_r} and panel~(d) in Fig.\
\ref{fig:ccc_model} for the distributions of $\Sigma_{\rm gas,dis }$
and $h_{\rm gas,disc}$, respectively). This upper limit to the energy
injection rate (per unit mass) from stellar feedback is clearly
smaller than the observed energy dissipation rate (per unit mass) of
clumps and the energy injection rate (per unit mass) from clump-clump
collisions, i.e.\
$\dot{\epsilon}_{\rm SN}<\dot{\epsilon}_{\rm
  diss,clump}\approx\dot{\epsilon}_{\rm
  inject,coll}\approx10$~km$^2$~s$^{-2}$~Myr$^{-1}$ (see panel~(c) in
Fig.~\ref{fig:ccc_model}). It therefore seems that feedback-driven
turbulence is much less important than collision-driven turbulence in
NGC~404.

\subsection{Clump migration towards the galaxy centre}
\label{sec:clump_migration}

Panels~(b) and (d) of Fig.~\ref{fig:ccc_model} show the clumps of
NGC~404 have relatively uniform properties (i.e.\ masses and sizes) as
a function of galactocentric distance, with the median clump masses
and sizes varying by at most a factor of $\approx4$ and $\approx2$,
respectively. In fact, the clumps of NGC~404 also seem to have similar
velocity dispersions (or similar energy dissipation rates; see
panels~(c) in Fig.~\ref{fig:ccc_model}) at different galactocentric
distances.
These surprisingly small variations are unlikely to be due to the
dendrogram decomposition algorithm, as the dendrogram approach should
allow to identify clumps with sizes varying by an order of magnitude
\citep[e.g.][]{colombo2014,henshaw2016,henshaw2019,wong2019,krieger2020}.

It is easy to understand why the clump masses and sizes do not vary
much across the molecular ring.
Indeed, we have demonstrated that these clump properties depend only
on the galactic properties in the molecular ring, specifically
$\Sigma_{\rm gas,disc}$ and $A$ (see Eqs.~\ref{eq:mc_crit} and
\ref{eq:predicted_clump_size}, respectively). As neither varies
significantly across the molecular ring (see
Figs.~\ref{fig:Sigma_gas_vs_r} and \ref{fig:a_vs_r}, respectively),
the clump masses and sizes are not expected to exhibit much variation
across this region either (see the red data points in panels~(b) and
panel~(d) of Fig.~\ref{fig:ccc_model}).

The question then is why do the clumps in the central region, where
galactic properties are significantly different, also share the same
properties as those in the molecular ring? One obvious possibility is
that the clumps currently in the central region migrated inward from
larger galactocentric distances (i.e.\ from the molecular ring). We
have shown in Section~\ref{sec:turbulence_collision_model} that
clump-clump collisions can extract kinetic energy from ordered
differential rotation and inject this energy into turbulence. Clumps
must therefore loose angular momentum and migrate toward the galaxy
centre \citep{gammie2001,vollmer2002,dekel2009b,namekata2011}. This scenario of
clump migration toward galactic centres is discussed at length in many
theoretical works
\citep{gammie2001,elmegreen2007,elmegreen2008,dekel2009b,krumholz2018},
numerical simulations of high-$z$ gaseous discs
\citep{aumer2010,bournaud2014,forbes2014,goldbaum2015,mandelker2016}
and observations of high-$z$ star-forming galaxies
\citep{genzel2008,genzel2012,swinbank2011,guo2012,genzel2020,tacconi2020}.
The migrating clumps may eventually form bulges, that stabilise the
systems against further fragmentation
\citep[e.g.][]{genzel2008,aumer2010}.

Unfortunately, we have not found direct evidence of clump migration
(i.e.\ inflow) in NGC~404, as its gas disc is viewed nearly face-on
($i\approx9\fdg3$ at $R_{\rm gal}\ge15$~pc; \citealt{davis2020}) and
detecting clear non-circular motions is very difficult. Nevertheless,
there seems to be some indirect evidence supporting inward clump
migration. First, the molecular ring of NGC~404 is connected to the
central region via a single arm-like structure (see
Fig.~\ref{fig:mom0_image}). The kinematics of this arm is complex and
fairly chaotic, with signs of strong streaming motions that may be due to
the funnelling gas from the outer regions inward
\citep{davis2020}. Second, the molecular ring is incomplete (see,
again, Fig.~\ref{fig:mom0_image}), implying it has recently been
disrupted, possibly by inward clump migration. A key requirement for
making inward clump migration a long-term phenomenon is a continuous,
rapid supply of cold gas, that would replenish the ring (or disc) as
it is being drained \citep{dekel2009}.
If the gas supplied to the molecular ring is reduced, the molecular
ring will gradually be destroyed, as seen in NGC~404.

If this migration scenario is true and clumps survive with physical
properties largely unchanged for a migration timescale (i.e.\ as they
migrate toward the galactic centre), one would naturally expect the
clumps in the central region and molecular ring to have similar clump
properties. We now calculate the timescale of clumps 
  migrating from the molecular ring to the central region, according
to standard accretion disc theory \citep[e.g.][]{lynden-bell1974}. The
timescale for clumps to migrate to the galaxy centre due to collisions
(i.e.\ viscosity) is
\begin{equation}
  t_{\rm migrate}=\frac{1}{4}\,L_{0}^2\,V_0^{-2\,}v^{-1}
\end{equation}
(see Section~3.1 of \citealt{lynden-bell1974}), where
$L_0=R_0^2\,\Omega_0$ is the angular momentum (per unit mass) of a
clump at galactocentric distance $R_0$, $V_0=R_0\,\Omega_0$ is the
orbital circular velocity of the clump, $\Omega_0$ is the angular
velocity of orbital circular rotation of the clump, and
$v=\frac{1}{3}v_{\rm shear}\lambda_{\rm coll}$ is the kinetic
viscosity. By adopting a galactocentric distance at the boundary
between the central region and molecular ring, i.e.\ $R_0=15$~pc, we
derive a migration timescale $t_{\rm migrate}\approx10$~Myr.
This rapid inward migration of clumps from the molecular
  ring to the central region
  ($t_{\rm migrate}\approx5.6~ t_{\rm orbit}$ at $R_0=15$~pc) suggests
  that the current molecular ring of NGC~404 is highly unstable and
  transient. Migrating clumps from the molecular ring may eventually
  merge once in the central region, forming a larger
  gravitationally-stable system where shear-driven clump-clump
  collisions (and thus clump migrations) can not be important.
  
Clump migration may also trigger turbulence by converting the kinetic
and gravitational energies of the clumps into turbulent energy
\citep[e.g.][]{krumholz2010,forbes2012,dekel2014,forbes2014}. Could it
be that the large disc velocity dispersion observed in the central
region of NGC~404 triggered by clump migration? The energy injection
rate (per unit mass) due to clump migration is \citep{dekel2014}
\begin{equation}
  \dot{\epsilon}_{\rm inject,migrate}=V_0^{2}/t_{\rm migrate}~.
\end{equation}
At a galactocentric distance of $R_0=15$~pc,
$\dot{\epsilon}_{\rm
  inject,migrate}\approx230$~km$^2$~s$^{-2}$~Myr$^{-1}$. If the disc
velocity dispersion is maintained by clump migration, one expects
\begin{equation}
  \dot{\epsilon}_{\rm inject,migrate}\approx\dot{\epsilon}_{\rm diss,disc}=\frac{1}{2}\sigma_{\rm gas,disc}^2/(2h_{\rm gas,disc}/\sigma_{\rm gas,disc})
\end{equation}
(see Eq.~6 in \citealt{dekel2014}), where
$\dot{\epsilon}_{\rm diss,disc} $ is the energy dissipation rate (per
unit mass) of the gas disc and
$2h_{\rm gas,disc}/\sigma_{\rm gas,disc}$ is the turbulence decay
timescale. Indeed, the central region (where
$\sigma_{\rm gas,disc}\approx10$~km~s$^{-1}$ and
$h_{\rm gas,disc}\approx1$~pc) has an average disc energy dissipation
rate (per unit mass)
$\dot{\epsilon}_{\rm diss,disc}\approx250$~km$^2$~s$^{-2}$~Myr$^{-1}$,
that is approximately equal to the calculated energy injection rate
(per unit mass) from clump migration. It thus seems that clump
migration may be the major mechanism driving turbulence in the central
region of NGC~404, where the observed large velocity dispersions
cannot be explained by clump-clump collisions or stellar feedback.
However, other mechanisms may also play a role given the potential
jet-ISM interaction in the core of this galaxy.

\section{Conclusions}
\label{sec:conclusions}

We studied the molecular structures (clumps and clouds) of the dwarf
lenticular galaxy NGC~404. We performed a dendrogram analysis of our
high-resolution ($\approx0.86\times0.51$~pc$^2$) ALMA $^{12}$CO(2-1)
data and identified a number of nested structures, including $953$
resolved clumps (i.e.\ leaves) and $1639$ resolved clouds (i.e.\
branches and trunks), whose radii range from $0.4$ -- $25$~pc. Our
main findings are as follows.

\begin{itemize}

\item Two distinct regions are identified: a gravitationally-stable
  central region (Toomre parameter $Q=3$ -- $30$ and gas fraction
  $f_{\rm gas}\approx10$ -- $40\%$) and a gravitationally-unstable
  molecular ring ($Q\lesssim 1$ and $f_{\rm gas}\approx50$ -- $70\%$).

\item The differential mass distribution functions of the clumps are
  best fitted by two power-laws with a turn-over or peak at
  $M_{\rm clump}\approx4000$~M$_\odot$.

\item The molecular structures of the central region have an unusually
  steep size -- linewidth relation
  $R_{\rm c}\propto\sigma_{\rm obs,los}^{0.82\pm0.11}$, while those of
  the molecular ring have a much shallower relation
  $R_{\rm c}\propto\sigma_{\rm obs,los}^{0.30\pm0.03}$ (with a
  flattening or turn-over at $R_{\rm c}\approx3$~pc). The latter is
  similar to the Kolmogorov law for turbulence
  ($\sigma_{\rm obs,los}\propto R_{\rm c}^{1/3}$;
  \citealt{kolmogorov1941}).
 
\item The molecular structures of the central region and the molecular
  ring have similar mass -- size relations (with power-law indices
  $D_{\rm m}=2.27\pm0.10$ and $D_{\rm m}=2.12\pm 0.01$,
  respectively). However, while in the central region, clumps and
  clouds have similar power-law indices ($D_{\rm m,clump}=2.07\pm0.16$
  versus $D_{\rm m,cloud}=2.22\pm0.10$), in the molecular ring, clumps
  have a much shallower power-law index than clouds
  ($D_{\rm m,clump}=1.63\pm0.04$ versus
  $D_{\rm m,cloud}=2.06\pm0.01$).

\item In the central region, both clumps (mean virial parameter
  $\langle\alpha_{\rm vir,clump}\rangle=1.52\pm0.11$) and clouds
  ($\langle\alpha_{\rm vir,cloud}\rangle=1.14\pm0.12$) appear to be in
  virial equilibria. In the molecular ring, however, while clumps are
  in rough virial equilibria
  ($\langle\alpha_{\rm vir,clump}\rangle=1.82\pm0.07$), clouds appear
  to be strongly gravitationally bound
  ($\langle\alpha_{\rm vir,cloud}\rangle=0.41\pm0.02$).
  The virial parameter of molecular structures in the molecular ring
  is in turn dependent on mass:
  $\alpha_{\rm vir,c}\propto M_{\rm c}^{-0.27\pm0.01}$.
\end{itemize}

We developed an analytical model of clump-clump collisions to explain
the clump properties and gas turbulence in the molecular ring. Our
model suggests that the collisions between clumps are driven by
gravitational instabilities coupled with galactic shear, that leads to
several results.

\begin{itemize}
  
\item The formation of clumps with
  $R_{\rm t,clump}\approx L_{\rm acc,clump}\approx\lambda_{\rm coll}$
  (where $\lambda_{\rm coll}\equiv G\Sigma_{\rm gas,disc}/2A^2$ is the
  critical collision length arising from our model), i.e.\ the tidal
  radii of clumps approximately equal to the average distance between
  neighbouring clumps.
 
\item A typical clump mass
  $M_{\rm clump}\approx\Sigma_{\rm gas,disc}\lambda_{\rm
    coll}^2\approx G^2\Sigma_{\rm gas,disc}^3/4A^4\equiv M_{\rm
    c,coll}$ and a typical clump size
  $R_{\rm clump}\approx0.38\lambda_{\rm coll}\equiv R_{\rm c,coll}$.

\item An energy injection rate (per unit mass) for the
  collision-induced turbulence
  $\dot{\epsilon}_{\rm inject,coll}\equiv G^2\Sigma_{\rm
    gas,disc}^2/2A\approx\dot{\epsilon}_{\rm diss,clump}$.

  
\item A size -- linewidth relation
  $\sigma_{\rm obs,los}\approx(2G^2\Sigma_{\rm
    gas,disc}^2/A)^{1/3}R_{\rm c}^{1/3}$ (assuming a Kolmogorov
  spectrum of turbulence), with a flattening at the turbulence driving
  scale $L_{\rm D}\approx\lambda_{\rm coll}$.
    
\item If the turbulence that supports the gas disc in vertical
  equilibrium is sustained by clump-clump collisions, a gas disc
  velocity dispersion
  $\sigma_{\rm max,coll}\equiv G\Sigma_{\rm
    gas,disc}/A\approx\sigma_{\rm gas,disc}$ and a gas disc scale
  height $h_{\rm gas,disc}\approx L_{\rm D}\approx\lambda_{\rm
    coll}$. Turbulence from clump-clump collisions can maintain the
  disc in a marginally gravitationally-stable state, i.e.\
  $Q\approx1$.
  
\item A virial parameter
  $\alpha_{\rm vir,clump}\approx\alpha_{\rm vir,crit}=2$ for clumps
  and $\alpha_{\rm vir,cloud}<1$ for clouds, as
  $M_{\rm clump}\approx M_{\rm Jeans}\approx M_{\rm c,coll}$ while
  $M_{\rm cloud}>M_{\rm Jeans}\approx M_{\rm
    c,coll}$. Collision-induced turbulence can thus maintain the
  clumps (but not the clouds) in rough virial equilibria.
 
\item A mass -- size relation for clumps
  $M_{\rm clump}=(5/\alpha_{\rm vir,crit}G)(2G^2\Sigma_{\rm
    gas,disc}^2/A)^{2/3}R_{\rm clump}^{5/3}$.
\end{itemize}

Our predictions above all match the observations very well in the
molecular ring of NGC~404, suggesting clump-clump collisions are the
dominant mechanism regulating clump properties and gas turbulence in
that region. As expected, the collision model fails to explain the
observations in the central region of NGC~404, where the gas disc is
strongly gravitationally stable ($Q=3$ -- $30$). It also seems that
clumps migrate inward from the molecular ring to the central region,
so that the clump properties do not change much between the two
regions. In turn, clump migration may be the major source of
turbulence in the central region, although other sources are possible
(e.g.\ AGN feedback and/or the magneto-rotational instability).

Our model may be relevant to the understanding of molecular gas discs
and star formation in high-$z$ disc galaxies, as the molecular ring of
NGC~404 resembles in many ways the star-forming gas discs observed at
high redshifts (e.g.\ clumpy morphology, low Toomre parameter and high
gas fraction). We note, however, that our findings are based on only
one galaxy. High-resolution observations of molecular gas in more
gravitationally-unstable gas discs are thus needed to further confirm
our model.

\section*{Acknowledgements} We thank the referee for providing
valuable comments that improved the paper. LL was supported by a
Hintze Fellowship, funded by the Hintze Family Charitable Foundation,
and by a DAWN Fellowship, funded by the Danish National Research
Foundation under grant No.\ 140. MB was supported by STFC consolidated
grant ``Astrophysics at Oxford'' ST/H002456/1 and ST/K00106X/1. GXL
acknowleges supports from NSFC grant W820301904 and 12033005. TAD
acknowledges support from a STFC Ernest Rutherford Fellowship. MDS
acknowledges support from a STFC DPhil studentship ST/N504233/1. This
publication arises from research funded by the John Fell Oxford
University Press Research Fund.

This paper makes use of the ALMA data. ALMA is a partnership of ESO
(representing its member states), NSF (USA) and NINS (Japan), together
with NRC (Canada) and NSC and ASIAA (Taiwan) and KASI (Republic of
Korea), in cooperation with the Republic of Chile. The Joint ALMA
Observatory is operated by ESO, AUI/NRAO and NAOJ. This paper also
makes use of observations made with the NASA/ESA Hubble Space
Telescope, and obtained from the Hubble Legacy Archive, which is a
collaboration between the Space Telescope Science Institute
(STScI/NASA), the Space Telescope European Coordinating Facility
(ST-ECF/ESA) and the Canadian Astronomy Data Centre
(CADC/NRC/CSA). This research has made use of the NASA/IPAC
Extragalactic Database (NED) which is operated by the Jet Propulsion
Laboratory, California Institute of Technology, under contract with
the National Aeronautics and Space Administration.

\section*{DATA AVALABILITY}
The data underlying this article are available in the ALMA archive
(\url{https://almascience.eso.org/asax/}) under project code: (i)
2015.1.00597.S, (ii) 2017.1.00572.S and (iii) 2017.1.00907.S.

\bibliographystyle{mnras}
\bibliography{references} 

\appendix
\section{Cumulative mass function of clumps}
\label{appendix:cumulative_mass_function}

The cumulative clump mass distribution function can be characterised
quantitatively by a power-law
\begin{equation}
  \label{eq:cumulative_mass_distribution_function}
  N_{\rm clump}(M^\prime_{\rm clump}>M_{\rm clump})
  =\left(\frac{M_{\rm clump}}{M_{0,{\rm clump}}}\right)^{\gamma_{\rm clump}+1}~,
\end{equation}
where $N_{\rm clump}(M^\prime_{\rm clump}>M_{\rm clump})$ is the
number of clumps with a mass greater than $M_{\rm clump}$,
$M_{0,{\rm clump}}$ sets the normalisation, and $\gamma_{\rm clump}$
is the power-law index.

\begin{figure*}
  \includegraphics[width=0.98\textwidth]{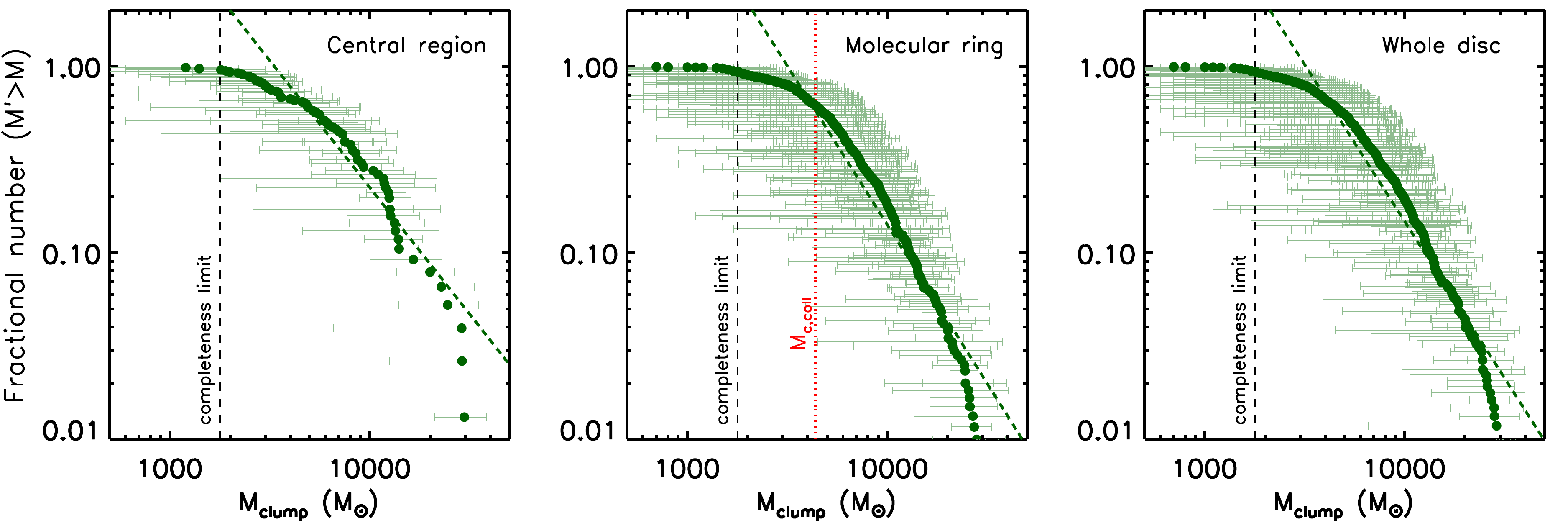}
  \caption{Normalised cumulative mass distribution function of the
    resolved clumps in the central region, molecular ring and whole
    disc of NGC~404, respectively. The power-law best fitting the
    cumulative mass distribution are overlaid as green dashed lines in
    each panel. Our mass completeness limit is indicated by a black
    vertical dashed line in each panel. The red dotted line in the
    middle panel indicates our model-predicted turn-over mass in the
    molecular ring ($M_{\rm c,coll}$; see
    Sections~\ref{sec:clump_mass_collision_model} and
    \ref{sec:clump_mass_function_collision_model} for more details). }
  \label{fig:clump_mass_distribution_cum}
\end{figure*}

Fig.~\ref{fig:clump_mass_distribution_cum} present the normalised
cumulative mass distribution function (and its power-law fit as a
dashed green line) of the resolved clumps in the the central region,
molecular ring and whole disc of NGC~404, respectively. The power-law
fits are only performed above the mass completeness limit
$M_{\rm comp}=1.8\times10^3$~M$_\odot$ (see Section\
\ref{sec:mass_function_clumps}).

We find power-laws fit the cumulative mass functions relatively well
in the mass regime $M_{\rm clump}\gtrsim4000$~M$_\odot$. The
best-fitting power-law index $\gamma_{\rm clump}$ of the cumulative mass
distribution function is $-2.36\pm0.10$, $-2.70\pm0.06$ and
$-2.67\pm0.07$ for the central region, molecular ring and whole disc,
respectively. These best-fitting power-law indexes are similar to
those of the differential mass distribution function in the high-mass regime
$\gamma^{+}_{\rm clump}$ ($-2.63\pm0.49$, $-2.87\pm0.13$ and
$-2.67\pm0.16$ for the central region, molecular ring and whole disc,
respectively; see Section\ \ref{sec:mass_function_clumps}).
 It thus seems that most of the molecular gas mass of
NGC~404 is located in low-mass clumps.

We note, however, that power-law functions only seem to fit well at
the high-mass ends of the cumulative clump mass distribution
functions, and \enquote{turn-overs}, that are break points in the
power-law functions, seem to be present in all the cumulative mass
distributions at the same clump mass
$M_{\rm clump}\approx4000$~M$_\odot$. Indeed, the cumulative mass
distribution functions show strong deviations from the best-fitting
power-laws (green dash lines in the 
Fig.~\ref{fig:clump_mass_distribution_cum}) below this turn-over mass.  As
this turn-over mass is much larger than the mass completeness limit,
these deviations are most likely real and probe (and thus inform on)
the underlying formation and destruction of clumps. The presence of
these turn-overs may in fact suggest less negative or even positive
power-law indices (i.e. ``slopes'') at the low-mass ends of the
cumulative clump mass distribution functions.

\section{Collision timescale}
\label{appendix:collision_timescale}

\begin{figure}
  \includegraphics[width=0.95\columnwidth]{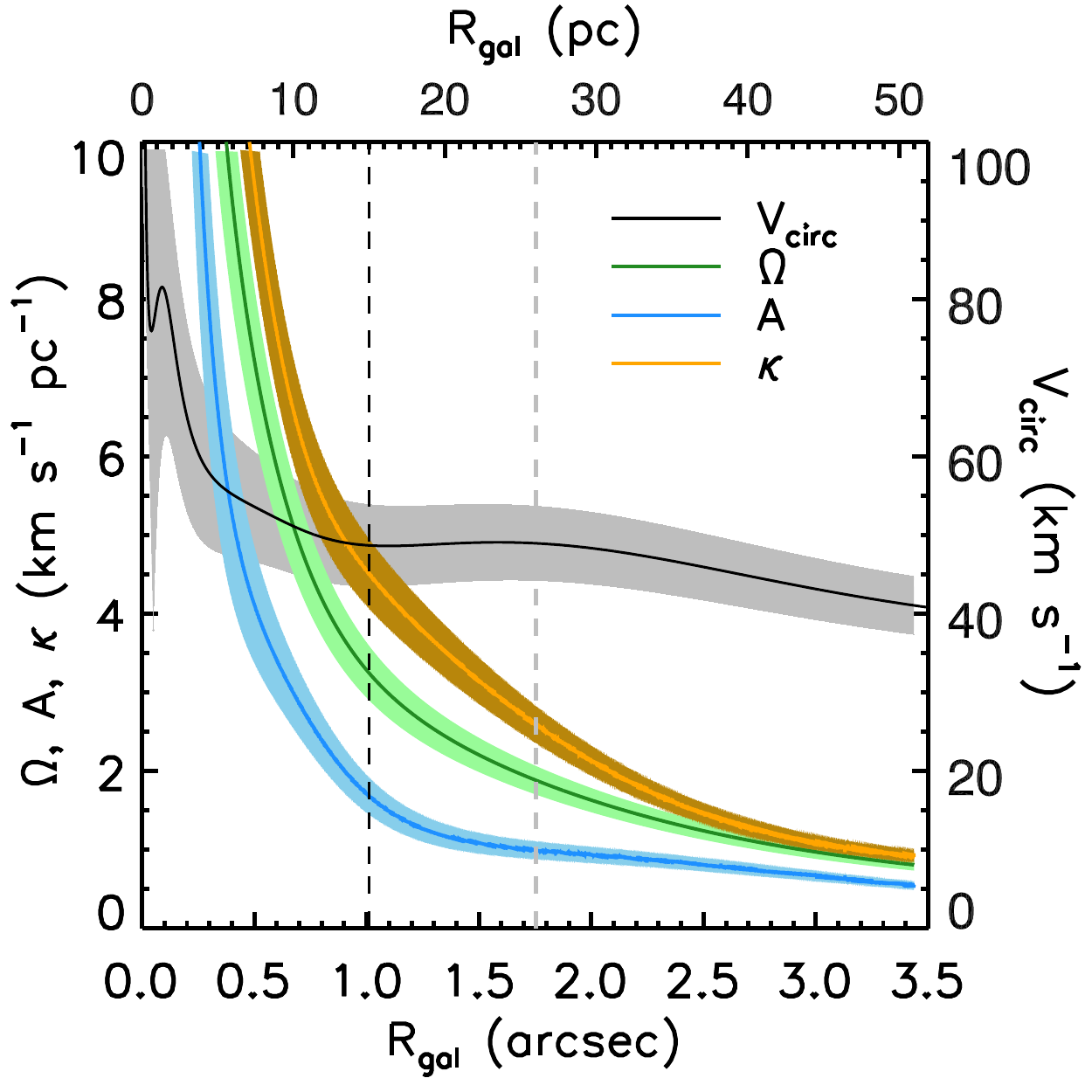}
  \caption{Galactocentric distance dependence of the orbital circular
    velocity $V_{\rm cir}$ (black curve), orbital circular angular
    velocity $\Omega$ (green curve), Oort's constants $A$ (blue curve)
    and epicyclic frequency $\kappa$ (orange curve) in NGC~404. The
    coloured envelope around each curve indicates the $\pm1$~$\sigma$
    uncertainties. The black vertical dashed line indicates the
    boundary ($R_{\rm gal}=15$~pc) between the central region and
    molecular ring, while the grey vertical dashed line indicates the
    galactocentric distance ($R_{\rm gal}=27$~pc) beyond which the
    molecular gas disc is no longer gravitationally stable (i.e.\
    Toomre parameter's $Q\le1$ at $R_{\rm gal}\ge27$~pc; see
    Section~\ref{sec:gravitational_instability_onset_collisions}). The
    circular velocity curve is nearly flat in the molecular ring.}
  \label{fig:a_vs_r}
\end{figure}

In this section, we derive the collision timescale of clumps from
first principles. Following \citet{tan2000}, we set the collision
velocity of clumps to be the shear velocity
\begin{equation}
  \label{eq:shear_velocity_app}
  \begin{split}
    v_{\rm shear} & \equiv v_{\rm shear}(R)=b\,(\Omega-\frac{dV_{\rm circ}}{dR})=2Ab
  \end{split}
\end{equation} 
(see Eq.~46 in \citealt{liu2021}), where
$A\equiv A(R)=-\frac{R}{2}\frac{d\Omega}{dR}$ is Oort's constant $A$
evaluated at the galactocentric distance $R$ of the clump
($R_{\rm gal}$ in Table~\ref{tab:gmc_properties}),
$\Omega\equiv\Omega(R)=V_{\rm circ}/R$ is the angular velocity of
orbital circular rotation, $V_{\rm circ}\equiv V_{\rm circ}(R)$ is the
circular velocity of the galaxy, and $b$ is the radial distance
between the orbits of the two colliding clumps. We note that the shear
velocity $v_{\rm shear}$ is derived using the shearing-sheet
approximation \citep[e.g.][]{binney2008}. Figure~\ref{fig:a_vs_r}
shows the dependence of $\Omega$ and Oort's constant $A$ on the
galactocentric distance $R$ in NGC~404.

We adopt the circular velocity curve of \citet{davis2020}, derived by
creating a gas dynamical model using the \texttt{Kinematic Molecular
  Simulation} (\texttt{KinMS}) package of \citet{davis2013}. Inputs to
the model include the stellar mass distribution, stellar mass-to-light
ratio, molecular gas mass, SMBH mass, as well as the disc orientation
(position angle and inclination) and position (spatially and
spectrally). The stellar mass distribution is parametrised by a
multi-Gaussian expansion (MGE; \citealt{emsellem1994,cappellari2002})
fit to {\it HST} images of the nucleus from \citet{nguyen2017}, while
the molecular gas mass distribution is parametrised by a MGE fit to
the ALMA CO(2-1) image of \citet{davis2020}. We use stellar
mass-to-light ratios calculated on a pixel-by-pixel basis from
multi-band imaging by \citet{davis2020}. The MGE model of the
molecular gas does not account for the flocculent sub-structure of the
molecular gas disc, but it does allow to quantify the contribution to
the potential of an ideal axisymmetric approximation of the observed
molecular gas disc \citep{davis2020}. The SMBH mass
($M_{\rm BH}\approx5.7\times10^5$~M$_\odot$), position angle
(${\rm PA}=37\fdg2$ at $R_{\rm gal}<15$~pc and ${\rm PA}=1^\circ$ at
$R_{\rm gal}\ge15$~pc) and inclination ($i=37\fdg1$ at
$R_{\rm gal}<15$~pc and $ i=9\fdg3$ at $R_{\rm gal}\ge15$~pc) were
then obtained from the best fit to the kinematics of the CO gas. Full
details of the fitting procedure can be found in \citet{davis2020}.

We set the radius of the effective collision cross section of a clump
to be its tidal radius $R_{\rm t}$ rather than its actual radius
$R_{\rm clump}$, as we consider clump-clump collisions to be any
mutual gravitational interaction and ultimate merging of clumps
(rather than exclusively physical collisions). We adopt the definition
of tidal radius from \citet{gammie1991} and \citet{tan2000}, whereby
the tidal radius marks the radial distance from a clump's centre at
which the shear velocity of the clump due to differential galactic
rotation is equal to its escape velocity.
%
This leads to
\begin{equation}
  \label{eq:tidal_radius_app}
  R_{\rm t}=(1-\beta_{\rm circ})^{-2/3}\left(\frac{2M_{\rm c}}{M_{\rm gal}}\right)^{1/3}R
\end{equation}
(see Eq.~8 in \citealt{tan2000}), where as before $M_{\rm c}$ is the
clump's mass,
$\beta_{\rm circ}\equiv\beta_{\rm circ}(R)=\frac{d\ln V_{\rm
    circ}}{d\ln R}$, and $M_{\rm gal}\equiv M_{\rm gal}(R)$ is the
total galaxy mass interior to $R$. Equation~\ref{eq:tidal_radius_app}
assumes a spherical mass distribution, i.e.\
$M_{\rm gal}(R)=V_{\rm circ}^2(R)R/G$, and can therefore be simplified
to
\begin{equation}
  \label{eq:tidal_radius2_app}
  R_{\rm t}=\left(\frac{G}{2A^2}\right)^{1/3}M_{\rm c}^{1/3}~.
\end{equation}
The tidal radius defined in this manner is the maximum size of a
gravitationally-bound cloud (of a given mass $M_{\rm c}$) allowed by
galactic rotational shear (quantified by $A$).
Normally, the tidal radius of a self-gravitating object is at least a
few times larger than its actual radius. We note that in our
formalism, a collision occurs between two clumps only when the radial
distance between their orbits $b$ is smaller than their tidal radius
$R_{\rm t}$.

Using the above-defined collision velocity (i.e.\ the shear velocity
$v_{\rm shear}$) and collision cross section (i.e.\ the tidal radius
$R_{\rm t}$), we can derive a clump-clump (or cloud-cloud) collision
rate
\begin{equation}
  \label{eq:collision_rate_tmp_app}
  \begin{split}
    Z_{\rm coll} \equiv Z_{\rm coll}(R) & =2\int^{R_{\rm t}}_0z_{\rm coll}\,db\\
    & =2\int^{R_{\rm t}}_0N_{\rm A}\,v_{\rm shear}\,db\\\
    & =4A\int^{R_{\rm t}}_0N_{\rm A}\,b\,db~,
  \end{split}
\end{equation}
where $z_{\rm coll}\equiv z_{\rm coll}(b)=N_{\rm A}\,v_{\rm shear}$ is
the collision rate per unit length,
$N_{\rm A}\equiv N_{\rm A}(b)=\Sigma_{\rm gas,disc}/M_{\rm c}$ is the
number surface density of molecular structures,
$\Sigma_{\rm gas,disc}\equiv\Sigma_{\rm gas,disc}(b)$ is the
coarse-grained gaseous mass surface density of the disc, and
$v_{\rm shear}$ is taken from Eq.~\ref{eq:shear_velocity_app}. It is
reasonable to define the number surface density of clumps $N_{\rm A}$
in this manner as we have assumed collisions occur only between clumps
of equal mass. The first factor of $2$ in
Eq.~\ref{eq:collision_rate_tmp_app} accounts for clumps/clouds either
catching up with others clumps/clouds at larger $R_{\rm gal}$ or being
caught up by other clumps/clouds at smaller $R_{\rm gal}$
\citep{tan2000}. If we assume clumps are approximately uniformly
distributed over a region of radius $\approx R_{\rm t}$ centered on
the clump, then $N_{\rm A}$ in Eq.~\ref{eq:collision_rate_tmp_app} (or
equivalently $\Sigma_{\rm gas,disc}$) is approximately constant, i.e.\
$N_{\rm A}(b)\approx N_{\rm A}(R)=\frac{1}{M_{\rm c}}\Sigma_{\rm
  gas,disc}(R)$, where $\Sigma_{\rm gas,disc}(R)$ is the
coarse-grained gaseous mass surface density of the disc evaluated at
the centre (galactocentric distance) of the reference clump
($R_{\rm gal}$ in Table~\ref{tab:gmc_properties}). This leads to
\begin{equation}
  \label{eq:collision_rate_app}
  \begin{split}
    Z_{\rm coll} & \approx4AN_{\rm A}\int^{R_{\rm t}}_0b\,db\\
    & \approx2AN_{\rm A}R_{\rm t}^2~.
  \end{split}
\end{equation}

\begin{figure}
  \includegraphics[width=0.95\columnwidth]{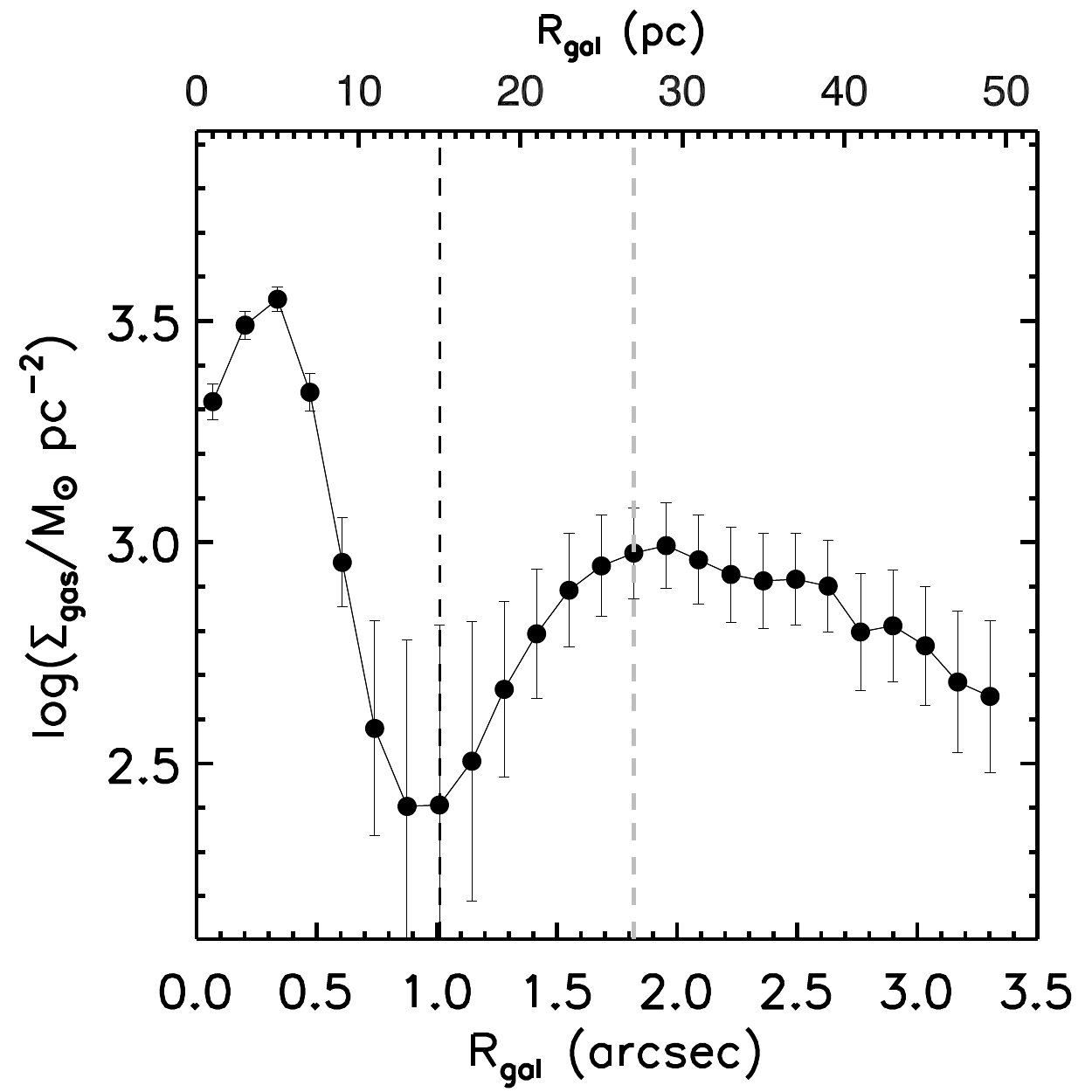}
  \caption{Galactocentric distance dependence of the coarse-grained
    deprojected molecular gas mass surface density of the NGC~404
    disc, with a radial bin size of $2$~pc. The error bars indicate
    the $1$~$\sigma$ scatter of the different mass surface densities
    within each radial bin (not the uncertainty on the mean within
    each bin, that is much smaller). The black vertical dashed line
    indicates the boundary ($R_{\rm gal}=15$~pc) between the central
    region and molecular ring, while the grey vertical dashed line
    indicates the galactocentric distance ($R_{\rm gal}=27$~pc) beyond
    which the molecular gas disc is no longer gravitationally stable
    (i.e.\ Toomre parameter's $Q\le1$ at $R_{\rm gal}\ge27$~pc; see
    Section~\ref{sec:gravitational_instability_onset_collisions}). }
\label{fig:Sigma_gas_vs_r}
\end{figure}

The clump-clump (or cloud-cloud) collision timescale is then
\begin{equation}
  \label{eq:collision_timescale_app}
  t_{\rm coll}\equiv t_{\rm coll}(R)=1/Z_{\rm coll}\approx\frac{1}{2AN_{\rm A}R_{\rm t}^2}\approx\frac{A^{1/3}M_{\rm c}^{1/3}}{2^{1/3}G^{2/3}\Sigma_{\rm gas,disc}}~,
\end{equation}
where the last expression assumes a spherical galaxy mass distribution
(via Eq.~\ref{eq:tidal_radius2_app}). Figure~\ref{fig:Sigma_gas_vs_r}
shows the deprojected molecular gas mass surface density
$\Sigma_{\rm gas,disc}$ as a function of the galactocentric distance
$R$ (i.e.\ $R_{\rm gal}$) in NGC~404. The mass surface density
$\Sigma_{\rm gas,disc}$ is computed by summing the molecular gas mass
within galactocentric annuli of increasing $R_{\rm gal}$ (with a
radial bin size of $2$~pc), corrected for inclination (where a fixed
position angle $PA=1^\circ$ and inclination angle $i=9\fdg3$ were
adopted). We note that our derived collision timescale above is
essentially the same as that derived by \citeauthor{tan2000}
(\citeyear{tan2000}; see their Eq.~13), except that we adopted
$R_{\rm t}$ rather than $1.6\,R_{\rm t}$ as the radius of the
effective collision cross section. We use the subscript \enquote{c}
when referring to the masses of the molecular structures in
Eqs.~\ref{eq:tidal_radius_app} -- \ref{eq:collision_timescale}, as the
equations apply equally to clump-clump collisions and cloud-cloud
collisions.

\bsp	
\label{lastpage}
\end{document}